\documentclass[a4paper,12pt]{article}

\usepackage{amsmath}
\usepackage[psamsfonts]{amssymb}
\usepackage{rsfs}
\usepackage{bm}

\usepackage{cite}
\usepackage[dvips]{graphicx}
\usepackage{color}

\makeatletter
\@addtoreset{equation}{section}
\makeatother

\begin{document} 

\begin{titlepage}

\baselineskip 10pt
\hrule 
\vskip 5pt
\leftline{}
\leftline{Chiba Univ. Preprint
          \hfill   \small \hbox{\bf CHIBA-EP-167}}
\leftline{\hfill   \small \hbox{July 2008}}
\vskip 5pt
\baselineskip 14pt
\hrule 
\vskip 1.0cm
\centerline{\Large\bf 
Reformulating $SU(N)$ Yang-Mills theory
} 
\vskip 0.3cm
\centerline{\Large\bf  
based on change of variables
}
\vskip 0.3cm
\centerline{\Large\bf  
}
\vskip 0.3cm
\centerline{\large\bf  
}

\vskip 0.5cm

\centerline{{\bf 
Kei-Ichi Kondo,$^{\dagger,{1}}$  
Toru Shinohara,$^{\ddagger,{2}}$
and  Takeharu Murakami$^{\ddagger,{3}}$
}}  
\vskip 0.5cm
\centerline{\it
${}^{\dagger}$Department of Physics, Graduate School of Science, 
Chiba University, Chiba 263-8522, Japan
}
\vskip 0.3cm
\centerline{\it
${}^{\ddagger}$Graduate School of Science and Technology, 
Chiba University, Chiba 263-8522, Japan
}
\vskip 1cm

\begin{abstract}
We propose a new version of $SU(N)$ Yang-Mills theory  reformulated in terms of new field variables which are obtained by a nonlinear change of variables from the original Yang-Mills gauge field. 
The reformulated Yang-Mills theory enables us to study the low-energy dynamics by explicitly extracting the topological degrees of freedom such as magnetic monopoles and vortices to clarify the mechanism for quark confinement. 
 The dual superconductivity in Yang-Mills theory is understood in a gauge-invariant manner,  as demonstrated recently by  a non-Abelian Stokes theorem for the Wilson loop operator, although the basic idea of this reformulation is based on the Cho-Faddeev-Niemi decomposition of the gauge potential. 
\end{abstract}

 
\vskip 0.5cm

\hrule  
\vskip 0.1cm
${}^1$ 
  E-mail:  {\tt kondok@faculty.chiba-u.jp}
  
${}^2$ 
  E-mail:  {\tt sinohara@graduate.chiba-u.jp}

${}^3$ 
  E-mail:  {\tt tom@fullmoon.sakura.ne.jp}

\par 
\par\noindent


\vskip 0.5cm

\newpage
\pagenumbering{roman}
\tableofcontents




\end{titlepage}


\pagenumbering{arabic}

\baselineskip 14pt
\section{Introduction}

\par
The original Yang-Mills theory \cite{YM54} is formulated in terms of the Yang-Mills gauge field.  This formulation is suitable for studying the high-energy dynamics of   Yang-Mills theory. For instance, it is well known that the perturbation theory in the coupling constant developed in terms of the Yang-Mills field is very powerful in the ultraviolet region due to its asymptotic freedom. 
In the low-energy region, however, we encounter the strong coupling problem,  and its description in terms of the Yang-Mills gauge field is no longer valid. 
In studying the low-energy dynamics of  Yang-Mills theory and quantum chromodynamics (QCD), it is important to specify the most relevant degrees of freedom for the  phenomenon in question. 
Quark confinement as a typical phenomenon in the low-energy dynamics caused by strong interactions is believed to be explained by topological defects including magnetic monopoles, vortices and merons.
This motivates us to devise another formulation of  Yang-Mills theory in terms of new variables reflecting the topological degrees of freedom.

For quark confinement, magnetic monopoles and center vortices are believed to be the dominant topological degrees of freedom.  However, Abelian magnetic monopoles \cite{tHooft81} or center vortices  in $SU(N)$ Yang-Mills theory have been obtained as gauge--fixing defects by partial gauge fixings called the maximal Abelian gauge \cite{tHooft81,KLSW87} $SU(N) \rightarrow U(1)^{N-1}$ or the maximal center gauge \cite{center} $SU(N) \rightarrow Z(N)$,  where $U(1)^{N-1}$ is the maximal torus (Cartan) subgroup and $Z(N)$ is the center subgroup of $SU(N)$.   
Therefore, the current method of constructing Abelian magnetic monopoles and center vortices cannot avoid charge of the gauge artifact.

In view of this, we take up the Cho-Faddeev-Niemi-Shabanov (CFNS) decomposition of the Yang-Mills field, which was first proposed in 1981 by Cho \cite{Cho80} and has been recently readdressed by Faddeev and Niemi \cite{FN98} and Shabanov \cite{Shabanov99}.  This machinery enables one to explain and understand some of the low-energy phenomena by separating the contributions due to topological defects in a {\it gauge-invariant manner}. 
This feature should be compared with other methods.  

By developing the approach founded in CFNS \cite{Cho80,FN98,Shabanov99}, we have succeeded in giving  a gauge-invariant description of the dual superconductivity in $SU(2)$ Yang-Mills theory in a continuum \cite{KMS06,Kondo04,Kondo06} and on a lattice \cite{KKMSSI05,IKKMSS06,SKKMSI07}.  
In particular, the Wilson loop operator \cite{Wilson74} is expressed exactly in terms of the gauge-invariant magnetic current \cite{Kondo08}, and the magnetic monopole can be defined in a gauge-invariant manner, which follows from a non-Abelian Stokes theorem for the Wilson loop operator, see \cite{KondoIV,KT99,Kondo08} and references therein. 
These results enable us to determine the role of a magnetic monopole in confinement. 

For the $SU(2)$ gauge group, the CFNS decomposition of the Yang-Mills field and the reformulation of  Yang-Mills theory in terms of the resulting new variables are essentially unique.  Therefore, we have nothing new to be added for $SU(2)$ gauge group. 
It is important to note that a unit vector field $\bm{n}(x)$ called the color field hereafter plays the key role. 

For the $SU(N)$ gauge group, the conventional approaches \cite{Cho80c,FN99a,FN99b,Shabanov02} introduce $r$ unit vector fields $\bm{n}_j(x)$ from the beginning, where $r=N-1$ is the rank of the $SU(N)$ group.  These $r$ color fields are used to define $r$ gauge-invariant magnetic monopoles. This feature should be compared with the  Abelian projection in which $r$ Abelian magnetic monopoles corresponding to every $U(1)$ of the maximal torus subgroup $U(1)^{r}$ are generated due to the partial gauge fixing $SU(N) \rightarrow U(1)^{r}$.
In this paper we  point out that such an approach is sufficient, but not necessary for reformulating  Yang-Mills theory and that there are other options.  In fact, we explicitly construct an option called the {\it minimal case}.
In connection to the minimal case, we call the conventional option [option corresponding to the conventional case] the {\it maximal case}.  
Even in the maximal case, we show that there are new options that yield the same conclusion. 

The conventional approach suggests that $r$ magnetic monopoles can be the dominant topological degrees of freedom for confinement.  
However, this is not the inevitable conclusion.  
As conjectured in \cite{KT99}, it is shown \cite{Kondo08}  using a non-Abelian Stokes theorem that  the $SU(N)$ Wilson loop operator can be rewritten in terms of the gauge-invariant magnetic current corresponding to a {\it single non-Abelian magnetic monopole} \cite{WY07} irrespective of $N$. 
Therefore, the Wilson loop operator does not require $r$ gauge-dependent Abelian and gauge-invariant Abelian-like magnetic monopoles.  Rather, the Wilson loop operator can probe only a single non-Abelian magnetic monopole. 
From this viewpoint, the minimal option is the most economical way of reformulating Yang-Mills theory, 
where a single unit color field $\bm{h}$ is introduced and the gauge-invariant magnetic monopole constructed in this option agrees exactly with the magnetic monopole derived from the Wilson loop operator through a non-Abelian Stokes theorem. 
The lattice versions of new reformulations for the $SU(N)$ gauge group given in this paper are constructed in \cite{KSSMKI08,SKKMSI07b}, which enable us to perform a numerical simulation on the lattice to study the nonperturbative phenomena in question. 

This paper is organized as follows. 
In section  2, we discuss how to construct the color field for $SU(N)$ gauge group.  In this construction we introduce the maximal and minimal options for $SU(N)$, $N \ge 3$.  
The distinction originates from the degeneracy associated with the diagonalization of the scalar field introduced for determining the color field, which undergoes an adjoint transformation under the gauge transformation. 
We examine how many degrees of freedom the color field carries from the viewpoint of the change of variables. 
Moreover, we point out that there are intermediate options other than the maximal and minimal options for $SU(N)$, $N \ge 4$. 

In section 3, 
we show how such a color field defined in section 2 is realized as a functional of the original Yang-Mills field for the maximal and minimal options.  
We obtain the defining equations, which enable us to specify the new variables as functionals of the original Yang-Mills field. This is important for identifying the color field as one of the new variables obtained by the change of variables from the original Yang-Mills field. 
We give a prescription called the reduction condition for obtaining the color field as a functional of the original Yang-Mills field.

In section 4, we discuss the physically interesting $SU(3)$ case to  explicitly demonstrate the general procedure given in the previous section. 

In section 5, we consider a quantum version of the reformulated Yang-Mills theory based on a functional integral method. 
The key issue is to obtain the Jacobian associated with the change of variables from the original Yang-Mills field to the new variables. 

The final section is devoted to a conclusion and discussion. Some of the technical content is summarized in the Appendices. 

\section{Construction of the color vector field}

In the conventional approach, the magnetic monopole in Yang-Mills theory  is extracted using the {\it Abelian projection} \cite{tHooft81}.  The Abelian projection selects a special direction in color space.  Consequently,  partial gauge fixing, e.g., the maximal Abelian (MA) gauge \cite{tHooft81,KLSW87} used so far for realizing the  Abelian projection, breaks the color symmetry explicitly. 
In our reformulation of   Yang-Mills theory, the color field ${\bf n}(x)$ plays the role of recovering the color symmetry that is lost in the conventional treatment of the Abelian projection, in which the maximal torus subgroup is selected from the original gauge group $G$.

First of all, we consider how to define the {\it color field} ${\bf n}(x)$ in $SU(N)$ Yang-Mills gauge theory.  The color field ${\bf n}(x)$ is used to specify only the color direction in the color space at each space-time point $x \in \mathbb{R}^D$ in $D$-dimensional space-time, and its magnitude is irrelevant for our purposes.  
Therefore, the  color field ${\bf n}(x)$ is introduced as a unit field;  its  length is equal to one, i.e., ${\bf n}(x) \cdot {\bf n}(x)=1$.

To express the color vector, we use two different notations in this paper: 
\\
\noindent
\underline{the vector form}: 
\begin{equation}
   {\bf n}(x)=(n_A(x))_{A=1}^{d}=(n_1(x),n_2(x), \cdots, n_d(x)) ,
  \quad
  d\equiv  {\rm dim}SU(N)=N^2-1 
,
\end{equation}
where $d$ is the dimension of the gauge group $G=SU(N)$.
\\
\noindent
\underline{Lie algebra form}: 
\begin{equation}
  \bm{n}(x)=n_A(x) T_A  \ (A=1,2, \cdots, d)
, 
\end{equation}
where $T_A$ are the generators  of the Lie algebra $\mathscr{G}=su(N)$ of the Lie group $G=SU(N)$.
The two notations are equivalent, since the color field is constructed in such a way that it transforms according to the adjoint representation under the action of the gauge group. 
Hereafter, we use these notations for other fields that transform according to the adjoint representation under the action of the gauge group, e.g.,
\begin{align}
\vec{\phi}(x) =& (\phi_A(x))_{A=1}^{d}=(\phi_1(x),\phi_2(x), \cdots, \phi_d(x)) ,
  \\
  \bm{\phi}(x)=& \phi_A(x) T_A  \ (A=1,2, \cdots, d)
 .
\end{align}
It should be understood that $T_A$ denotes the generator in the fundamental representation, unless otherwise stated.

The product of two Lie algebra valued functions, $\mathscr X=\mathscr X^A T^A$ and $\mathscr Y=\mathscr Y^A T^A$, is rewritten as the sum of three types of products, which is written in the vector form as 
\begin{equation}
  \mathscr X  \mathscr Y = \mathscr X^A \mathscr Y^B  T^A T^B 
  = \frac{1}{2N} (\mathbf X \cdot \mathbf Y) {\bf 1}+ \frac12 i  \mathbf X \times \mathbf Y + \frac12  \mathbf X * \mathbf Y 
 ,
\end{equation}
where the three types of products are defined by 
\begin{subequations}
\begin{align}
\mathbf X \cdot \mathbf Y
 :=& X^AY^A,
\\
\mathbf X\times\mathbf Y
 :=& f^{ABC}X^BY^C,
\\
\mathbf X*\mathbf Y
 :=& d^{ABC}X^BY^C
 .
\end{align}
\end{subequations}
Here the structure constants $f^{ABC}$ of the Lie algebra $su(N)$ are defined by 
\begin{equation}
 f^{ABC}=-2i{\rm Tr}([T^A,T^B]T^C)  , 
 \quad (A,B,C\in\{1,2,\ldots N^2-1\})
\end{equation}
from the commutators among the generators:
\begin{equation}
[T^A,T^B]=if^{ABC}T^C
 .
\end{equation}
On the other hand, the anticommutator of the generators:
\begin{equation}
\{T_A,T_B\}
 =\frac1N \delta^{AB}\bm 1
  +d^{ABC}T_C   
 ,
 \label{anti-com}
\end{equation}
defines  completely symmetric symbols:
\begin{equation}
d^{ABC}=2{\rm Tr}(\{T^A,T^B\}T^C)
  .
  \label{d-def}
\end{equation}
In this paper we adopt the following normalization for the generators of $su(N)$:
\begin{equation}
{\rm Tr}(T^AT^B)=\frac12 \delta^{AB} . 
\quad (A,B \in\{1,2,\ldots N^2-1\})
\end{equation}
In other words, the  Lie algebra is closed under the products $\cdot, \times$ and $*$.

\subsection{$SU(2)$ case}

First of all, we consider the $SU(2)$ case as an introductory problem.  
For $SU(2)$,  a Hermitian matrix 
\begin{equation}
 \bm{\phi}(x)=\phi_A(x) T_A  , \quad T_A=\frac12 \sigma_A , \ ( A=1, 2,  3)
\end{equation}
with 2 by 2 Pauli matrices $\sigma_A$ can be cast into the diagonal form  using a suitable unitary matrix $U \in SU(2) (\subset U(2))$ as
\begin{equation}
  U(x) \bm{\phi}(x) U^\dagger(x) = {\rm diag}(\Lambda_1(x),\Lambda_2(x)) :=\bm \Lambda(x)
 .
\end{equation}
The eigenvalues $\Lambda_1$ and $\Lambda_2$ must be real due to the Hermiticity of $\bm{\phi}$, i.e., $\bm{\phi}^\dagger=\bm{\phi}$. 
The traceless condition ${\rm tr}(\bm{\phi})={\rm tr}(\bm \Lambda)=\Lambda_1+\Lambda_2=0$ yields
$\Lambda_1=-\Lambda_2$.  Hence, $\bm{\phi}(x)$ is written as 
\begin{equation}
   \bm{\phi}(x)  = U^\dagger(x) {\rm diag}(\Lambda_1(x),-\Lambda_1(x))U(x) 
   = \Lambda_1(x) U^\dagger(x) \sigma_3 U(x) 
 .
\end{equation}
For the explicit representation of the matrix
\begin{equation}
 \bm{\phi}(x)=\phi_A(x) T_A  
 = \frac12 \begin{pmatrix}
  \phi_3(x) & \phi_1(x)-i\phi_2(x) \cr
  \phi_1(x)+i\phi_2(x) & -\phi_3(x) \cr
  \end{pmatrix}
,  
\end{equation}
its eigenvalues are easily obtained as 
\begin{equation} 
\Lambda_1(x),\Lambda_2(x) = \pm \frac12  \sqrt{\phi_1(x)^2+\phi_2(x)^2+\phi_3(x)^2} 
= \pm \frac12 \sqrt{ \vec{\phi}(x) \cdot \vec{\phi}(x)}
:= \pm \frac12 |\vec{\phi}(x)| 
,
\end{equation}
since
$\det(\bm{\phi}-\lambda {\bf 1}) = \lambda^2 - \frac14 (\phi_1^2+\phi_2^2+\phi_3^2)$.

Imposing the condition of unit length $|\phi(x)|=1$, we obtain $\Lambda_1(x)=\pm 1/2$, and therefore a unit color field $\bm{n}(x)$ is obtained in the form 
\begin{equation}
  \bm{n}(x)= \pm U^\dagger(x) T_3 U(x)  
 .
 \label{aor}
\end{equation}
The unit color field $\bm{n}(x)$ is also expressed as
\begin{equation}
  \bm{n}(x) = \bm{\phi}(x)/|\bm{\phi}(x)| , \quad 
  |\bm{\phi}(x)| := \sqrt{2{\rm tr}(\bm{\phi}(x)^2)} = \sqrt{\vec{\phi}(x) \cdot \vec{\phi}(x)} 
 .  
\end{equation}
Note that $\bm{\phi}(x)$ is invariant under the $U(1)$ local transformation in the sense that for $U_\theta(x)=e^{iT_3 \theta}U(x)$, 
\begin{align}
  U_\theta(x) \bm{\phi}(x) U_\theta^\dagger(x) 
  = \Lambda_1(x) e^{iT_3 \theta}  \sigma_3 e^{-iT_3 \theta} 
  =  \Lambda_1(x)  \sigma_3 
 .
\end{align}
Here we consider the Weyl group as a discrete subgroup of SU(2), i.e., the permutation group $\mathscr{P}_2$ with 2 elements corresponding to the permutations of the 2 bases in the fundamental representation.
An element of the Weyl group is expressed as a constant (or uniform) off-diagonal matrix, e.g., 
\begin{align}
 W 
=& \exp \left[ i\pi \left( \frac{\sigma_1}{2} \cos \varphi + \frac{\sigma_2}{2} \sin \varphi \right) \right] 
\nonumber\\
=& i\left(  \sigma_1  \cos \varphi +  \sigma_2 \sin \varphi \right)
\nonumber\\
=& i \begin{pmatrix}
    0 & e^{-i\varphi} \cr
    e^{+i\varphi}  & 0 \cr
\end{pmatrix}
 \in \mathscr{P}_2 \subset SU(2) 
 .
\end{align}
Therefore, $\bm{\phi}(x)$ changes its sign under the Weyl transformation: For $U_W(x)=WU(x)$,  we have 
\begin{align}
  U_W(x) \bm{\phi}(x) U_W^\dagger(x) 
  = \Lambda_1(x) W \sigma_3 W^\dagger
  = - \Lambda_1(x)  \sigma_3 
 .
\end{align}
Changing the sign of $\Lambda(x)$ globally corresponds to changing the order of the  diagonal components $\Lambda_k(x)$ globally.
We can fix the discrete global symmetry for the Weyl group by imposing an additional condition such as $\Lambda(x) \ge 0$ or the ordering condition $\Lambda_1(x) \ge \Lambda_2(x)$.

Now we consider the situation in which the two eigenvalues become degenerate at a certain space-time point $x_0 \in \mathbb{R}^D$, i.e., $\Lambda_1(x_0)=\Lambda_2(x_0)$.  This is realized only when $\Lambda_1(x_0)=\Lambda_2(x_0)=0$.  Therefore, the three conditions 
$\phi_1(x_0)=\phi_2(x_0)=\phi_3(x_0)=0$
must be satisfied simultaneously at the degenerate point $x_0$. Hence, a set of degenerate points forms a 0-dimensional (pointlike) manifold in $\mathbb{R}^3$ or a one-dimensional (linelike) manifold in $\mathbb{R}^4$.  
Note that the conditions, $\phi_1(x_0)=\phi_2(x_0)=\phi_3(x_0)=0$,
are $SU(2)$ rotation-invariant. 
The rotation-invariant condition $\bm{\phi}(x)=0$ for $\bm{\phi}(x)$ can be regarded as the singularity condition for the unit vector field $\bm{n}(x)= \bm{\phi}(x)/|\bm{\phi}(x)|$. 
Therefore, the degenerate point of $\bm{\phi}(x)$ appears as the singular point of $\bm{n}(x)$.
In fact,  it is known that a magnetic monopole in Yang-Mills theory can appear as the hedgehog configuration at such a singular point, see, for example, \cite{Kondo08}.

For $G=SU(2)$, the color field is given by the  three-dimensional unit vector  
\begin{equation}
  {\bf n}(x)=(n_1(x),n_2(x),n_3(x))  
 .
\end{equation}
The unit vector ${\bf n}(x)=(n_1(x),n_2(x),n_3(x))$ specifies a two-sphere $S^2$ at $x$, which is isomorphic to $SU(2)/U(1)$, since $n_1(x)^2+n_2(x)^2+n_3(x)^2=1$. 
The maximal torus group of $SU(2)$ is $U(1)$ and the  $U(1)$ local rotation corresponds to the local rotation around the direction of the color  field ${\bf n}(x)$, which does not change (the direction of) the color vector ${\bf n}(x)$ at $x$. Here, by the local rotation, we mean that the angle of rotation is arbitrary at each spacetime point. 
This enables us to decompose the group $SU(2)_{\rm local}$ as
\begin{equation}
 SU(2)_{\rm local}=[SU(2)/U(1)]_{\rm local}\times U(1)_{\rm local} 
 .
\end{equation}
With shown in the above, the color field in the $SU(2)$ case can be constructed using the so-called adjoint orbit representation:
\begin{equation}
  \bm{n}(x)=U^\dagger(x) T_3 U(x)  , \quad U(x) \in SU(2)
 .
 \label{aor2}
\end{equation}
It is easy to check for example using the Euler-angle representation of the group element $U(x)$ that the diagonal $U(1)$ part of the group element $U(x)$ does not affect the right-hand side of the adjoint orbit representation.  Therefore, $\bm{n}(x)$ defined in this way indeed reflects the $[SU(2)/U(1)]_{\rm local}$ symmetry. 
The composition of $U(x) \rightarrow U(x)U_\omega(x)$ yields the adjoint local rotation 
\begin{equation}
  \bm{n}(x) \rightarrow \bm{n}_\omega(x)=U_\omega^\dagger(x) U^\dagger(x) T_3 U(x)U_\omega(x)  =U_\omega^\dagger(x) \bm{n}(x) U_\omega(x)  
 ,
\end{equation}
whose infinitesimal form is given for $U_\omega(x)=\exp(ig\bm{\omega}(x))=\exp(igT_A \omega_A(x))$ by
\begin{equation}
  \delta_\omega  \bm{n}(x) := \bm{n}_\omega(x) - \bm{n}(x) = ig[\bm{n}(x),\bm{\omega}(x)]
 .
\end{equation}
This is indeed the desired transformation property for the color field under the local rotation.

The conventional Abelian projection in the $SU(2)$ case corresponds to taking the special limit of the color  field; the color field is chosen to be uniform in the third direction at whole space-time points:
\begin{equation}
 {\bf n}(x) \equiv {\bf n}_{\infty}:=(0,0,1) 
\Longleftrightarrow \bm{n}(x) \equiv \bm{n}_{\infty} :=T_3 
 .
\end{equation}
Even in this limit, we have still  the $U(1)$ local symmetry for the $U(1)$ local rotation  around the uniform color  field 
${\bf n}_{\infty}:=(0,0,1)$, which does not change  the uniform color vector.  This limit is identified with the partial gauge fixing corresponding to the Abelian projection   
\begin{equation}
 SU(2)_{\rm local} \rightarrow U(1)_{\rm local} 
 .
\end{equation}

\subsection{$SU(3)$ case}

We now consider the $SU(3)$ case. 
We consider a real scalar field $\bm{\phi}(x)$ taking its value in the Lie algebra $su(3)$ of $SU(3)$:
\begin{equation}
 \bm{\phi}(x)=\phi_A(x) T_A  , \quad \phi_A(x) \in \mathbb{R}  . 
\ ( A=1, 2, \cdots, 8 )
\end{equation}
Thus, $\bm{\phi}$ is a traceless and Hermitian matrix for real $\phi_A$, i.e., ${\rm tr}(\bm{\phi})=0$ and $\bm{\phi}^\dagger=\bm{\phi}$, since we have chosen   the traceless generators to be Hermitian; ${\rm tr}(T_A)=0$ and $(T_A)^\dagger=T_A$.   The Hermitian matrix $\bm{\phi}$ can be cast into the diagonal form $\bm\Lambda$ using a suitable unitary matrix $U \in SU(3)( \subset U(3))$:
\begin{equation}
  U(x) \bm{\phi}(x) U^\dagger(x) = {\rm diag}(\Lambda_1(x),\Lambda_2(x),\Lambda_3(x)) := \bm{\Lambda}(x)
 ,
 \label{diag}
\end{equation}
with real elements $\Lambda_1,\Lambda_2,\Lambda_3$.
The traceless condition leads to   
\begin{equation}
  {\rm tr}(\bm{\phi}(x)) = {\rm tr}(\bm{\Lambda}(x)) = \Lambda_1(x)+\Lambda_2(x)+\Lambda_3(x)=0 
 .
\end{equation}
The diagonal matrix $\bm{\Lambda}$ is expressed as a linear combination of two diagonal  generators  $H_1$ and $H_2$ belonging to the Cartan subalgebra as
\begin{equation}
  U(x) \bm{\phi}(x) U^\dagger(x) 
  = a(x) H_1 + b(x) H_2
  .
\end{equation}
From (\ref{diag}), we obtain the relation 
\begin{align}
    2 {\rm tr}(\bm{\phi}^2) 
  =& 2 \phi_A \phi_B {\rm tr}(T_A T_B)=\phi_A \phi_B \delta^{AB} = \phi_A \phi_A  
= \vec{\phi} \cdot \vec{\phi}
 \quad ( A,B=1, 2, \cdots, 8 ) 
  \nonumber\\
  =& 2  {\rm tr}(\bm{\Lambda}^2) 
  =2 (\Lambda_1^2+\Lambda_2^2+\Lambda_3^2) 
  = a^2+b^2 
 .
\end{align}

At this stage, the color vector field ${\bf n}(x)$ defined by
$\bm{n}(x)= \bm{\phi}(x)/|\bm{\phi}(x)|$
is an  eight-dimensional unit vector, i.e., 
\begin{equation}
 {\bf n}(x) \cdot {\bf n}(x) = n_A(x) n_A(x) =2 {\rm tr}(\bm{n}(x)^2)=  1
   \ ( A=1, 2, \cdots, 8 )
    .
\end{equation}
In other words,  ${\bf n}$ belongs to the 7-sphere, ${\bf n} \in S^7$, or the target space of the map is $S^7$; 
$\bm{n}:\mathbb{R}^D \rightarrow S^7$.
Hence,  $\bm{n}(x)$ has 7 independent degrees of freedom at each $x$.
However, the group $G$ does not act transitively on the manifold of the target space  $S^7$.
\footnote{
We say that the group $G$ acts transitively on the manifold $M$  if any two elements of $M$ are connected by a group transformation.  
} 
If the stationary subgroup $\tilde{H}$ of $G$ is nontrivial,%
\footnote{
We define the stationary subgroup as a subgroup of $G$ that consists of all the group elements $h$ that leave the reference state (or highest weight state) $|\Lambda>$ invariant up to a phase factor: 
$h |\Lambda >=|\Lambda>e^{i\phi(h)}$.
}
 then the transitive target space $M$ is identified with the left coset space, $G/\tilde{H}$, where $G/\tilde{H} \subset M$. 
Thus, we have
\begin{equation}
x \in \mathbb{R}^D \rightarrow  \bm{n}(x) := \bm{\phi}(x)/|\bm{\phi}(x)| \in G/\tilde{H} \subset S^7 
  .
\end{equation} 
Then the unit color  field $\bm{n}(x) \in su(3)$
is expressed as 
\begin{align}
  \bm{n}(x) 
  =& (\cos \vartheta(x))\bm{n}_1(x)  + (\sin \vartheta(x)) \bm{n}_2(x) ,
\nonumber\\
\bm{n}_1 (x)  :=& U^\dagger(x)  H_1 U(x) , \quad 
\bm{n}_2 (x)  := U^\dagger(x)  H_2 U(x)   
\label{su3-n}
 ,
\end{align}
where  $a^2+b^2=1$ is used to rewrite $a$ and $b$ in terms of an angle $\vartheta$:
$
\cos \vartheta(x) = a(x)
$, 
$
\sin \vartheta(x)  = b(x)
 .
$
Note that $\bm{n}_1(x)$ and $\bm{n}_2(x)$  are  Hermitian and traceless unit fields: 
\begin{equation}
  \bm{n}_j^\dagger(x)=\bm{n}_j(x) , \quad 
  {\rm tr}(\bm{n}_j(x))=0 ,  \quad
 2 {\rm tr}(\bm{n}_j(x)^2)=1 \ (j \in \{1,2\}) 
 .
\end{equation}
This is also the case for the color field $\bm{n}(x)$:
\begin{equation}
  \bm{n}^\dagger(x)=\bm{n}(x) , \quad 
  {\rm tr}(\bm{n}(x))=0 ,  \quad
 2 {\rm tr}(\bm{n}(x)^2)=1 
 .
\end{equation}
It should be remarked that $U$ used in the expression for $\bm{n}_1$ and $\bm{n}_2$ is regarded as an element of $SU(3)$ rather than $U(3)$, since a diagonal generator of a unit matrix in $U(3)$ commutes trivially with all the other generators of $U(3)$.
It is easy to see that the $SU(2)$ case considered in the previous subsection is reproduced in a similar way to the $SU(3)$ case we have just considered. 

For concreteness, we can adopt the Gell-Mann matrices $\lambda_A$  for $SU(3)$ to denote the generators $T_A=\lambda_A/2$ ($A=1,\cdots,8$):
\begin{align}
 \lambda ^1=& \left(
  \begin{array}{ccc}
   0 & 1 & 0 \\
   1 & 0 & 0 \\
   0 & 0 & 0 \\
  \end{array}
 \right),\quad
 \lambda ^2=\left(
  \begin{array}{ccc}
   0 & -i & 0 \\
   i & 0 & 0 \\
   0 & 0 & 0 \\
  \end{array}
 \right),\quad
 \lambda ^3=\left(
  \begin{array}{ccc}
   1 & 0 & 0 \\
   0 & -1 & 0 \\
   0 & 0 & 0 \\
  \end{array}
 \right) ,
 \nonumber\\
 \lambda ^4=& \left(
  \begin{array}{ccc}
   0 & 0 & 1 \\
   0 & 0 & 0 \\
   1 & 0 & 0 \\
  \end{array}
 \right),\quad
 \lambda ^5=\left(
  \begin{array}{ccc}
   0 & 0 & -i \\
   0 & 0 & 0 \\
   i & 0 & 0 \\
  \end{array}
 \right),\quad
 \nonumber\\
 \lambda ^6=& \left(
  \begin{array}{ccc}
   0 & 0 & 0 \\
   0 & 0 & 1 \\
   0 & 1 & 0 \\
  \end{array}
 \right),\quad
 \lambda ^7=\left(
  \begin{array}{ccc}
   0 & 0 & 0 \\
   0 & 0 & -i \\
   0 & i & 0 \\
  \end{array}
 \right),\quad
 \lambda ^8=\frac{1}{\sqrt{3}}\left(
  \begin{array}{ccc}
   1 & 0 & 0 \\
   0 & 1 & 0 \\
   0 & 0 & -2 \\
  \end{array}
 \right)
 .
\end{align}
For the Gell-Mann matrices, we have
\begin{equation}
 \begin{array}{l}
  f_{123}=1  ,
\\
  f_{147}=f_{246}=f_{257}=f_{345}= -f_{156}= -f_{367} = \frac{1}{2} ,
\\
  f_{458}=f_{678}=\frac{\sqrt{3}}{2} ,
 \end{array}
\end{equation}
and
\begin{equation}
 \begin{array}{l}
  d_{118}=d_{228}=d_{338}=-d_{888}=\frac{1}{\sqrt{3}} ,
\\
  d_{146}=d_{157}=d_{256}=d_{344}=d_{355}=\frac{1}{2} ,
\\
  d_{247}=d_{366}=d_{377}=-\frac{1}{2} ,
\\
  d_{448}=d_{558}=d_{668}=d_{778}=-\frac{1}{2\sqrt{3}} .
 \end{array}
\end{equation}

For this choice:  $H_1=T_3=\lambda_3/2$ and $H_2=T_8=\lambda_8/2$,
\begin{equation}
 a(x)= 2\Lambda_1(x)+  \Lambda_3(x) =-2\Lambda_2(x)-  \Lambda_3(x) , \quad b(x)=- \sqrt{3}  \Lambda_3(x)
 .
\end{equation}
Here we use the notation $\bm{n}_3,\bm{n}_8$  instead of $\bm{n}_1,\bm{n}_2$:
\begin{align}
\bm{n}_3 (x) = U^\dagger(x)  T_3 U(x) =\bm{n}_1 (x)   , \quad 
\bm{n}_8 (x)  = U^\dagger(x)  T_8 U(x)  =\bm{n}_2 (x)   
 .
\end{align}

\subsubsection{Independent degrees of freedom carried by the color field $\bm{n}$}

From the viewpoint of reformulating the theory, it is important to recognize which are the independent degrees of freedom to describe the theory in question. 
For $SU(3)$ with a rank of two, it appears to be convenient to introduce two unit vector fields ${\bf n}_3(x)$ and ${\bf n}_8(x)$. 
In fact, this choice  has been adopted as a conventional  approach  by many authors. 
However, we point out that this choice is not necessarily unique from the viewpoint of reformulating the theory, as shown below.

First, we observe that under the three types of products, ${\bf n}_k$ ($k=3,8$) satisfy
\begin{subequations}
\begin{align}
{\bf n}_3\cdot{\bf n}_3=1,
\quad
{\bf n}_3\cdot{\bf n}_8=0,\quad 
{\bf n}_8\cdot{\bf n}_8=1,
\quad
\label{nSU3pro-a}
\\
{\bf n}_3\times{\bf n}_3=0,\quad 
{\bf n}_3\times{\bf n}_8=0, \quad 
{\bf n}_8\times{\bf n}_8=0 
 ,
\label{nSU3pro-b}
\end{align}
and 
\begin{align}
{\bf n}_3*{\bf n}_3=\frac1{\sqrt3}{\bf n}_8,
\quad
{\bf n}_3*{\bf n}_8=\frac1{\sqrt3}{\bf n}_3,
\quad
{\bf n}_8*{\bf n}_8=\frac{-1}{\sqrt3}{\bf n}_8
 ,
 \label{nSU3pro-c}
\end{align}
\end{subequations}
where
the equations (\ref{nSU3pro-c}) are also written as
$
{\bf n}_i*{\bf n}_j  =d^{ijk}{\bf n}_k 
$, 
$
i,j,k \in \{3,8\}
$.
\footnote{
This is equivalent to the relationship between $\bm{n}_3$ and $\bm{n}_8$ of
\begin{subequations}
\begin{equation}
\{\bm{n}_i,\bm{n}_j\}
 =\frac13\delta^{ij}\bm 1
  +d^{ijk}\bm{n}_k,  \quad i,j,k\in\{3,8\}
  ,
\end{equation}
following from the identity  
\begin{equation}
\{T_i,T_j\}
 =\frac13\delta^{ij}\bm 1
  +d^{ijk}T_k,
\quad i,j,k\in\{3,8\} 
 ,
\label{eq:anti-commutator of TT}
\end{equation}
\end{subequations}
where we have used the identity for the anticommutator of the generators (\ref{anti-com}) 
and  
$d^{ija}=2{\rm Tr}(\{T_i,T_j\}T_a)=0$
($a\in\{1,2,4,5,6,7\}$)  from (\ref{d-def}).
}
For $SU(3)$, we find from   (\ref{nSU3pro-a}),   (\ref{nSU3pro-b}) and  (\ref{nSU3pro-c}) that ${\bf n}_3$ and ${\bf n}_8$ constitute a closed set of variables under all multiplications: $\cdot$, $\times$, and $*$. 
In particular, it is important to note that ${\bf n}_8$ can be constructed from ${\bf n}_3$ by the  multiplication $*$, i.e., ${\bf n}_8={\sqrt3} {\bf n}_3*{\bf n}_3$. (In other words, ${\bf n}_8$ is a composite of ${\bf n}_3$.)  In view of this, only the field ${\bf n}_3$ is an independent variable. 
This interpretation does not necessarily agree with the conventional approach. 
 This case will be called the \textit{maximal case}.

On the other hand, we find that the field ${\bf n}_8$ is closed under self-multiplication: 
\begin{subequations}
\begin{align}
{\bf n}_8\cdot{\bf n}_8=&1,
\quad
\\
{\bf n}_8\times{\bf n}_8=&0 
 ,
\\
{\bf n}_8*{\bf n}_8=&\frac{-1}{\sqrt3}{\bf n}_8 
 .
\end{align}
\end{subequations}
This constitutes a novel option, which will be called the \textit{minimal case}.

This observation suggests that  in the maximal or minimal case  for $SU(3)$, one of the fields, ${\bf n}_3$ or ${\bf n}_8$, is sufficient as a representative of the color field $\bm{n}$  to rewrite   Yang-Mills theory based on the change of variables, and that the choice of  ${\bf n}_3$ or ${\bf n}_8$ as a fundamental variable can be used an alternative for the equivalent reformulation of Yang-Mills theory.  This statement will be verified  in the next section.

It should be noted, however, that ${\bf n}_3$ carry 6 degrees of freedom, while ${\bf n}_8$ carry 4 degrees of freedom. Therefore, the other fields to be introduced for rewriting Yang-Mills theory must provide the remaining degrees of freedom in each case.   In particular, for the $SU(2)$ gauge group,  $\bm{n}$ carries always 2 degrees of freedom and there is no distinction between the maximal and minimal cases. 

For $N \ge 4$, there exist intermediate cases other than the maximal and minimal cases. To see this, it is instructive to consider the $SU(4)$ case explicitly, as outlined in the next section.  This also clarifies the meaning of this type of classification and helps to correct a misunderstanding in the previous work \cite{Shabanov99} regarding the number of independent degrees of freedom. 



\subsubsection{Maximal and minimal cases defined by  degeneracies}

We now show that the unit color field $\bm{n}$ is classified into two categories, maximal and minimal, according to the degeneracies of the eigenvalues.
By taking into account the fact that the three eigenvalues obey two equations,
\begin{equation}
\Lambda_1(x)+\Lambda_2(x)+\Lambda_3(x)=0 ,
\quad
\Lambda_1(x)^2+\Lambda_2(x)^2+\Lambda_3(x)^2=\frac12 
 ,
\end{equation}
we find that only one degree of freedom is independent, which corresponds to the choice of $\vartheta$.  
Therefore, the category to which  $\bm{n}$ belongs is specified by the value of $\vartheta$.

\begin{enumerate}
\item[(I)]
 Maximal case in which three eigenvalues $\Lambda_1,\Lambda_2,\Lambda_3$ of $\bm\Lambda$ are distinct.  In the maximal case, the stationary subgroup of $\bm{n}$ is 
\begin{equation}
 \tilde{H}=U(1) \times U(1)
 ,
\end{equation}
and 
  $\bm{n}$ covers the six-dimensional internal or target space $SU(3)/(U(1)\times U(1))$, i.e., the flag space $F_2$:
\begin{equation}
\bm{n}  \in G/\tilde{H} =SU(3)/(U(1)\times U(1)) = F_2
 .
\end{equation}
The maximal cases are realized for any angle $\vartheta \in [0,2\pi)$ except for 6 values of   $\vartheta$ in the minimal case, i.e., $\vartheta \notin \{ \frac{1}{6}\pi,\frac{1}{2}\pi,\frac{5}{6}\pi,\frac{7}{6}\pi,\frac{3}{2}\pi,\frac{11}{6}\pi \}$.

\item[(II)]
 Minimal case in which two of the three eigenvalues are  equal. In the minimal case, the stationary subgroup of $\bm{n}$ becomes 
\begin{equation}
 \tilde{H}=U(2)  
 ,
\end{equation}
and  
 $\bm{n}$ covers only the four dimensional internal or target space $SU(3)/U(2)$, i.e., the complex projective space $P^2(\mathbf{C})$:
\begin{equation}
 \bm{n}   \in G/\tilde{H} = SU(3)/U(2) =P^2(\mathbf{C}) 
 .
\end{equation}

The minimal cases are exhausted by 6 values of $\vartheta$, i.e., 
$\vartheta = \frac{(2n-1)}{6} \pi \ (n=1,2,\cdots,6) 
$ or
$\vartheta \in \{ \frac{1}{6}\pi,\frac{1}{2}\pi,\frac{5}{6}\pi,\frac{7}{6}\pi,\frac{3}{2}\pi,\frac{11}{6}\pi \}$

\end{enumerate}

These definitions for maximal and minimal cases based on the degeneracies of eigenvalues are equivalent to those given in the previous subsection based on multiplication properties, as shown in the following.


\subsubsection{Minimal case}

According to (II) in the above definition, the minimal cases are given by $\vartheta = \frac{(2n-1)}{6} \pi \ (n=1,2,\cdots,6) 
$.

\begin{itemize}
\item $\Lambda_2=\Lambda_3$: 
$\Lambda_2=\Lambda_3=-\frac{1}{2\sqrt{3}} < \Lambda_1= \frac{1}{\sqrt{3}} 
 \Longrightarrow 
 (a,b) =  \left( \frac{\sqrt{3}}{2},  \frac{1}{2}\right) \Longrightarrow  \vartheta=\frac{1}{6}\pi 
$

\item $\Lambda_3=\Lambda_1$: 
$\Lambda_3=\Lambda_1=-\frac{1}{2\sqrt{3}} < \Lambda_2= \frac{1}{\sqrt{3}} 
 \Longrightarrow 
 (a, b) =  \left(-\frac{\sqrt{3}}{2}, \frac{1}{2} \right) \Longrightarrow  \vartheta=\frac{5}{6}\pi
$

\item $\Lambda_1=\Lambda_2$: 
$\Lambda_1=\Lambda_2=-\frac{1}{2\sqrt{3}} < \Lambda_3=\frac{1}{\sqrt{3}} 
 \Longrightarrow 
 (a,b) = (0, -1) \Longrightarrow  \vartheta=\frac{3}{2}\pi
$
\end{itemize}
and their Weyl reflections (rotation by  angle $\pi$) in the weight diagram: 
\begin{itemize}
\item $\Lambda_2=\Lambda_3$: 
$\Lambda_2=\Lambda_3=\frac{1}{2\sqrt{3}} > \Lambda_1=-\frac{1}{\sqrt{3}} 
 \Longrightarrow 
 (a, b) = \left( -\frac{\sqrt{3}}{2},  -\frac{1}{2} \right)  \Longrightarrow  \vartheta=\frac{7}{6}\pi 
$

\item $\Lambda_3=\Lambda_1$: 
$\Lambda_3=\Lambda_1=\frac{1}{2\sqrt{3}} > \Lambda_2=-\frac{1}{\sqrt{3}} 
 \Longrightarrow 
 (a,b)=\left(   \frac{\sqrt{3}}{2},  -\frac{1}{2} \right) \Longrightarrow  \vartheta=\frac{11}{6}\pi
$

\item $\Lambda_1=\Lambda_2$: 
$\Lambda_1=\Lambda_2=\frac{1}{2\sqrt{3}} > \Lambda_3=-\frac{1}{\sqrt{3}} 
 \Longrightarrow 
 (a, b) = (0, 1) \Longrightarrow  \vartheta=\frac{\pi}{2} 
$ .

\end{itemize}
Note that each of three two-dimensional vectors $(a,b)$ is proportional to  a weight vector for the fundamental representations of $3$ and $3^*$ in the weight diagram, 
$(2/\sqrt{3})\Lambda_j$, see, for example \cite{Kondo08}.  The total set is invariant under the action of the Weyl group. 

In particular, for $(a,b)=(0,\pm1)$ or equivalently $\vartheta=\pi/2, 3\pi/2$, the color field $\bm{n}(x)$ can be written using only $\bm{n}_8(x)$, i.e., $\bm{n}(x)=\pm \bm{n}_8(x)$, and $\bm{n}_3(x)$ disappears from the color field. 
In other minimal cases, $\bm{n}_3(x)$ reappears in the color field $\bm{n}(x)$, and the $\vartheta=\pi/2, 3\pi/2$ cases may appear to be special and distinct from other cases.   
As the structure of the degenerate matrix $\Lambda$ indicates, however, the color field $\bm{n}(x)$ has the same degrees of freedom in all the minimal cases. 
Therefore, the difference is apparent due to the special choice of the Gell-Mann matrices.  
In fact, we find that $\bm{n}_8(x)$ covers only the four dimensional internal space $SU(3)/U(2)$, i.e., the complex projective space $P^2(\mathbf{C})$. Also, 
\begin{equation}
 \bm{n}_8 =U^\dagger T_8 U = U^\dagger H_2 U \in SU(3)/U(2) =P^2(\mathbf{C}) 
 ,
\end{equation}
in agreement with the above definition (II). 
This originates from the fact that the $SU(2)\times U(1)_8=U(2)$ rotations caused by four generators $\{T_1,T_2,T_3,T_8\}$  do not change (the direction of) $\bm{n}_8^\infty:=T_8$, since the only generators commuting with $T_8$ are  the four generators $\{T_1,T_2,T_3,T_8\}$ specified by the standard Gell-Mann matrices:
\begin{equation}
[T_8,T_A]=0  .
\quad (A=1,2,3,8) 
\end{equation}
 
Thus, all choices of $\vartheta \in \{ \frac{1}{6}\pi,\frac{1}{2}\pi,\frac{5}{6}\pi,\frac{7}{6}\pi,\frac{3}{2}\pi,\frac{11}{6}\pi \}$ should be treated on  an equal footing, but the apparently simplest way to define the color field in the minimal case is to choose $\vartheta=\pi/2$, i.e., $\bm{n}(x)=\bm{n}_8(x)$.

\subsubsection{Maximal case}

In particular, for $(a,b)=(\pm 1,0)$ or equivalently $\vartheta=0, \pi$, the color field $\bm{n}(x)$ can be  written using only $\bm{n}_3(x)$, i.e., $\bm{n}(x)=\pm \bm{n}_3(x)$, and $\bm{n}_8(x)$ disappears from the expression of the color field $\bm{n}(x)$. 
In the other maximal cases with $\vartheta \ne 0, \pi$,  $\bm{n}(x)$ contains both $\bm{n}_3(x)$ and $\bm{n}_8(x)$. 
The appearances of the representation $\bm{n}(x)$ are considerably different from each other in the two cases.  However, both reveal  the same physical situation  
because ${\bf n}_8(x)$ is constructed from ${\bf n}_3(x)$ as ${\bf n}_8(x)={\sqrt3} {\bf n}_3(x)*{\bf n}_3(x)$. 
Therefore, it is sufficient for us to consider $\bm{n}_3(x)$ to define the color  field $\bm{n}(x)$. 
In fact, $\bm{n}_3(x)$ covers the six-dimensional internal or target space $SU(3)/(U(1)\times U(1))$, i.e., the flag space $F_2$:
\begin{equation}
\bm{n}_3= U^\dagger T_3U = U^\dagger H_1 U \in SU(3)/(U(1)\times U(1)) = F_2
 ,
\end{equation}
in agreement with the above definition (I). 
This is understood from the fact that the $U(1)\times U(1)$ rotations caused by the two diagonal generators $\{T_3,T_8\}$ do not change (the direction of) $\bm{n}_3^\infty:=T_3$, since  the only generators commuting with $T_3$ are the two generators $\{T_3,T_8\}$:
\begin{equation}
[T_3,T_3]=0,\quad
[T_3,T_8]=0
 .
\end{equation}
Thus the easiest way to define the color field in the maximal case is to choose $\vartheta=0$, i.e., $\bm{n}(x)=\bm{n}_3(x)$. 
Of course, this does not prohibit the  introduction of  both $\bm{n}_3(x)$ and $\bm{n}_8(x)$ for convenience. 
\footnote{
In the original approach for decomposing the $SU(3)$ Yang-Mills gauge field \cite{Cho80c}, two fields $\bm{n}_3$ and $\bm{n}_8$ are introduced from the beginning as they were fundamental variables.  
Therefore, this option is included in the maximal case. 
}



\subsection{$SU(N)$ case}

\subsubsection{General consideration}

We consider the $su(N)$-valued field 
\begin{equation}
 \bm{\phi}(x)=\phi_A(x) T_A   . 
\ ( A=1, 2, \cdots, N^2-1 )
\end{equation}
Thus, $\bm{\phi}$ is a traceless and Hermitian matrix for real $\phi_A$, i.e., ${\rm tr}(\bm{\phi})=0$ and $\bm{\phi}^\dagger=\bm{\phi}$, since ${\rm tr}(T_A)=0$ and $(T_A)^\dagger=T_A$.   
Therefore, it can be cast into the diagonal form $\bm\Lambda$ using a unitary matrix $U \in SU(N) \subset U(N)$:
\begin{equation}
  U(x) \bm{\phi}(x) U^\dagger(x) = {\rm diag}(\Lambda_1(x),\Lambda_2(x), \cdots, \Lambda_N(x)) := \bm\Lambda(x)
 ,
\end{equation}
where $\Lambda_a$ ($a=1,\cdots,N$) are real and satisfy
\begin{equation}
{\rm tr}(\bm{\phi}(x))={\rm tr}(\Lambda(x))=\sum_{j=1}^{N}\Lambda_a(x)=0 
 .
\end{equation}
Therefore, the diagonal matrix  $\Lambda$  is expressed as a linear combination of $r=N-1$ diagonal  generators  $H_j$ ($j=1,2,\cdots, r$) belonging to the Cartan subalgebra 
\begin{align}
  U(x) \bm{\phi}(x) U^\dagger(x) 
  =& \sum_{j=1}^{r} a_j(x) H_j  
 ,
\end{align}
where $a_j(x)$ is obtained as a linear combination of $\Lambda_j(x)$.

By taking into account the relation 
\begin{align}
\vec{\phi} \cdot \vec{\phi} =&  \phi_A \phi_A 
  \nonumber\\ =&  2 {\rm tr}(\bm{\phi}^2) =  2 {\rm tr}(U \bm{\phi} U^\dagger U \bm{\phi} U^\dagger ) 
   \sum_{j,k=1}^{r} 2 a_j a_k   {\rm tr}(H_j H_k)
  =  \sum_{j=1}^{r} a_j^2 
  \nonumber\\
  =&  2{\rm tr}(\Lambda^2) = 2 \sum_{j=1}^{r}  \Lambda_j^2   
 ,
\end{align}
the unit color  field $\bm{n}(x) \in su(N)$ satisfying 
\begin{equation}
{\bf n}(x) \cdot {\bf n}(x) = n_A(x) n_A(x) =2 {\rm tr}(\bm{n}(x)^2)=  1
\end{equation}
is expressed as 
\begin{align}
  \bm{n}(x) 
  =& \sum_{j=1}^{r} a_j(x) \bm{n}_j(x)   ,
\quad 
\bm{n}_j (x)  :=  U^\dagger(x)  H_j U(x)  
\label{general-n}
 ,
\end{align}
where each coefficient $a_j$ belongs to the $(r-1)$-sphere $S^{r-1}$: 
\begin{align}
 \sum_{j=1}^{r} a_j(x)^2 =  {\bf n}(x) \cdot {\bf n}(x) = n_A(x) n_A(x) =2 {\rm tr}(\bm{n}(x)^2)=  1
 .
\end{align}
Since each field $\bm{n}_j(x)$  ($j=1, \cdots, r)$ is a  traceless and Hermitian unit field: 
\begin{equation}
  \bm{n}_j^\dagger(x)=\bm{n}_j(x) , \quad 
  {\rm tr}(\bm{n}_j(x))=0 ,  \quad
 2 {\rm tr}(\bm{n}_j(x)^2)=1 \ (j=1, \cdots, r) 
 ,
\end{equation}
the color field $\bm{n}(x)$ satisfies the same property:
\begin{equation}
  \bm{n}^\dagger(x)=\bm{n}(x) , \quad 
  {\rm tr}(\bm{n}(x))=0 ,  \quad
 2 {\rm tr}(\bm{n}(x)^2)=1 
 .
\end{equation}
Note that the $U$ used for the expression for $\bm{n}_j$  is regarded as an element of the $SU(N)$ group rather than $U(N)$, since a diagonal generator in $U(N)$ commutes with all other generators of $U(N)$.

At this stage, the color field $\bm{n}(x)$ is an  $(N^2-2)$-dimensional vector, i.e., $n_A n_A = 1$ ($A=1, 2, \cdots, N^2-1$), or $\bm{n}(x)$ has $(N^2-2)$ degrees of freedom at each $x$. 
Therefore, $\bm{n}_j$ ($j=1, \cdots, r$) must have $N^2-2-(N-2)=N^2-N$ degrees of freedom in total, since $a_j$ ($j=1, \cdots, r$) are responsible for the remaining $(N-2)$ degree of freedom. 
See Appendix~\ref{appendix:SU(N)-group} for the fundamental properties of the $SU(N)$ group.

The maximal and minimal cases for $SU(N)$ are defined as follows. 
\begin{enumerate}
\item[(I)]
 Maximal case in which all eigenvalues $\Lambda_1, \dots,\Lambda_N$ of $\bm{\Lambda}$ are distinct.  The stationary subgroup of $\bm{n}$ is $\tilde{H}=U(1)^{N-1}$, and 
  $\bm{n}$ covers the $(N^2-N)$-dimensional internal or target space $SU(N)/U(1)^{N-1}$, i.e., the flag space $F_2$:
\begin{equation}
\bm{n}  \in SU(N)/U(1)^{N-1}  = F_{N-1}
 .
\end{equation}

\item[(II)]
 Minimal case in which $N-1$  out of $N$ eigenvalues  of $\bm{\Lambda}$ are equal. The  stationary subgroup of $\bm{n}$ becomes $\tilde{H}=U(N-1)$, and  
 $\bm{n}$ covers only the $2(N-1)$-dimensional internal or target space $SU(N)/U(N-1)$, i.e., the complex projective space $P^{N-1}(\mathbb{C})$, which is a submanifold of $F_{N-1}$:
\begin{equation}
 \bm{n}   \in SU(N)/U(N-1)=CP^{N-1} =P^{N-1}(\mathbb{C}) 
 .
\end{equation}

\end{enumerate}


For the minimal case of $SU(N)$, we consider 
\begin{equation}
 \bm{n}_r 
=U^\dagger H_r U \in SU(N)/U(N-1) 
=CP^{N-1} , \quad {\rm dim}CP^{N-1}=2(N-1)
 .
\end{equation}
The $SU(N-1)\times U(1)_{N^2}=U(N-1)$ rotations caused by $(N-1)^2$ generators $\{T_1,T_2,\cdots, T_{(N-1)^2-1},T_{N^2-1} \}$  do not change the direction of $\bm{n}_r$, as only the $(N-1)^2$  generators commute with  $H_r$. 
Consequently,  
$\bm{n}_r$ runs over only the $(2N-2)$-dimensional internal space $SU(N)/U(N-1)=P^{N-1}(\mathbb{C})$.
Therefore, 
$\bm{n}_r$ covers the smallest target space. This is the minimal case. 

To obtain the largest internal or target space of the color field $\bm{n}$, i.e.,  the $(N^2-N)$-dimensional space $SU(N)/U(1)^{N-1}$, i.e., the flag space $F_{N-1}$, 
we must determine the diagonal matrix $T$ in the form 
\begin{equation}
 \bm{n} = U^\dagger T U \in SU(N)/U(1)^{N-1} = F_{N-1},  
\quad  {\rm dim}F_{N-1}=N(N-1) 
 ,
\end{equation}
such that all the generators commuting with $T$ are exhausted solely by $r=N-1$ diagonal generators $H_j$ (including itself): 
\begin{equation}
[H_j,H_k]=0 
,
\end{equation}
and the $U(1)^{r}=U(1)^{N-1}$ rotations caused by the $r=N-1$ diagonal generators $H_j$ do not change the direction of $\bm{n}$ (among the $N^2-1$ generators $T_A$ for  the general unitary transformation $U=\exp(i\sum_{A=1}^{N^2-1} \theta_AT_A)$). 
The diagonal matrix $T$ can be constructed by a linear combination of all diagonal generators $H_j$ such that all diagonal elements in $T$ have distinct values. This is the maximal case.

In addition, there are intermediate cases in which some of the $N$ eigenvalues are degenerate.  
To see the situation concretely, we examine the $SU(4)$ case.  

\subsubsection{$SU(4)$}

The $SU(4)$ group has $4^2-1=15$ generators $T_A=\lambda_A/2$ ($A=1,2,\cdots,15$). Among them,  3 generators $T_3, T_8, T_{15}$ are diagonal in the Gell-Mann representation: 
\begin{align}
 \lambda ^1=& \left(
  \begin{array}{cccc}
   0 & 1 & 0 & 0 \\
   1 & 0 & 0 & 0 \\
   0 & 0 & 0 & 0 \\
   0 & 0 & 0 & 0 \\
  \end{array}
 \right),\quad
 \lambda ^2=\left(
  \begin{array}{cccc}
   0 & -i & 0 & 0 \\
   i & 0 & 0 & 0 \\
   0 & 0 & 0 & 0 \\
   0 & 0 & 0 & 0 \\
  \end{array}
 \right),\quad
 \lambda ^3=\left(
  \begin{array}{cccc}
   1 & 0 & 0 & 0 \\
   0 & -1 & 0 & 0 \\
   0 & 0 & 0 & 0 \\
   0 & 0 & 0 & 0 \\
  \end{array}
 \right) ,
 \nonumber\\
 \lambda ^4=& \left(
  \begin{array}{cccc}
   0 & 0 & 1 & 0 \\
   0 & 0 & 0 & 0 \\
   1 & 0 & 0 & 0 \\
   0 & 0 & 0 & 0 \\
  \end{array}
 \right),\quad
 \lambda ^5=\left(
  \begin{array}{cccc}
   0 & 0 & -i & 0 \\
   0 & 0 & 0 & 0 \\
   i & 0 & 0 & 0 \\
   0 & 0 & 0 & 0 \\
  \end{array}
 \right), 
 \nonumber 
\end{align}
\begin{align}
 \lambda ^6=& \left(
  \begin{array}{cccc}
   0 & 0 & 0 & 0 \\
   0 & 0 & 1 & 0 \\
   0 & 1 & 0 & 0 \\
   0 & 0 & 0 & 0 \\
  \end{array}
 \right),\quad
 \lambda ^7=\left(
  \begin{array}{cccc}
   0 & 0 & 0 & 0 \\
   0 & 0 & -i & 0 \\
   0 & i & 0 & 0 \\
   0 & 0 & 0 & 0 \\
  \end{array}
 \right),\quad
 \lambda ^8=\frac{1}{\sqrt{3}}\left(
  \begin{array}{cccc}
   1 & 0 & 0 & 0 \\
   0 & 1 & 0 & 0 \\
   0 & 0 & -2 & 0 \\
   0 & 0 & 0 & 0 \\
 \end{array}
 \right) ,
 \nonumber\\ 
 \lambda ^9=& \left(
  \begin{array}{cccc}
   0 & 0 & 0 & 1 \\
   0 & 0 & 0 & 0 \\
   0 & 0 & 0 & 0 \\
   1 & 0 & 0 & 0 \\
  \end{array}
 \right),\quad
 \lambda ^{10}=\left(
  \begin{array}{cccc}
   0 & 0 & 0 & -i \\
   0 & 0 & 0 & 0 \\
   0 & 0 & 0 & 0 \\
   i & 0 & 0 & 0 \\
  \end{array}
 \right), 
 \nonumber 
\end{align}
\begin{align}
 \lambda ^{11}=& \left(
  \begin{array}{cccc}
   0 & 0 & 0 & 0 \\
   0 & 0 & 0 & 1 \\
   0 & 0 & 0 & 0 \\
   0 & 1 & 0 & 0 \\
  \end{array}
 \right),\quad
 \lambda ^{12}=\left(
  \begin{array}{cccc}
   0 & 0 & 0 & 0 \\
   0 & 0 & 0 & -i \\
   0 & 0 & 0 & 0 \\
   0 & i & 0 & 0 \\
  \end{array}
 \right),  
 \nonumber\\ 
 \lambda ^{13}=& \left(
  \begin{array}{cccc}
   0 & 0 & 0 & 0 \\
   0 & 0 & 0 & 0 \\
   0 & 0 & 0 & 1 \\
   0 & 0 & 1 & 0 \\
  \end{array}
 \right),\quad
 \lambda ^{14}=\left(
  \begin{array}{cccc}
   0 & 0 & 0 & 0 \\
   0 & 0 & 0 & 0 \\
   0 & 0 & 0 & -i \\
   0 & 0 & i & 0 \\
  \end{array}
 \right),\quad
 \lambda ^{15}=\frac{1}{\sqrt{6}}\left(
  \begin{array}{cccc}
   1 & 0 & 0 & 0 \\
   0 & 1 & 0 & 0 \\
   0 & 0 & 1 & 0 \\
   0 & 0 & 0 & -3 \\
 \end{array}
 \right)
 .
\end{align}
The weight vectors of the fundamental representation $\bf 4$ of $SU(4)$ are given by
\begin{align}
  \nu^1 &=  \left({1 \over 2}, {1 \over 2\sqrt{3}},   {1
\over 2\sqrt{6}} \right) ,
  \quad
  \nu^2  =  \left(-{1 \over 2}, {1 \over 2\sqrt{3}},  {1
\over 2\sqrt{6}} \right) ,
  \nonumber\\
  \nu^3 &=  \left(0, -{1 \over \sqrt{3}}, {1
\over 2\sqrt{6}} \right) ,
  \quad
  \nu^4  =  \left(0, 0,   {-3 \over 2\sqrt{6}}
\right) .
\end{align}

Since ${\rm rank} \ SU(4)=3$, the root space of $SU(4)$ forms the three-dimensional space $(T_3,T_8,T_{15})$ in which there are 12 root vectors shown as follows. 
First, in the $T_3-T_8$ plane, 6 root vectors are the same as those in $SU(3)$:
\begin{align}
 (-\frac{1}{2}, \frac{\sqrt{3}}{2},0) =: \alpha^2, &\quad
 (\frac{1}{2}, \frac{\sqrt{3}}{2},0) =\alpha^1+\alpha^2, 
 \nonumber\\ 
 (-1,0,0)=-\alpha^1,  \quad\quad\quad\quad & \quad\quad\quad\quad  ( 1,0,0) =: \alpha^1 ,  
 \nonumber\\ 
 (-\frac{1}{2}, -\frac{\sqrt{3}}{2},0) =-\alpha^1-\alpha^2, & \quad
 (\frac{1}{2}, -\frac{\sqrt{3}}{2},0) = -\alpha^2
 .  
\end{align}
The other 6 root vectors have nonvanishing components in the $T_{15}$ direction:
\begin{align}
 (\frac{1}{2},  \frac{1}{2\sqrt{3}},\frac{2}{\sqrt{6}}) =\alpha^1+\alpha^2+\alpha^3,&  
 \quad\quad\quad\quad\quad\quad
 ( -\frac{1}{2},   \frac{1}{2\sqrt{3}}, \frac{2}{\sqrt{6}}) = \alpha^2+\alpha^3,   
 \nonumber\\& \quad
(0,  -\frac{1}{\sqrt{3}},\frac{2}{\sqrt{6}}) =: \alpha^3 
, 
\end{align}
with their inversions with respect to the origin 
\begin{align}
& (0,   \frac{1}{\sqrt{3}},-\frac{2}{\sqrt{6}})=- \alpha^3, 
 \nonumber\\ 
 (-\frac{1}{2},  -\frac{1}{2\sqrt{3}},-\frac{2}{\sqrt{6}})= -\alpha^2-\alpha^3, & \quad   \quad\quad\quad\quad
 (- \frac{1}{2},  -\frac{1}{2\sqrt{3}},-\frac{2}{\sqrt{6}})=-\alpha^1-\alpha^2-\alpha^3
  .   
\end{align}
The angle between two root vectors is $2\pi/3$ or $\pi/2$,  as easily seen by calculating their inner product, since all the root vectors have the unit length.  
Which root vector is positive depends on the definition of the positive root, while the numbers of  positive roots and simple roots do not depend on their definitions. 
For example, the positive roots according to our definition \cite{KT99} are as follows: 
\begin{align}
 & ( 1,0,0),  \quad
(\frac{1}{2}, \frac{\sqrt{3}}{2},0), \quad
 (-\frac{1}{2}, \frac{\sqrt{3}}{2},0),   \quad
 \nonumber\\ &
 (\frac{1}{2},  \frac{1}{2\sqrt{3}},\frac{2}{\sqrt{6}}),  \quad
 (0,   -\frac{1}{\sqrt{3}}, \frac{2}{\sqrt{6}}), \quad
 ( \frac{1}{2},  \frac{1}{2\sqrt{3}},\frac{2}{\sqrt{6}})  
 .  
\end{align}
The simple roots are $\alpha^1, \alpha^2, \alpha^3$:
\begin{equation}
 \alpha^1 := ( 1,0,0)   , \quad
 \alpha^2 := (-\frac{1}{2}, \frac{\sqrt{3}}{2},0) , \quad
 \alpha^3 := (0,  -\frac{1}{\sqrt{3}},\frac{2}{\sqrt{6}})  
, 
\end{equation}
and other roots are expressed as linear combinations of the simple roots as indicated above. 
In fact, $\alpha^1, \alpha^2$  and $\alpha^3$ satisfy 
\begin{equation}
 \alpha^1 \cdot  \alpha^1 = \alpha^2 \cdot  \alpha^2 = \alpha^3 \cdot  \alpha^3 = 1 ,
\quad 
 \alpha^1 \cdot  \alpha^2 = \alpha^2 \cdot  \alpha^3 =- \frac12 ,
\quad
 \alpha^1 \cdot  \alpha^3 = 0
 . 
\end{equation}

For $SU(4)$, we find that the following six relations hold for the multiplication $*$:
\begin{align}
{\bf n}_1 * {\bf n}_1  =& \frac{1}{\sqrt{3}} {\bf n}_2 + \frac{1}{\sqrt{6}} {\bf n}_3  ,
\nonumber\\
{\bf n}_1 * {\bf n}_2 =& \frac{1}{\sqrt{3}} {\bf n}_1 = {\bf n}_2 * {\bf n}_1 ,
\nonumber\\
{\bf n}_1  * {\bf n}_3 =&  \frac{1}{\sqrt{6}}  {\bf n}_1 = {\bf n}_3 * {\bf n}_1 ,
\nonumber\\
{\bf n}_2 * {\bf n}_2 =& \frac{-1}{\sqrt{3}} {\bf n}_2 + \frac{1}{\sqrt{6}} {\bf n}_3 , 
\nonumber\\
{\bf n}_2 * {\bf n}_3 =&  \frac{1}{\sqrt{6}}  {\bf n}_2 = {\bf n}_3 * {\bf n}_2 ,
\nonumber\\
{\bf n}_3 * {\bf n}_3 =& \frac{-2}{\sqrt{6}} {\bf n}_3
 .
\label{nSU4}
\end{align}
The last equation shows that the field ${\bf n}_3$ is closed under self-multiplication by $*$. 
This is the minimal case. 
The last three equations in (\ref{nSU4}) show that the field ${\bf n}_2$ and ${\bf n}_3$ are closed under  multiplication. This corresponds to  intermediate cases. 
The first three equations in (\ref{nSU4}) show that ${\bf n}_1$, ${\bf n}_2$ and ${\bf n}_3$ are closed under  multiplication. This constitutes the maximal case. 
It should be remarked that ${\bf n}_1$, ${\bf n}_2$ and ${\bf n}_3$ in the maximal case are constructed from the single color field $\bm n$, which is sufficient for our purposes. 

In all cases (maximal, intermediate and minimal), the single color field $\bm{n}(x)$   is constructed by choosing the appropriate matrix $T$  such that
\begin{equation}
  \bm{n}(x) = U^\dagger(x) T U(x)   \in G/\tilde{H}
 .
\end{equation}
The minimal case occurs when three of the four eigenvalues are equal, which corresponds to the largest stability subgroup $\tilde{H}=U(3)$,  e.g.,
\begin{equation}
 T_{15} = \frac12 \lambda ^{15}=\frac{1}{2\sqrt{6}}\left(
  \begin{array}{cccc}
   1 & 0 & 0 & 0 \\
   0 & 1 & 0 & 0 \\
   0 & 0 & 1 & 0 \\
   0 & 0 & 0 & -3 \\
 \end{array}
 \right) 
, \quad
 \tilde{H}= \frac{U(3) \times U(1)}{U(1)} = U(3) 
  .
  \label{SU4-min-T}
\end{equation}
The two intermediate cases are obtained for the stability subgroup $\tilde{H}=U(2) \times SU(2)$, e.g.,
\begin{equation}
 T =  \frac{1}{2\sqrt{2}}\left(
  \begin{array}{cccc}
   1 & 0 & 0 & 0 \\
   0 & 1 & 0 & 0 \\
   0 & 0 & -1 & 0 \\
   0 & 0 & 0 & -1 \\
 \end{array}
 \right)
 = \frac{\sqrt{2}}{\sqrt{3}} \frac{\lambda_8}{2} + \frac{1}{\sqrt{3}} \frac{\lambda_{15}}{2}  
 , \quad
 \tilde{H}=\frac{U(2) \times U(2)}{U(1)} = U(2) \times SU(2)
  .
\end{equation}
and $\tilde{H}=U(2) \times U(1)$, e.g.,
\begin{equation}
 T_3 = \frac12 \lambda ^{3}=\frac{1}{2}\left(
  \begin{array}{cccc}
   1 & 0 & 0 & 0 \\
   0 & -1 & 0 & 0 \\
   0 & 0 & 0 & 0 \\
   0 & 0 & 0 & 0 \\
 \end{array}
 \right)
 , \quad
 \tilde{H}=\frac{U(1) \times U(1) \times  U(2)}{U(1)}
 = U(1) \times  U(2) 
  ,
  \label{SU4-inter1-T}
\end{equation}
or
\begin{equation}
 T_8 = \frac12 \lambda ^{8}=\frac{1}{2\sqrt{3}}\left(
  \begin{array}{cccc}
   1 & 0 & 0 & 0 \\
   0 &  1 & 0 & 0 \\
   0 & 0 & -2 & 0 \\
   0 & 0 & 0 & 0 \\
 \end{array}
 \right)
 , \quad
 \tilde{H}=\frac{U(2) \times U(1) \times U(1)}{U(1)}
 = U(2) \times U(1)
  .
  \label{SU4-inter2-T}
\end{equation}
The maximal case occurs when all (four) eigenvalues are distinct, which corresponds to the smallest stability subgroup $\tilde{H}=U(1)^3$, e.g., 
\begin{equation}
 T^\prime =  \frac{1}{2\sqrt{5}}\left(
  \begin{array}{cccc}
   1 & 0 & 0 & 0 \\
   0 & 2 & 0 & 0 \\
   0 & 0 & -1 & 0 \\
   0 & 0 & 0 & -2 \\
 \end{array}
 \right)
 = \frac{-\sqrt{5}}{10} \frac{\lambda_3}{2} + \frac{\sqrt{15}}{6} \frac{\lambda_8}{2} + \frac{2\sqrt{30}}{15} \frac{\lambda_{15}}{2} 
 , \quad
 \tilde{H}= \frac{U(1)^{4}}{U(1)} = U(1)^{3}
  .
  \label{SU4-max-T}
\end{equation}
Here the matrix has been chosen such that the absolute value of the elements is simple and as small as possible.

In this example, we can see the role of the three fields $\bm{n}_1$, $\bm{n}_2$, and $\bm{n}_3$ in constructing the three cases. 
Note that ${\rm dim}SU(4)=15$ and ${\rm dim}G/\tilde{H}={\rm dim}G-{\rm dim}\tilde{H}$. 

First, 
we find that $T_{15}$ commutes with $T_1, T_2, \cdots, T_8$ and $T_{15}$ which constitute 9 generators of $U(3)$. Therefore, $\bm{n}_3$ spans the $15-9=6$ dimensional space:
\begin{equation}
\bm{n}_3= U^\dagger H_3 U = U^\dagger T_{15} U \in SU(4)/U(3) = P^3(\mathbb{C}) 
 ,
\end{equation}
which generates the color field $\bm{n}$ in the minimal case of $SU(4)$, as shown in   (\ref{SU4-min-T}).

Second, 
we find that $T_{8}$ commutes with $T_1, T_2, T_3$ and $T_{8}$, which constitute 4 generators of $U(2)$, in addition to $T_{15}$ as a generator of $U(1)$. Therefore, $\bm{n}_2$ spans the $15-5=10$-dimensional   space: 
\begin{equation}
\bm{n}_2= U^\dagger H_2 U = U^\dagger T_{8} U\in SU(4)/(U(2) \times U(1))  
  ,
\end{equation}
which can be used to define the color field $\bm{n}$ in  an intermediate case of $SU(4)$  as shown in (\ref{SU4-inter2-T}).

Third,  
we find that $T_3$ commutes with $T_3$, $T_8$, $T_{13}$, $T_{14}$ and $T_{15}$.  Then $T_{13}$, $T_{14}$ and two linear combinations of $T_{8}$ and $T_{15}$ (i.e., $\frac{1}{3} \sqrt{6}T_{15}-\frac{1}{3} \sqrt{3}T_{8}$ and $-\frac{1}{3} \sqrt{6}T_{15}-\frac{2}{3} \sqrt{3}T_{8}$) constitute 4 generators of $U(2)$, in addition to $T_{3}$ as a generator of $U(1)$.  Thus, $\bm{n}_1$ spans the $15-5=10$-dimensional space:
\begin{equation}
\bm{n}_1= U^\dagger H_1 U = U^\dagger T_{3} U \in SU(4)/(U(2)\times U(1))
 ,
\end{equation}
which can be adopted as the color field $\bm{n}$ in an intermediate case of $SU(4)$ as shown in (\ref{SU4-inter1-T}). 

The color field $\bm{n}$ in the maximal case of $SU(4)$ is obtained by the following linear combination of $\bm{n}_1$, $\bm{n}_2$ and $\bm{n}_3$ as exemplified in (\ref{SU4-max-T}): 
\begin{equation}
\bm{n} = \frac{-\sqrt{5}}{10}   \bm{n}_1  + \frac{\sqrt{15}}{6}  \bm{n}_2 + \frac{2\sqrt{30}}{15} \bm{n}_3  \in SU(4)/U(1)^3 = F_3 
 ,
\end{equation}
which is consistent with the above consideration based on (\ref{nSU4}). 
 
For larger $N$, there may exist more intermediate cases, although we do not discuss such cases here.

\section{Reduction condition and change of variables}

\begin{figure}[tbp]
\begin{center}
\includegraphics[height=6.5cm]{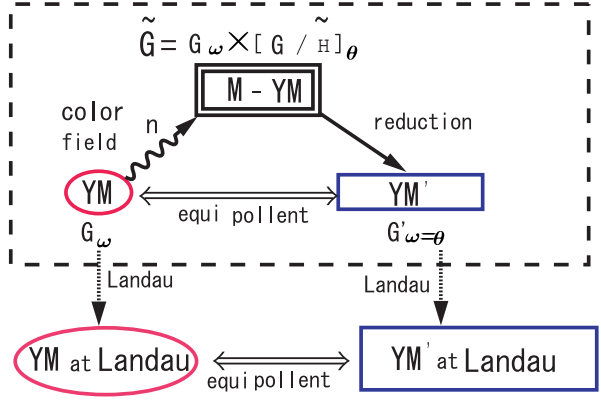}
\caption{\small 
The relationship between the original Yang-Mills (YM) theory and the reformulated Yang-Mills (YM') theory.  A single color field $\bm n$ is introduced to enlarge the original Yang-Mills theory with a gauge group $G$ into the master Yang-Mills (M-YM) theory  with the enlarged gauge symmetry $\tilde{G}=G \times G/\tilde{H}$.  The reduction conditions are imposed to reduce the master Yang-Mills theory to the reformulated Yang-Mills theory with the equipollent gauge symmetry $G^\prime$. 
}
\label{fig:enlarged-YM}
\end{center}
\end{figure}

Our strategy for reformulating Yang-Mills theory is shown in Fig.~\ref{fig:enlarged-YM}. 
By introducing a single color field $\bm n$ in the original  Yang-Mills theory written in terms of the variable $\mathscr{A}_\mu$ with an original gauge group $G=SU(N)$, we obtain a gauge theory called the {\it master Yang-Mills theory} with the enlarged gauge symmetry $\tilde G =G_{\mathscr A}\times [G/\tilde{H}]_{\bm n}$, where the degrees of freedom $[G/\tilde{H}]_{\bm n}$ are possessed by the color field $\bm n$. 
 By imposing a sufficient number of constraints, say, the {\it reduction conditions}, 
\footnote{
In previous papers, we called the reduction condition the new maximal Abelian gauge (new MAG).  However, this is misleading for $SU(N)$, $N \ge 3$. Therefore, we do not use this terminology in this paper.
}
to eliminate the extra degrees of freedom, 
the master Yang-Mills theory is reduced to the gauge theory  reformulated in terms of new variables  with the gauge symmetry $G^\prime=SU(N)$, say the {\it equipollent gauge symmetry},%
\footnote{
In previous papers, we called the equipollent gauge transformation the gauge transformation II. 
}
 which is respected by the new variables. 
\begin{equation}
  G \nearrow (\text{enlargement}) \quad  \tilde G = G_{\mathscr A}\times [G/\tilde{H}]_{\bm n} \quad \searrow (\text{reduction}) \quad G^\prime
   .
\end{equation}
The reformulated Yang-Mills theory is written in terms of new variables, i.e., the color field $\bm n(x)$ and the other new fields.


In the reformulated Yang-Mills theory, the color field $\bm n(x)$   is given as a functional of the $SU(N)$ Yang-Mills gauge field $\mathscr A_\mu(x)$:
\begin{equation}
 \bm n(x) = \bm n_{[\mathscr A]}(x) 
\label{nAx}
 .
\end{equation}
Other new variables are also obtained from the original $\mathscr A_\mu(x)$ by a change of variables thanks to the existence of this color field. 
In fact, we shall show later that  imposing the reduction condition is  a relevant prescription for obtaining $\bm n_{[\mathscr A]}(x)$.  For the moment, therefore, we do not ask how this is achieved and we omit the subscript $\mathscr A$ of $\bm n_{[\mathscr A]}$ to simplify the notation, keeping the above comments in mind.

We require that the original $SU(N)$ gauge field $\mathscr{A}_\mu(x)$ is decomposed into two parts, $\mathscr{V}_\mu(x)$ and $\mathscr{X}_\mu(x)$:
\begin{equation}
 \mathscr{A}_\mu(x)
=\mathscr{V}_\mu(x)+\mathscr{X}_\mu(x)
 ,
\end{equation}
such that  
$\mathscr{V}_\mu(x)$ transforms under the $SU(N)$ gauge transformation $\Omega(x)$ identically to the original gauge field $\mathscr{A}_\mu(x)$, while $\mathscr{X}_\mu(x)$ transforms identical to an adjoint matter field:
\begin{subequations}
\begin{align}
  \mathscr{V}_\mu(x) & \rightarrow \mathscr{V}_\mu^\prime(x) = \Omega(x) [\mathscr{V}_\mu(x) + ig^{-1} \partial_\mu] \Omega^{\dagger}(x) 
 ,
 \label{V-ctransf}
  \\
  \mathscr{X}_\mu(x) & \rightarrow \mathscr{X}_\mu^\prime(x) = \Omega(x)  \mathscr{X}_\mu(x) \Omega^{\dagger}(x) 
 \label{X-ctransf}
 , 
\end{align}
by way of a \textit{single}  $\bm{n}$ field, which transforms according to the adjoint representation: 
\begin{equation}
 \bm{n}(x)   \rightarrow \bm{n}^\prime(x) = \Omega(x)  \bm{n}(x) \Omega^{\dagger}(x)
 .
\end{equation}
 \label{gauge-transf}
\end{subequations}
The gauge transformation for the new variables is the  {\it equipollent gauge transformation}, which should be compared with the original gauge transformation for $\mathscr{A}_\mu(x)$:
\begin{align}
  \mathscr{A}_\mu(x) & \rightarrow \mathscr{A}_\mu^\prime(x) = \Omega(x) [\mathscr{A}_\mu(x) + ig^{-1} \partial_\mu] \Omega^{\dagger}(x) 
 . 
\end{align}

In the following, we use the decomposition of the Lie algebra valued function $\mathscr F$ into two parts, an $\tilde{H}$-commutative part $\mathscr F_{\tilde{H}}$  and the remaining part $\mathscr F_{G/\tilde{H}}$:
\begin{align}
  \mathscr F(x) 
:= \mathscr F_{\tilde{H}}(x) + \mathscr F_{G/\tilde{H}}(x) 
 \quad 
\mathscr F_{\tilde{H}}(x) \in \mathscr{L}(\tilde{H}), \ 
\mathscr F_{G/\tilde{H}}(x) \in \mathscr{L}(G/\tilde{H}), 
 .
\end{align}
where
\begin{align}
  [ \mathscr F_{\tilde{H}}(x) , \bm{n}(x)  ] = 0 
 .
\end{align}
In this decomposition, it should be remarked that $\tilde{H}$ does not necessarily agree with the maximal torus subgroup $H$ for $SU(N)$, $N \ge 3$. 
For $SU(2)$,  $\tilde{H}=H$ and the two parts are uniquely specified:
\begin{align}
  \mathscr F 
= \mathscr F_{H}  + \mathscr F_{G/H} 
 , \quad 
\mathscr F_{H} \in \mathscr{L}(H), \ 
\mathscr F_{G/H} \in \mathscr{L}(G/H)  
 .
\end{align}
This is also written as
\begin{align}
  \mathscr F 
:= \mathscr F_\parallel + \mathscr F_\perp 
 ,  
\end{align}
which corresponds to the well-known decomposition of a vector $\bm{F}$ into the parallel part $\bm{F}_\parallel$  and perpendicular part  $\bm{F}_\perp$ in the vector form:
\begin{align}
  \bm{F} 
=  \bm{F}_\parallel +  \bm{F}_\perp 
= \bm{n} (\bm{n} \cdot \bm{F}) + \bm{n} \times (\bm{F} \times \bm{n})
 ,
\end{align}
which follows from the simple identity,
$
 \bm{n} \times (\bm{n} \times \bm{F}) = \bm{n} (\bm{n} \cdot \bm{F}) - (\bm{n} \cdot \bm{n}) \bm{F}.
$

\subsection{Maximal case for $SU(N)$}

In the maximal case, it is convenient to introduce a set of unit fields  $\bm n_j(x)$ ($j=1, \cdots, r$) using the adjoint orbit representation:
\begin{align}
  \bm{n}_j(x) = U^{\dagger}(x) H_j U(x) , \quad j \in \{ 1, 2, \cdots, r \} 
 ,
\label{ador}
\end{align}  
where  $r:=\text{rank} SU(N)=N-1$ and $H_j$ are the Cartan subalgebra.
The fields $\bm{n}_j(x)$ defined in this way are indeed unit vectors, since
\begin{align}
  (\bm{n}_j(x),\bm{n}_k(x)) &= 2 {\rm tr}(\bm{n}_j(x)\bm{n}_k(x))
=  2 {\rm tr}(U^{\dagger}(x) H_j U(x) U^{\dagger}(x) H_k U(x)) 
\nonumber\\&
=  2 {\rm tr}( H_j  H_k ) 
= (H_j ,H_k) = \delta_{jk}
 .
\end{align}
These unit vectors mutually commute, 
\begin{align}
  [ \bm{n}_j(x), \bm{n}_k(x)] = 0, \quad j,k \in \{ 1, 2, \cdots, r \} 
 ,
\label{com}
\end{align}
since $H_j$ are the Cartan subalgebra obeying 
\begin{align}
  [ H_j ,H_k ] = 0, \quad j,k \in \{ 1, 2, \cdots, r \} 
 .
\end{align}

\subsubsection{Decomposition}

In the maximal case, 
the decomposed fields $\mathscr V_\mu(x)$ and $\mathscr X_\mu(x)$ are specified by two defining equations (conditions):

(I) all $\bm{n}_j(x)$ are covariant constants in the background $\mathscr{V}_\mu(x)$:
\begin{align}
  0 = D_\mu[\mathscr{V}] \bm{n}_j(x) 
:=\partial_\mu \bm{n}_j(x) -  ig [\mathscr{V}_\mu(x), \bm{n}_j(x)] ,
\quad (j=1,2, \cdots, r) 
\label{defVL}
\end{align}

(II)  $\mathscr{X}^\mu(x)$ does not have the $H$-commutative part, i.e., $\mathscr{X}^\mu(x)_{H}=0$. In other words,   $\mathscr{X}_\mu(x)$ is orthogonal to all $\bm{n}_j(x)$:
\begin{align}
  (\bm{n}_j(x), \mathscr{X}_\mu(x))   := 2{\rm tr}(\bm{n}_j(x) \mathscr{X}_\mu(x))  = n_j^A(x)  \mathscr{X}_\mu^A(x)= 0  
 . 
\quad (j=1,2, \cdots, r) 
\label{defXL}
\end{align}
It should be remarked that the defining equations are invariant under the gauge transformation (\ref{gauge-transf}).

First, we determine the $\mathscr{X}_\mu$ field by solving the defining equations. 
For this purpose, we take into account the following identity: 
Any $su(N)$ Lie algebra valued function $\mathscr F$ can be  decomposed into the $H$-commutative part $\mathscr F_H$ and the remaining $G/H$ part $\mathscr F_{G/H}$ \cite{Shabanov99,Shabanov02}:
\footnote{
This identity is equivalent to the identity \cite{FN99a}
\begin{align}
 \delta^{AB}  = n_j^A n_j^B  -f^{ACD} n_j^C f^{DEB} n_j^E  
 .
\end{align}
}
\begin{align}
  \mathscr F 
:= \mathscr F_{H} + \mathscr F_{G/H} 
:= \sum_{j=1}^{r=N-1}  \bm{n}_j(\bm{n}_j, \mathscr F) +   \sum_{j=1}^{r=N-1}  [\bm{n}_j, [\bm{n}_j, \mathscr F]] 
 .
\label{idv}
\end{align}
The derivation of this identity is given in Appendix \ref{appendix:idv}. 
In proving this identity,  we have used the identification (\ref{ador}).

We apply the identity (\ref{idv}) to $\mathscr{X}_\mu$ and use the second defining equation (\ref{defXL}) to obtain   
\begin{align}
 \mathscr{X}_\mu = \sum_{j=1}^{r} (\mathscr{X}_\mu,\bm{n}_j)\bm{n}_j +   \sum_{j=1}^{r}  [\bm{n}_j, [\bm{n}_j, \mathscr{X}_\mu]]
=  \sum_{j=1}^{r}   [\bm{n}_j, [\bm{n}_j, \mathscr{X}_\mu]]
 .
\end{align}
Then we take into account the first defining equation: 
\begin{align}
 \mathscr{D}_\mu[\mathscr{A}]\bm{n}_j 
 =& \partial_\mu \bm{n}_j - ig[\mathscr{A}_\mu, \bm{n}_j]
\nonumber\\
=& \mathscr{D}_\mu[\mathscr{V}]\bm{n}_j - ig [\mathscr{X}_\mu, \bm{n}_j] 
\nonumber\\ 
=& - ig [\mathscr{X}_\mu, \bm{n}_j] =  ig [ \bm{n}_j, \mathscr{X}_\mu] 
 .
\end{align}
Thus,  $\mathscr{X}_\mu(x)$ is expressed in terms of $\mathscr{A}_\mu(x)$ and $\bm{n}_j(x)$ as
\begin{align}
 \mathscr{X}_\mu(x) = -ig^{-1} \sum_{j=1}^{r}  [\bm{n}_j(x), \mathscr{D}_\mu[\mathscr{A}]\bm{n}_j(x) ]
 .
\end{align}

Next, the $\mathscr{V}_\mu$ field is expressed in terms of $\mathscr{A}_\mu(x)$ and $\bm{n}_j(x)$: 
\begin{align}
  \mathscr{V}_\mu(x) 
  =& \mathscr{A}_\mu(x)  - \mathscr{X}_\mu(x)  
  \nonumber\\
  =& \mathscr{A}_\mu(x)  + ig^{-1} \sum_{j=1}^{r}  [\bm{n}_j(x), \mathscr{D}_\mu[\mathscr{A}]\bm{n}_j(x) ] 
  \nonumber\\
  =& \mathscr{A}_\mu(x) - \sum_{j=1}^{r}  [\bm{n}_j(x), [ \bm{n}_j(x), \mathscr{A}_\mu(x) ] ]
+ ig^{-1} \sum_{j=1}^{r}    [\bm{n}_j(x) , \partial_\mu  \bm{n}_j(x) ]
 .
\label{defV11}
\end{align}
We now apply the identity (\ref{idv}) to $\mathscr{A}_\mu$ to obtain a simpler form:
\begin{align}
\mathscr{V}_\mu(x)
=\sum_{j=1}^{r}(\mathscr{A}_\mu(x),\bm{n}_j(x))
\bm{n}_j(x)+ig^{-1} \sum_{j=1}^{r} [\bm{n}_j(x) , \partial_\mu  \bm{n}_j(x)].
\label{defV}
\end{align}
Thus, $\mathscr{V}_\mu(x)$ and $\mathscr{X}_\mu(x)$ are  written in terms of $\mathscr A_\mu(x)$ once $\bm n_j(x)$ are given.

It should be remarked that the background field $\mathscr{V}_\mu(x)$ contains a part $\mathscr{C}_\mu(x)$ which commutes with all $\bm{n}_j(x)$:
\begin{align}
    [ \mathscr{C}_\mu(x), \bm{n}_j(x) ] = 0 
 .
    \quad (j=1,2, \cdots, r=N-1) 
\label{specC}
\end{align}
Such an $H$-commutative part (or a parallel part in the vector form) $\mathscr{C}_\mu(x)$  in  $\mathscr{V}_\mu(x)$ is not determined uniquely from the first defining equation (\ref{defVL}) alone. However, it is determined by the second defining equation as shown above. 
In view of this,  we further decompose $\mathscr V_\mu(x)$ into $\mathscr C_\mu(x)$ and $\mathscr B_\mu(x)$:
\begin{align}
 \mathscr V_\mu(x)
 =\mathscr C_\mu(x)
  +\mathscr B_\mu(x)
 .
\end{align}
Applying the identity (\ref{idv}) to $\mathscr{C}_\mu(x)$ and by taking into account (\ref{specC}), we obtain
\begin{align}
 \mathscr{C}_\mu(x)
 = \sum_{j=1}^{r}  (\mathscr{C}_\mu(x),\bm{n}_j(x)) \bm{n}_j(x) 
 . 
\end{align}
If the remaining part $\mathscr{B}_\mu(x)$, which is not $H$-commutative, i.e., $[\mathscr{B}_\mu(x),\bm{n}_j(x)]  \ne 0$, is perpendicular to all $\bm{n}_j(x)$:
\begin{align}
 (\mathscr{B}_\mu(x),  \bm{n}_j(x))
= 2{\rm tr}(\mathscr{B}_\mu(x) \bm{n}_j(x) )=0 
 ,
\quad (j=1,2, \cdots, r)
\label{defBL}
\end{align}
then we have   
\begin{align}
  (\mathscr{A}_\mu(x),\bm{n}_j(x)) 
=  (\mathscr{V}_\mu(x),\bm{n}_j(x)) 
= (\mathscr{C}_\mu(x),\bm{n}_j(x)) 
 . 
\end{align}
Consequently, the $H$-commutative part $\mathscr{C}_\mu(x)$ is
\begin{align}
 \mathscr{C}_\mu(x)
 = \sum_{j=1}^{r}  (\mathscr{A}_\mu(x),\bm{n}_j(x)) \bm{n}_j(x)
 . 
\end{align}
and the remaining part $\mathscr{B}_\mu(x)$ is determined as
\footnote{
The SU(2) version in the vector form  is expressed as 
$\mathbf{B}_\mu(x)
=g^{-1}  \partial_\mu  {\bf n}(x) \times  {\bf n}(x)$.
} 
\begin{align}
  \mathscr{B}_\mu(x) 
  =   ig^{-1}  \sum_{j=1}^{r} [ \bm{n}_j(x), \partial_\mu  \bm{n}_j(x) ]  
 .
\label{B}
\end{align}
In fact, it is easy to verify that this expression indeed satisfies (\ref{defBL}) and  that 
\begin{align}
 D_\mu[\mathscr{B}] \bm{n}_j(x) 
=\partial_\mu \bm{n}_j(x) -  ig [\mathscr{B}_\mu(x), \bm{n}_j(x)] = 0
\quad (j=1,2, \cdots, r) 
 .
\label{defBL2}
\end{align}
There are other ways of deriving the same result.
\footnote{
For example, the same expression for $\mathscr{V}_\mu$ is also obtained by solving the defining equations as follows. 
Taking into account the commutator of the first defining equation (\ref{defVL}) with $\bm n_j$, we have 
$
  ig^{-1}   [ \bm{n}_j(x), \partial_\mu  \bm{n}_j(x) ] 
= ig^{-1}    [ \bm{n}_j(x), ig[\mathscr{V}_\mu(x), \bm{n}_j(x)] ] 
=     [  \bm{n}_j(x), [\bm{n}_j(x),\mathscr{V}_\mu(x)]  ] 
 .
$
Then we obtain the relation
$
  \mathscr{V}_\mu(x) 
=  \sum_{j=1}^{r}  (\mathscr{V}_\mu(x),\bm{n}_j(x)) \bm{n}_j(x) + ig^{-1} \sum_{j=1}^{r}    [ \bm{n}_j(x) , \partial_\mu  \bm{n}_j(x) ] 
 .
$
The second defining equation (\ref{defXL}) leads to 
$
(\mathscr{A}_\mu(x),\bm{n}_j(x)) = ( \mathscr{V}_\mu(x),\bm{n}_j(x))
 ,
$
and hence we have (\ref{defV}).
}

Thus, once a full set of color fields $\bm n_j(x)$ is given, the original gauge field  has the following decomposition in the Lie algebra form: 
\begin{subequations}
\begin{align}
\mathscr A_\mu(x)
 =\mathscr V_\mu(x)
  +\mathscr X_\mu(x)
 =\mathscr C_\mu(x)
  +\mathscr B_\mu(x)
  +\mathscr X_\mu(x)
,
\end{align}
where each part is expressed  in terms of $\mathscr{A}_\mu(x)$ and $\bm{n}_j(x)$ as
\begin{align}
  \mathscr{C}_\mu(x) =&   \sum_{j=1}^{N-1}( \mathscr{A}_\mu(x),\bm{n}_j(x))  \bm{n}_j(x) 
= \sum_{j=1}^{N-1}  c_\mu^j(x) \bm{n}_j(x) ,
\\
  \mathscr{B}_\mu(x) =&   
   ig^{-1} \sum_{j=1}^{N-1} [\bm{n}_j(x), \partial_\mu  \bm{n}_j(x)] ,
\\
 \mathscr{X}_\mu(x) =& -ig^{-1}   \sum_{j=1}^{N-1}  [\bm{n}_j(x), \mathscr{D}_\mu[\mathscr{A}]\bm{n}_j(x) ]
 .
\end{align}
\label{NLCV-maximal}
\end{subequations}
In what follows, the summation over the index $j$ should be understood when it is repeated, unless otherwise stated.

Equivalently, the decomposition (\ref{NLCV-maximal})  is written in the vector form  as 
\begin{subequations}
\begin{align}
  \mathbf{A}_\mu(x) 
= \mathbf{V}_\mu(x)  + \mathbf{X}_\mu(x)  
= \mathbf{C}_\mu(x) + \mathbf{B}_\mu(x) + \mathbf{X}_\mu(x)  
 ,
\end{align}
where each part is expressed in terms of $\mathbf{A}_\mu(x)$ and ${\bf n}_j(x)$ as
\begin{align}
  \mathbf{C}_\mu(x) :=& \sum_{j=1}^{N-1} ( \mathbf{A}_\mu(x)  \cdot {\bf n}_j(x)) {\bf n}_j(x)  = \sum_{j=1}^{N-1} c_\mu^j(x) {\bf n}_j(x)  ,
\\
  \mathbf{B}_\mu(x) :=&   g^{-1} \sum_{j=1}^{N-1}  (\partial_\mu  {\bf n}_j(x) \times  {\bf n}_j(x)) 
 ,
\\
 \mathbf{X}_\mu(x)  =& g^{-1}  \sum_{j=1}^{N-1}  ({\bf n}_j(x) \times \mathscr{D}_\mu[\mathbf{A}]{\bf n}_j(x))   
 .
\label{eqV}
\end{align}
\label{NLCV-maximal2}
\end{subequations}
All $\bm{n}_j(x)$ fields in the maximal case are constructed from a single color field $\bm{n}(x)$.  
Therefore, a single color field $\bm{n}(x)$ is sufficient to specify the decomposition. See section \ref{subsection:maximalSU3} for the explicit form.

Finally, we point out that another equivalent expression is obtained in a slightly different way. 
The differentiation of $\bm{n}_k(x)$ yields the relation 
\begin{align}
 \partial_\mu \bm{n}_k(x) 
= [\partial_\mu  U^{\dagger}(x) U(x), \bm{n}_k(x)]  
 . 
\label{deln}
\end{align}
If we require the covariant constantness  for all  $\bm{n}_k(x)$ in the background of $\mathscr{V}_\mu(x)$:
\begin{align}
  D_\mu[\mathscr{V}] \bm{n}_k(x) 
:=\partial_\mu \bm{n}_k(x) -  ig [\mathscr{V}_\mu(x), \bm{n}_k(x)]
 = 0 
 ,
\label{DBn1}
\end{align}
the following relation must hold for all $\bm{n}_k(x)$ by taking into account the relation (\ref{deln}).  
\begin{align}
[\mathscr{V}_\mu(x) +  ig^{-1} \partial_\mu U^{\dagger}(x) U(x) , \bm{n}_k(x)]
 = 0 
 .
\label{DBn2}
\end{align}
This shows that $\partial_\mu  U^{\dagger}(x) U(x) -ig \mathscr{V}_\mu(x) \in su(N)$ commutes with all $\bm{n}_k(x)$ and that it can be written as a linear combination of all $\bm n_j$:
\begin{align}
  \mathscr{V}_\mu(x) +  ig^{-1} \partial_\mu U^{\dagger}(x) U(x) 
=  \sum_{j=1}^{r} c_\mu^j(x) \bm{n}_j(x) \in U(1)^{r}
 ,
\end{align}
or
\begin{align}
 \mathscr{V}_\mu(x) 
=  \sum_{j=1}^{r} \bm{n}_j(x) c_\mu^j(x)  -ig^{-1} \partial_\mu U^{\dagger}(x) U(x) 
=  \sum_{j=1}^{r} \bm{n}_j(x) c_\mu^j(x)  + ig^{-1} U^{\dagger}(x) \partial_\mu U(x)
 .
\end{align}
Here  $\mathscr{V}_\mu$ is determined up to the terms parallel to $\bm{n}_j$. 
It is clear that $\mathscr{B}_\mu(x)$ corresponds to the pure gauge form.

\subsubsection{Reduction in the maximal case}

We wish to regard the new variables $\bm{n}^A, c_\mu^j, \mathscr{X}_\mu^A$ as those obtained by  the change of variables from the original gauge field: 
\begin{align}
  \mathscr{A}_\mu^A &\Longrightarrow  ( \bm{n}^A, c_\mu^j, \mathscr{X}_\mu^A )   .
\quad (A=1, \cdots, N^2-1; j=1, \dots, N-1; \mu=1, \cdots, D)
\end{align}
In the maximal case,  the naive counting of independent degrees of freedom is as follows. 
\begin{itemize}
\item
$\mathscr A_\mu \in \mathscr{L}(G)=su(N)  \rightarrow (N^2-1)D$ ,
\item
$\bm{n}  \in \mathscr{L}(G/H)=su(N)-[u(1)+ \cdots +u(1)] \rightarrow N^2-1-(N-1)=N^2-N$ ,
\item
$c_\mu \in \mathscr{L}(H)=u(1)+ \cdots +u(1) \rightarrow  (N-1)D$,

\item
$\mathscr X_\mu \in \mathscr{L}(G/H)=su(N)-u(N-1) \rightarrow (N^2-1)D- (N-1)D=(N^2-N)D $.

\end{itemize}
In the decomposition just given, therefore, there is an issue of mismatch for the independent degrees of freedom. 
In fact, the new variables carry $N^2-N$  extra degrees of freedom after the decomposition.
Therefore, we must eliminate $N^2-N$  degrees of freedom. 
For this purpose,  we intend to impose $N^2-N$ constraints to eliminate  the extra degrees of freedom. 


The transformation properties of the decomposed fields $\mathbf B_\mu, \mathbf C_\mu, \mathbf X_\mu$ are uniquely determined, once we specify those for $\mathbf A_\mu$ and ${\bf n}$,   as in the $SU(2)$ case discussed in the previous paper \cite{KMS06}.  
We consider the  infinitesimal version of the enlarged gauge transformation $\delta_{\omega,\theta}$,  which is obtained by combining the  local transformations for $\delta_\omega{\mathbf A}_\mu$ and $\delta_\theta{\bf n}$:
\begin{equation}
\delta_\omega {\mathbf A}_\mu(x)=D_\mu[\mathbf A]\bm\omega(x),
\quad
\delta_\theta {\bf n}(x) =g{\bf n}(x) \times\bm\theta_\perp(x) 
 ,
\label{egt1}
\end{equation}
where $\bm\theta_\perp \in \mathscr{L}(G/H)$.

We propose a constraint,  which we call the reduction condition, as follows.
To find the reduction condition for $G=SU(N)$, we calculate the $SU(N)$ version of $\mathbf X^\mu$ squared  as suggested from the $SU(2)$ case: 
\begin{align}
 g^2  \mathbf X^\mu\cdot\mathbf X_\mu
 &=\{ {\bf n}_j \times  D_\mu[\mathbf A]{\bf n}_j  \} \cdot \{ {\bf n}_k \times  D_\mu[\mathbf A]{\bf n}_k  \}
   \nonumber\\
 &= D_\mu[\mathbf A]{\bf n}_k \cdot [({\bf n}_j \times  D_\mu[\mathbf A]{\bf n}_j) \times {\bf n}_k]  
\nonumber\\
 &=  D_\mu[\mathbf A]{\bf n}_k \cdot [({\bf n}_k \times  D_\mu[\mathbf A]{\bf n}_j) \times {\bf n}_j]   
\nonumber\\
 &= D_\mu[\mathbf A]{\bf n}_k \cdot [({\bf n}_j \times  D_\mu[\mathbf A]{\bf n}_k) \times {\bf n}_j]
\nonumber\\
 &= D_\mu[\mathbf A]{\bf n}_k \cdot [D_\mu[\mathbf A]{\bf n}_k - ({\bf n}_j \cdot  D_\mu[\mathbf A]{\bf n}_k) {\bf n}_j]
\nonumber\\
 &= (D_\mu[\mathbf A]{\bf n}_j)^2
 ,
\end{align}
where the summation is over $j,k$ and 
 we have used the Jacobi identity in the third and fourth equalities and ${\bf n}_j \cdot D_\mu[\mathscr A]{\bf n}_k=0$ in the last step. 

We show that a reduction condition is obtained by minimizing the functional  
\begin{align}
 R[\mathbf A, \{ \bm n_j \}]
 :=   \int d^Dx \frac12 (D_\mu[\mathbf A] {\bf n}_j)^2
  ,
\end{align}
under the enlarged gauge transformation.
In fact, the transformation of the integrand 
$
(D_\mu[\mathbf A]\bm n_j)^2
$
under the infinitesimal enlarged gauge transformation is as follows: 
\begin{align}
& \delta_{\omega,\theta} \left\{ \frac12(D_\mu[\mathbf A]{\bf n}_j)^2 \right\} 
   \nonumber\\
 &=(D_\mu[\mathbf A]{\bf n}_j)\cdot\delta_{\omega,\theta} (D_\mu[\mathbf A]{\bf n}_j)
   \nonumber\\
 &=(D_\mu[\mathbf A]{\bf n}_j)\cdot
   \{D_\mu[\mathbf A]\delta_{\omega,\theta} {\bf n}_j
     +g\delta_{\omega,\theta} \mathbf A_\mu\times{\bf n}_j\}
   \nonumber\\
 &=(D_\mu[\mathbf  A]{\bf n}_j)\cdot
   \{D_\mu[\mathbf A](g{\bf n}_j\times\bm\theta_\perp)
     +gD_\mu[\mathbf A]\bm\omega\times{\bf n}_j\}
   \nonumber\\
 &=(D_\mu[\mathbf A]{\bf n}_j)\cdot
   \{ g(D_\mu[\mathbf A]{\bf n}_j) \times\bm\theta_\perp
   + g{\bf n}_j\times D_\mu[\mathbf A]\bm\theta_\perp
     +gD_\mu[\mathbf A]\bm\omega\times{\bf n}_j\}
   \nonumber\\
 &=g(D_\mu[\mathbf A]{\bf n}_j)\cdot
   \{D_\mu[\mathbf A](\bm\omega-\bm\theta_\perp)\times{\bf n}_j\}
   \nonumber\\
 &=g(D_\mu[\mathbf A]{\bf n}_j)\cdot
   \{D_\mu[\mathbf  A](\bm\omega_\perp-\bm\theta_\perp)\times{\bf n}_j\},
   \nonumber\\
 &=g({\bf n}_j\times D_\mu[\mathbf A]{\bf n}_j)\cdot
   D_\mu[\mathbf A](\bm\omega_\perp-\bm\theta_\perp) 
 ,
\end{align}
where we have used 
$D_\mu[\mathbf A]\bm\omega_\parallel
=D_\mu[\mathbf A](\omega_\parallel {\bf n})
= \partial_\mu \omega_\parallel {\bf n} + \omega_\parallel (D_\mu[\mathbf A] {\bf n})
$ to obtain the sixth equality.

Therefore, $(D_\mu[\mathbf A]{\bf n}_j)^2$ is invariant under the the subset $\bm\omega_\perp=\bm\theta_\perp$ of the enlarged gauge transformation (\ref{egt1}).
The infinitesimal variation of the functional is
\begin{align}
\delta_{\omega,\theta} R[\mathbf A, \{ {\bf n}_j \}]
&=\int d^Dx  g({\bf n}_j\times D_\mu[\mathbf A]{\bf n}_j)\cdot
   D_\mu[\mathbf A](\bm\omega_\perp-\bm\theta_\perp) 
   \nonumber\\
 &=- \int d^Dx (\bm\omega_\perp-\bm\theta_\perp) \cdot D_\mu[\mathbf A] g({\bf n}_j\times D_\mu[\mathbf A]{\bf n}_j)
   \nonumber\\
&=- \int d^Dx (\bm\omega_\perp-\bm\theta_\perp) \cdot  g({\bf n}_j\times D_\mu[\mathbf A]D_\mu[\mathbf A]{\bf n}_j)
 .
\end{align}
Thus, we obtain the differential form of the reduction condition:
\begin{equation}
\bm\chi[ \mathbf A,{\bf n} ]
 :={\bf n}_j\times D_\mu[\mathbf A]D_\mu[\mathbf A]{\bf n}_j
 \equiv0 
  .
\label{eq:nMAG_minimal_diff}
\end{equation}
Using the Leibnitz rule for the covariant derivative $D_\mu[\mathbf A]$, 
\begin{equation}
\bm\chi[ \mathbf A,{\bf n} ]
 ={\bf n}_j\times D_\mu[\mathbf A]D_\mu[\mathbf A]{\bf n}_j
 =  D_\mu[\mathbf A] \{ {\bf n}_j \times D_\mu[\mathbf A]{\bf n}_j \}
  ,
\end{equation}
the differential reduction condition can  also be expressed in terms of  $\mathbf V_\mu$   and $\mathbf X_\mu$ in the vector form:
\begin{equation}
\bm\chi[{\bf n},\mathbf C,\mathbf X]
 :=D^\mu[\mathbf V]\mathbf X_\mu
 \equiv0
 ,
\end{equation}
or in the Lie algebra form:
\begin{align}
0= \bm{\chi}[\bm n,\mathscr A_\mu] 
&:= D_\mu[\mathscr{V}] \mathscr{X}_\mu(x) \equiv 
 \partial_\mu \mathscr{X}_\mu(x) -  ig [\mathscr{V}_\mu(x), \mathscr{X}_\mu(x)]
\nonumber\\&
=\partial_\mu \mathscr{X}_\mu - ig c_\mu^j [\bm{n}_j, \mathscr{X}_\mu] 
-  [[ \partial_\mu \bm{n}_j, \bm{n}_j], \mathscr{X}_\mu]  
 .
\end{align}
Note that $\bm\chi \in SU(N)/U(1)^r$ and the number of conditions for $\bm{\chi}=(\chi^A) =0$ $(A=1, \cdots, N^2-1)$ is  $N^2-1-(N-1)=N^2-N$ as expected, 
since $\bm\chi$ is subject to  $N-1$ orthogonality conditions: 
\begin{align}
(\bm{n}_j(x), \bm\chi(x))
= (\bm{n}_j(x), D_\mu[\mathscr{V}] \mathscr{X}_\mu(x)) 
= 0 \quad (j=1, \cdots, r=N-1)  
 . 
\end{align}
This  follows from the defining equations (\ref{defXL}) and (\ref{defVL}) as 
\begin{align}
 (\bm{n}_j , D_\mu[\mathscr{V}] \mathscr{X}_\mu ) 
&=  (\bm{n}_j , \partial_\mu \mathscr{X}_\mu ) 
-ig (\bm{n}_j , [\mathscr{V}_\mu , \mathscr{X}_\mu ]) 
\nonumber\\&
= \partial_\mu  (\bm{n}_j , \mathscr{X}_\mu ) 
- (\partial_\mu \bm{n}_j , \mathscr{X}_\mu ) 
-ig (\bm{n}_j , [\mathscr{V}_\mu , \mathscr{X}_\mu ]) 
\nonumber\\&
= -ig ([\mathscr{V}_\mu , \bm{n}_j] , \mathscr{X}_\mu ) 
-ig (\bm{n}_j , [\mathscr{V}_\mu , \mathscr{X}_\mu ]) 
= 0 
 .
\end{align}
By solving the differential reduction condition for a given $\mathbf A_\mu(x)$, the color field $\bm n(x)$ is  in principle obtained, thereby, ${\bf n}(x)$ is obtained as a functional of the original gauge field $\mathbf A_\mu(x)$.


\subsection{Minimal case for $SU(N)$}

We now discuss general features of the minimal case for $SU(N)$. 
In this case,  $\mathscr A_\mu$ is decomposed into 
$\mathscr{V}_\mu(x)$ and $\mathscr{X}_\mu(x)$, i.e.,  
$
 \mathscr{A}_\mu(x) = \mathscr{V}_\mu(x) + \mathscr{X}_\mu(x) 
$, 
using  only a single color  field $\bm h(x)$ without other  fields $\bm n_j(x)$.
\footnote{
 In the Gell-Mann representation of the generators, we adopt the last diagonal matrix $T_{N^2-1}=H_{N-1}$ for defining a single color field:
\begin{align}
 \bm h(x) :=\bm n_r(x) =U^\dagger(x) H_r U(x)
  .
\end{align}
}
Here we require that $\mathscr{V}_\mu(x)$ and $\mathscr{X}_\mu(x)$ are expressed in terms of $\mathscr{A}_\mu(x)$ and $\bm h(x)$ so as to obey the expected transformation property 
for the given gauge transformations of $\mathscr{A}_\mu(x)$ and $\bm h(x)$.

\subsubsection{Decomposition}

The respective components $\mathscr V_\mu(x)$ and $\mathscr X_\mu(x)$ are specified by two defining equations (conditions):

\noindent
(I)  $\bm{h}(x)$ is a covariant constant in the background $\mathscr{V}_\mu(x)$:
\begin{align}
  0 = D_\mu[\mathscr{V}] \bm{h}(x) 
:=\partial_\mu \bm{h}(x) -  ig [\mathscr{V}_\mu(x), \bm{h}(x)]
 ;
\label{defVL2}
\end{align}
(II)  $\mathscr{X}^\mu(x)$  does not have the $\tilde{H}$-commutative part, i.e., $\mathscr{X}^\mu(x)_{\tilde{H}}=0$:
\begin{align}
  \mathscr{X}^\mu(x)_{\tilde{H}} := \left( {\bf 1} -   2\frac{N-1}{N}  [\bm{h} , [\bm{h} ,  \cdot ]]
\right) \mathscr{X}^\mu(x)   = 0  
\label{defXL2}
 . 
\end{align}
Note that  condition (II) is different from the orthogonality to $\bm{h}(x)$:
$
  (\mathscr{X}_\mu(x), \bm{h}(x)) 
 := 2{\rm tr}(\mathscr{X}_\mu(x) \bm{h}(x) ) 
 = \mathscr{X}_\mu^A(x) \bm{h}^A(x)  = 0  
$, 
which is not sufficient for characterizing the $\tilde{H}$-commutative part, in contrast to the $H$-commutative part.
This is understood from an identity used in the minimal case, 
see Appendix.~\ref{section:minimal-id}.

First, we apply the second  defining equation  (\ref{defXL2}) to $\mathscr{X}^\mu(x)$:
\begin{align}
  \mathscr{X}_\mu(x) 
=& 
\mathscr{X}_\mu(x){}_{\tilde{H}}
+   \frac{2(N-1)}{N}  [\bm{h} , [\bm{h} , \mathscr{X}_\mu(x)]]
\nonumber\\
=&   \frac{2(N-1)}{N}  [\bm{h} , [\bm{h} , \mathscr{X}_\mu(x)]]
 .
\end{align}
By taking into account the first defining equation: 
\begin{align}
 \mathscr{D}_\mu[\mathscr{A}]\bm{h}  
=& \mathscr{D}_\mu[\mathscr{V}]\bm{h}  - ig [\mathscr{X}_\mu, \bm{h} ] 
\nonumber\\ 
=& 
 ig [ \bm{h} , \mathscr{X}_\mu] 
 ,
\end{align}
 $\mathscr{X}_\mu(x)$ is expressed in terms of $\mathscr{A}_\mu(x)$ and $\bm{h}(x)$ as
\begin{align}
 \mathscr{X}_\mu(x) 
=   -ig^{-1}  \frac{2(N-1)}{N}  [\bm{h}(x), \mathscr{D}_\mu[\mathscr{A}]\bm{h}(x) ]
 .
\end{align}
Next, the $\mathscr{V}_\mu$ field is expressed in terms of $\mathscr{A}_\mu(x)$ and $\bm{h}(x)$: 
\begin{align}
  \mathscr{V}_\mu(x) 
  =& \mathscr{A}_\mu(x)  - \mathscr{X}_\mu(x)  
  \nonumber\\
  =& \mathscr{A}_\mu(x)  + ig^{-1}  \frac{2(N-1)}{N}  [\bm{h}(x), \mathscr{D}_\mu[\mathscr{A}]\bm{h}(x) ]
 .
\label{defV12}
\end{align}
Thus, $\mathscr{V}_\mu(x)$ and $\mathscr{X}_\mu(x)$ are  written in terms of $\mathscr A_\mu(x)$ once $\bm h(x)$ is given as a functional of $\mathscr A_\mu(x)$.

We further decompose $\mathscr V_\mu(x)$ into the $\tilde{H}$-commutative part $\mathscr C_\mu(x)$ and the remaining part $\mathscr B_\mu(x)$:
\begin{align}
 \mathscr V_\mu(x)
 =\mathscr C_\mu(x)
  +\mathscr B_\mu(x)
 . 
\end{align}
We rewrite (\ref{defV12}) as
\begin{align}
\mathscr{V}_\mu(x)
  =& \mathscr{A}_\mu(x)   - \frac{2(N-1)}{N}   [\bm{h}(x), [ \bm{h}(x), \mathscr{A}_\mu(x) ] ]
  \nonumber\\
  &+ ig^{-1} \frac{2(N-1)}{N}   [\bm{h}(x) , \partial_\mu  \bm{h}(x) ]
 .
\label{defV2}
\end{align}
The first two terms on the right-hand side of (\ref{defV2})  together  constitute the $\tilde{H}$-commutative part of $\mathscr{A}_\mu(x)$, i.e., $\mathscr{A}_\mu(x)_{\tilde H}$.  
Therefore, we obtain
\begin{subequations}
\begin{align}
  \mathscr{C}_\mu(x) :=& \mathscr{A}_\mu(x){}_{\tilde{H}}
=  \left( {\bf 1} -   2\frac{N-1}{N}  [\bm{h}(x) , [\bm{h} (x),  \cdot ]]
\right) \mathscr{A}_\mu(x)
\nonumber\\
=& \mathscr{A}_\mu(x) - \frac{2(N-1)}{N}   [\bm{h}(x), [ \bm{h}(x), \mathscr{A}_\mu(x) ] ]
,
\\
 \mathscr{B}_\mu(x)
=& ig^{-1} \frac{2(N-1)}{N}[\bm{h}(x) , \partial_\mu  \bm{h}(x) ]
 .
\end{align}
\end{subequations}
In fact, $\mathscr{C}_\mu(x)$ commutes with $\bm{h}(x)$, as we show in (\ref{Ch}):
\begin{align}
    [ \mathscr{C}_\mu(x), \bm{h}(x) ] = 0 
 ,
    \quad (j=1,2, \cdots, r=N-1) 
\label{specC2}
\end{align}
and $\mathscr{B}_\mu(x)$ is noncommutative, $[\mathscr{B}_\mu(x),\bm{h}(x)]  \ne 0$, and is orthogonal to  $\bm{h}(x)$:
\begin{align}
 (\mathscr{B}_\mu(x),  \bm{h}(x))
= 2{\rm tr}(\mathscr{B}_\mu(x) \bm{h}(x) )=0 
 .
\label{defBL22}
\end{align}

Thus, once a single color field $\bm{h}(x)$ is given, we have the decomposition 
\begin{subequations}
\begin{align}
\mathscr A_\mu(x)
 =& \mathscr V_\mu(x)
  +\mathscr X_\mu(x)
 =\mathscr C_\mu(x)
  +\mathscr B_\mu(x)
  +\mathscr X_\mu(x)
 ,
\\
  \mathscr{C}_\mu(x)
=& \mathscr{A}_\mu(x) - \frac{2(N-1)}{N}   [\bm{h}(x), [ \bm{h}(x), \mathscr{A}_\mu(x) ] ]
,
\\
 \mathscr{B}_\mu(x)
=& ig^{-1} \frac{2(N-1)}{N}[\bm{h}(x) , \partial_\mu  \bm{h}(x) ]
,
\\
 \mathscr{X}_\mu(x) =& -ig^{-1}  \frac{2(N-1)}{N}  [\bm{h}(x), \mathscr{D}_\mu[\mathscr{A}]\bm{h}(x) ]
 .
\end{align}
 \label{NLCV-minimal}
\end{subequations}
Thus, all the new variables have been written in terms of $\bm h$ and $\mathscr A_\mu$.

It turns out that $\mathscr X_\mu$ constructed in this way belongs to the coset 
\begin{align}
  \mathscr X_\mu \in \mathscr{G} - \tilde{\mathscr{H}} = su(N)-u(N-1) 
,
\end{align}
since for an arbitrary element $\tilde{{\bm h}} \in \tilde{H}$
\begin{align}
 (\tilde{{\bm h}}, \mathscr{X}_\mu) 
=& 2{\rm tr}(\tilde{{\bm h}} \mathscr{X}_\mu) 
\nonumber\\
=&  -i \frac{2(N-1)}{gN} 2{\rm tr}( \tilde{{\bm h}}  [ {\bm h}, D_\mu{[\mathscr A]} {\bm h} ] )
\nonumber\\
=&  -i \frac{2(N-1)}{gN} 2{\rm tr}( [ \tilde{{\bm h}} , {\bm h}]  D_\mu{[\mathscr A]} {\bm h} )
\nonumber\\
=&   0 
,
\end{align}
where in the last step we have  used 
$[ \tilde{{\bm h}} , {\bm h}]=0$.
Similarly, it is shown that
\begin{align}
  \mathscr B_\mu \in \mathscr{G} - \tilde{\mathscr{H}} = su(N)-u(N-1) 
 .
\end{align}
Moreover, we can show that
\begin{align}
  \mathscr C_\mu \in  \tilde{\mathscr{H}} =  u(N-1) 
 ,
\end{align}
since
\begin{align}
 [ {\bm h} , \mathscr{C}_\mu ]
=& [ {\bm h} , \mathscr{A}_\mu ]  - \frac{2(N-1)}{N} [ {\bm h} , [ [\mathscr A_\mu  ,{\bm h}],  {\bm h}  ] ]
\nonumber\\
=&  [ {\bm h} , \mathscr{A}_\mu ] - \frac{2(N-1)}{N} \left( \frac{2}{N} [ {\bm h} ,  \mathscr{A}_\mu] + \frac{2(2-N)}{\sqrt{2N(N-1)}} [ {\bm h} ,  \{  {\bm h}  , \mathscr{A}_\mu \} ] - [ {\bm h} , \{ {\bm h}, \{ {\bm h}, \mathscr{A}_\mu \} \} ]  \right) 
\nonumber\\
=&  [ {\bm h} , \mathscr{A}_\mu ] - \frac{2(N-1)}{N} \left( \frac{2}{N} [ {\bm h} ,  \mathscr{A}_\mu] + \frac{2(2-N)^2}{2N(N-1)}  [ {\bm h} , \mathscr{A}_\mu  ] - \frac{(2-N)^2}{2N(N-1)}  [ {\bm h} ,  \mathscr{A}_\mu   ]  \right) 
\nonumber\\
=&  [ {\bm h} , \mathscr{A}_\mu ] - [ {\bm h} , \mathscr{A}_\mu ]
\nonumber\\
=& 0
,
\label{Ch}
\end{align}
where we have used
\begin{align}
  [[\mathscr A  ,{\bm h}],  {\bm h}  ]  
=& \{ \mathscr A, \{ {\bm h}, {\bm h} \} \} - \{ {\bm h}, \{ {\bm h}, \mathscr A \} \} 
\nonumber\\
=&  \frac{2}{N} \mathscr A + \frac{2(2-N)}{\sqrt{2N(N-1)}} \{ \mathscr A , {\bm h} \} - \{ {\bm h}, \{ {\bm h}, \mathscr A \} \}   
,
\end{align}
and
\begin{align}
  [ {\bm h} ,  \{ {\bm h}, \mathscr{A}_\mu  \} ]
=&   [ {\bm h}{\bm h} ,    \mathscr{A}_\mu  ]
= \frac{(2-N)}{\sqrt{2N(N-1)}} [ {\bm h} , \mathscr A ]  
 .
\end{align}
Thus,  new variables constructed in this way indeed satisfy the desired property 
\begin{align}
 D_\mu[\mathscr{V}] {\bm h}  
 = D_\mu[\mathscr{B}] {\bm h} -ig[ \mathscr{C}_\mu ,{\bm h}  ]
 = 0
 .
\end{align}
It is instructive to note that the above $\mathscr C_\mu$ is written in the form
\begin{equation}
 \mathscr C_\mu = u_\mu^k \bm n_k, \quad u_\mu^k = (\bm n_k, \mathscr A_\mu), \quad \bm n_k = U^\dagger T_k U \ (T_k \in su(N-1))
 ,
\end{equation}
where $k$ runs over $k=1, \cdots, (N-1)^2$ and $N^2-1$.
Note that these $\bm n_k$ for $k=1, \cdots, (N-1)^2$ are not uniquely defined.  This is because $\bm n_k$ ($k=1, \cdots, (N-1)^2$) can be changed using the rotation within $\tilde H=U(N-1)$ without changing $\bm n_r$, while $\bm n_r$ is invariant under the $\tilde H=U(N-1)$ rotation. 
This feature is discussed for $SU(3)$ in more detail.

\subsubsection{Reduction in the minimal case}

The minimal version of the reduction condition is obtained as follows.
For $G=SU(N)$, the stability group is $\tilde H=U(N-1)$.  Therefore, the respective field variable has the following degrees of freedom at each space-time point:
\begin{itemize}
\item
$\mathscr A_\mu \in \mathscr{L}(G)=su(N)  \rightarrow (N^2-1)D$ ,
\item
$\bm h \in \mathscr{L}(G/\tilde{H})=su(N)-u(N-1) \rightarrow (N^2-1)-(N-1)^2=2(N-1)$ ,
\item
$\mathscr C_\mu \in \mathscr{L}(\tilde{H})=u(N-1) \rightarrow (N-1)^2D$,

\item
$\mathscr X_\mu \in \mathscr{L}(G/\tilde{H})=su(N)-u(N-1) \rightarrow 2(N-1)D$.

\end{itemize}
If we wish to regard the new variables $\mathscr C_\mu$, $\mathscr X_\mu$ and $\bm h$ as those obtained from the original field variable $\mathscr A_\mu$ by the non-linear change of variables
\begin{equation}
 \mathscr A_\mu  \Longrightarrow (\bm h, \mathscr C_\mu  ,  \mathscr X_\mu )  
 ,
\end{equation}
we must give a procedure for eliminating the $2(N-1)$ extra degrees of freedom. 
Here we do not include the variable $\mathscr B_\mu$ in this counting, since it is written only in terms of $\bm h$. 
For this purpose, we impose  $2(N-1)$ reduction conditions $\bm\chi=0$:
\begin{itemize}
\item
$\bm\chi\in{\cal G}/{\cal H}=su(N)/u(N-1) \rightarrow 2(N-1)$.
\end{itemize}
By introducing $\bm h$ in the original $SU(N)$ Yang-Mills theory, the gauge symmetry is enlarged to $SU(N)_{\mathscr A}\times[SU(N)/U(N-1)]_{\bm h}$.  The resulting theory is called the master Yang-Mills theory.  By imposing an appropriate constraint, say, the minimal version of the reduction condition, the master Yang-Mills theory is reduced to the gauge theory with the gauge symmetry $SU(N)$, say, the equipollent gauge symmetry, which is respected by the new variables.

Recall that $\mathbf X_\mu$ is transformed according to the adjoint representation under the equipollent gauge transformation.
As in the $SU(2)$ case, therefore, it is expected that such a constraint is given by minimizing the  functional 
\begin{align}
  \int d^Dx  \frac12 g^2  \mathbf X_\mu\cdot\mathbf X^\mu
  =& \frac{2(N-1)^2}{N^2}\int d^Dx(\bm h\times D_\mu[\mathbf A]\bm h)^2
 \nonumber\\
  =& \frac{N-1}{N}\int d^Dx(D_\mu[\mathbf A]\bm h)^2,
\end{align}
with respect to the enlarged gauge transformation: 
\begin{equation}
\delta\mathbf A_\mu=D_\mu[\mathbf A]\bm\omega,
\quad
\delta\bm h
 =g\bm h\times\bm\theta
 =g\bm h\times\bm\theta_\perp 
 .
\quad
(\bm\omega \in \mathscr{L}(G) ,\bm\theta_\perp \in \mathscr{L}(G/\tilde{H}))
\end{equation}
In fact, the enlarged gauge transformation of the functional $R[\mathbf A, \bm h]$:
\begin{align}
R[\mathbf A, \bm h]
 :=   \int d^Dx \frac12 (D_\mu[\mathbf A]\bm h)^2,
\end{align}
 is given by
\begin{align}
\delta R[\mathbf A, \bm h]
 &= 
   \int d^Dx
   D_\mu[\mathbf A]\bm h
   \cdot\delta(D_\mu[\mathbf A]\bm h)
   \nonumber\\
 &= 
   \int d^Dx
   D_\mu[\mathbf A]\bm h
   \cdot
   (D_\mu[\mathbf A]\delta\bm h
    +g\delta\mathbf A_\mu\times\bm h)
   \nonumber\\
 &= 
   \int d^Dx
   D_\mu[\mathbf A]\bm h
   \cdot
   (D_\mu[\mathbf A](g\bm h\times\bm\theta_\perp)
    +gD_\mu[\mathbf A]\bm\omega\times\bm h)
   \nonumber\\
 &=g
   \int d^Dx
   D_\mu[\mathbf A]\bm h
   \cdot
   D_\mu[\mathbf A]\{\bm h\times(\bm\theta_\perp-\bm\omega)\}
   \nonumber\\
 &=-g
    \int d^Dx
    D^\mu[\mathbf A]D_\mu[\mathbf A]\bm h
   \cdot
   \{\bm h\times(\bm\theta_\perp-\bm\omega_\perp)\}
   \nonumber\\
 &=g
   \int d^Dx
   (\bm\theta_\perp-\bm\omega_\perp)
   \cdot\left(\bm h\times D^\mu[\mathbf A]D_\mu[\mathbf A]\bm h\right)
 ,
\end{align}
where $\bm\omega_\perp$ denotes the component of $\bm\omega$ in the direction $\mathscr{L}(G/\tilde{H})$. 
Thus, we obtain the differential form of the reduction condition:
\begin{equation}
\bm\chi[\mathbf A,\bm h]
 :=\bm h\times D^\mu[\mathbf A]D_\mu[\mathbf A]\bm h
 \equiv0 
  ,
\label{eq:nMAG_minimal_differential}
\end{equation}
which is also expressed in terms of $\mathbf X_\mu$ and $\mathbf V_\mu$ ($\bm h$ and $\mathbf C_\mu$):
\begin{equation}
\bm\chi[\bm h,\mathbf C,\mathbf X]
 :=D^\mu[\mathbf V]\mathbf X_\mu
 \equiv0
 .
\end{equation}
The form of the constraint (\ref{eq:nMAG_minimal_differential}) tells us that 
$\bm\chi\in \mathscr{L}(G/\tilde{H})$, namely, $\bm\chi$ has  no component in the direction $\mathscr{L}(\tilde{H})$.  Therefore, $\bm\chi$  gives $(N^2-1)-(N-1)-2=2(N-1)$ independent conditions as desired.


To determine which part of the symmetry is left after imposing the minimal version of the reduction condition (\ref{eq:nMAG_minimal_differential}), we perform the gauge transformation on the enlarged gauge-fixing functional $\bm\chi$:
{
\allowdisplaybreaks
\begin{align}
\delta\bm\chi
 &=\delta\bm h\times D^\mu[\mathbf A]D_\mu[\mathbf A]\bm h
   +\bm h\times D^\mu[\mathbf A]D_\mu[\mathbf A]\delta\bm h
   \nonumber\\
 &\quad
   +\bm h\times(g\delta\mathbf A_\mu\times D_\mu[\mathbf A]\bm h)
   +\bm h\times D^\mu[\mathbf A](g\delta\mathbf A_\mu\times\bm h)
   \nonumber\\
 &=(g\bm h\times\bm\theta_\perp)
   \times D^\mu[\mathbf A]D_\mu[\mathbf A]\bm h
   +\bm h
    \times D^\mu[\mathbf A]D_\mu[\mathbf A](g\bm h\times\bm\theta_\perp)
   \nonumber\\
 &\quad
   +\bm h\times(gD_\mu[\mathbf A]\bm\omega\times D_\mu[\mathbf A]\bm h)
   +\bm h\times D^\mu[\mathbf A](gD_\mu[\mathbf A]\bm\omega\times\bm h)
   \nonumber\\
 &=g(\bm h\times D^\mu[\mathbf A]D_\mu[\mathbf A]\bm h)
   \times\bm\theta_\perp
   +g(D^\mu[\mathbf A]D_\mu[\mathbf A]\bm h\times\bm\theta_\perp)
    \times \bm h
   \nonumber\\
 &\quad
   +g\bm h
    \times(D^\mu[\mathbf A]D_\mu[\mathbf A]\bm h\times\bm\theta_\perp)
   +2g\bm h
    \times(D^\mu[\mathbf A]\bm h\times D_\mu[\mathbf A]\bm\theta_\perp)
   \nonumber\\
 &\qquad\quad
   +g\bm h
    \times(\bm h\times D^\mu[\mathbf A]D_\mu[\mathbf A]\bm\theta_\perp)
   +g\bm h\times(D^\mu[\mathbf A]\bm\omega\times D_\mu[\mathbf A]\bm h)
   \nonumber\\
 &\quad
   +g\bm h\times(D^\mu[\mathbf A]D_\mu[\mathbf A]\bm\omega\times\bm h)
   +g\bm h\times(D_\mu[\mathbf A]\bm\omega\times D^\mu[\mathbf A]\bm h)
   \nonumber\\
 &=g(\bm h\times D^\mu[\mathbf A]D_\mu[\mathbf A]\bm h)
   \times\bm\theta_\perp
   \nonumber\\
 &\quad
   +2g\bm h
    \times(D^\mu[\mathbf A]\bm h\times D_\mu[\mathbf A]\bm\theta_\perp)
   +g\bm h
    \times(\bm h\times D^\mu[\mathbf A]D_\mu[\mathbf A]\bm\theta_\perp)
   \nonumber\\
 &\quad
   +2g\bm h\times(D^\mu[\mathbf A]\bm\omega\times D_\mu[\mathbf A]\bm h)
   +g\bm h\times(D^\mu[\mathbf A]D_\mu[\mathbf A]\bm\omega\times\bm h)
   \nonumber\\
 &=g(\bm h\times D^\mu[\mathbf A]D_\mu[\mathbf A]\bm h)
   \times\bm\theta_\perp
   \nonumber\\
 &\quad
   +2g\bm h
    \times\{D^\mu[\mathbf A]\bm h
            \times D_\mu[\mathbf A](\bm\theta_\perp-\bm\omega_\perp)\}
   +g\bm h
    \times\{\bm h\times D^\mu[\mathbf A]D_\mu[\mathbf A]
                            (\bm\theta_\perp-\bm\omega_\perp)\}
   \nonumber\\
 &\quad
   +2g\bm h\times(D^\mu[\mathbf A]\bm\omega_\parallel
                    \times D_\mu[\mathbf A]\bm h)
   +g\bm h\times(D^\mu[\mathbf A]D_\mu[\mathbf A]\bm\omega_\parallel
                   \times\bm h)
   \nonumber\\
 &=g(\bm h\times D^\mu[\mathbf A]D_\mu[\mathbf A]\bm h)
   \times\bm\theta_\perp
   \nonumber\\
 &\quad
   +2g\bm h
    \times\{D^\mu[\mathbf A]\bm h
            \times D_\mu[\mathbf A](\bm\theta_\perp-\bm\omega_\perp)\}
   +g\bm h
    \times\{\bm h\times D^\mu[\mathbf A]D_\mu[\mathbf A]
                            (\bm\theta_\perp-\bm\omega_\perp)\}
   \nonumber\\
 &\quad
   +g\bm h\times(D^\mu[\mathbf A]D_\mu[\mathbf A]\bm h
                   \times\bm\omega_\parallel)
   \nonumber\\
 &=g(\bm h\times D^\mu[\mathbf A]D_\mu[\mathbf A]\bm h)
   \times(\bm\theta_\perp+\bm\omega_\parallel)
   \nonumber\\
 &\quad
   +2g\bm h
    \times\{D^\mu[\mathbf A]\bm h
            \times D_\mu[\mathbf A](\bm\theta_\perp-\bm\omega_\perp)\}
   +g\bm h
    \times\{\bm h\times D^\mu[\mathbf A]D_\mu[\mathbf A]
                            (\bm\theta_\perp-\bm\omega_\perp)\} ,
\label{eq:dchi}
\end{align}
}
\noindent
where $\bm\omega_\parallel$ and $\bm\omega_\perp$ denote the components  of $\bm\omega$ in $\mathscr{L}(\tilde{H})$ and $\mathscr{L}(G/\tilde{H})$, respectively.
($\bm\omega=\bm\omega_\parallel+\bm\omega_\perp$).
Here we have used 
\begin{align}
0&=D^\mu[\mathbf A]D_\mu[\mathbf A](\bm\omega_\parallel\times\bm h)
   \nonumber\\
 &=D^\mu[\mathbf A]D_\mu[\mathbf A]\bm\omega_\parallel\times\bm h
   +2D^\mu[\mathbf A]\bm\omega_\parallel\times D_\mu[\mathbf A]\bm h
   +\bm\omega_\parallel\times D^\mu[\mathbf A]D_\mu[\mathbf A]\bm h
 .
\end{align}
This result shows that the minimal version of the reduction  condition $\bm\chi\equiv0$ leaves 
$\bm\theta_\perp=\bm\omega_\perp$ intact. When $\bm\theta_\perp=\bm\omega_\perp$, we find from  (\ref{eq:dchi})
\begin{equation}
 \delta\bm\chi=g\bm\chi\times\bm\alpha 
 , \quad 
\bm\alpha=(\bm\alpha_\parallel,\bm\alpha_\perp)
=(\bm\omega_\parallel,\bm\omega_\perp=\bm\theta_\perp) 
.
\end{equation}

\section{$SU(3)$ case}

We now describe the $SU(3)$ Yang-Mills field explicitly for practical purposes.

\subsection{Maximal case for $SU(3)$}\label{subsection:maximalSU3}

\subsubsection{Decomposition for $SU(3)$: maximal case}

It is obvious that the maximal case of $SU(3)$ goes in the same way as the $SU(N)$ case given above.
We introduce two fields:
\begin{align}
 \bm n_3 := U^\dagger T_3 U , \quad
 \bm n_8 :=  U^\dagger T_8 U , \quad U\in G =SU(3) 
 .
\end{align}
As already demonstrated in the $SU(N)$ case, the key identity for deriving the decomposition is given by
\begin{equation}
\mathbf{V}
 =  {\bf n}_3(\mathbf{V}\cdot{\bf n}_3) + {\bf n}_8(\mathbf{V}\cdot{\bf n}_8) + {\bf n}_3 \times (\mathbf{V} \times {\bf n}_3) + {\bf n}_8 \times (\mathbf{V} \times {\bf n}_8)
  .
\label{eq:Projection OP vector}
\end{equation}
The decomposition of the $SU(3)$ gauge field in the vector notation is
\begin{align}
\mathbf A_\mu
 &=\mathbf V_\mu
  +\mathbf X_\mu 
  =\mathbf C_\mu
  +\mathbf B_\mu
  +\mathbf X_\mu
 ,
\nonumber\\
& \mathbf C_\mu
 :=({\bf n}_3 \cdot\mathbf A_\mu){\bf n}_3
    +({\bf n}_8 \cdot\mathbf A_\mu){\bf n}_8 ,
    \nonumber\\
& \mathbf B_\mu
 :=g^{-1} \partial_\mu{\bf n}_3 \times{\bf n}_3
    +g^{-1} \partial_\mu{\bf n}_8 \times{\bf n}_8 ,
    \nonumber\\
& \mathbf X_\mu
 :=g^{-1}  {\bf n}_3 \times D_\mu[\mathbf A]{\bf n}_3
    + g^{-1}  {\bf n}_8 \times D_\mu[\mathbf A]{\bf n}_8
 .
\end{align} 
It is important to note that $\bm{n}_3$ and $\bm{n}_8$ are introduced simply for convenience and that the independent dynamical degrees of freedom are provided by  a single color field $\bm{n}$. For instance, the simplest choice in the maximal case is given by
\begin{equation}
 {\bf n}_3(x) = {\bf n}(x), \quad 
 {\bf n}_8(x)={\sqrt3} {\bf n}_3(x)*{\bf n}_3(x)
 = {\sqrt3} {\bf n}(x)*{\bf n}(x)
 . 
\end{equation}
For more details, see Appendix~\ref{appendix:moreSU3}.

In a similar way to the $SU(2)$ case discussed in the previous paper \cite{KMS06}, the transformation properties of the decomposed fields $\mathbf B_\mu, \mathbf C_\mu, \mathbf X_\mu$ are uniquely determined once we specify those for $\mathbf A_\mu$ and ${\bf n}$. 
For  independent transformations of $\mathbf{A}_\mu$ and ${\bf n}$:
\begin{align}
 \delta_\omega \mathbf{A}_\mu(x) =&  D_\mu[\mathbf A] \bm \omega(x) , \quad 
\delta_\theta {\bf n}(x)
 =g{\bf n}(x) \times \bm\theta(x)
 =g{\bf n}(x) \times \bm\theta_\perp(x) ,
\nonumber\\
& (\bm\omega \in SU(3), \ \bm\theta_\perp \in SU(3)/(U(1)\times U(1)))
\label{enlarge-gt}
,
\end{align}
the new variables $\mathbf{V}_\mu$ and $\mathbf{X}_\mu$ are transformed under the enlarged gauge transformation as
\begin{align}
  \delta_{\omega,\theta} \mathbf{V}_\mu(x) 
  =&  D_\mu[\mathbf{V}] \bm{\omega}_\parallel(x)  + D_\mu[\mathbf{V}] \bm{\theta}_\perp(x)  + g \mathbf{X}_\mu(x) \times (\bm{\omega}_\perp(x)-\bm{\theta}_\perp(x))
 ,
\nonumber\\
  \delta_{\omega,\theta} \mathbf{X}_\mu(x) 
  =& g \mathbf{X}_\mu(x) \times  (\bm{\omega}_\parallel(x)+\bm{\theta}_\perp(x)) + D_\mu[\mathbf{V}](\bm{\omega}_\perp(x)-\bm{\theta}_\perp(x))
  .
\label{enlarge-gt2}
\end{align}
If $\bm{\omega}_\perp(x)=\bm{\theta}_\perp(x)$, in particular, $ \mathbf{V}_\mu$ and $ \mathbf{X}_\mu$ are transformed as
\begin{align}
 \delta_{\omega} \mathbf{V}_\mu(x) = D_\mu[\mathbf V] \bm \omega(x)  , \quad
 \delta_{\omega} \mathbf{X}_\mu(x) =  g\mathbf{X}_\mu(x) \times \bm \omega(x)
  \label{gt2}
,
\end{align}
which agrees with the equipollent gauge transformation to be obtained by the reduction condition. 

In $D$-dimensional space-time, the original gauge field $\mathbf A_\mu$ for $G=SU(3)$ has $8D$ components.   The   field ${\bf n}_3$ belongs to  $G/\tilde{H}=SU(3)/(U(1)\times U(1))=F_2$ and hence has $8-2=6$ degrees of freedom. 
As already pointed out, ${\bf n}_8$ is constructed from ${\bf n}_3$ by 
$
 {\bf n}_8 = \sqrt3 {\bf n}_3*{\bf n}_3 
  .
$
Note that $\mathbf B_\mu$ is uniquely determined  once ${\bf n}_3$ is fixed. 
The maximal Abelian potential $\mathbf C_\mu$ commutes with both ${\bf n}_3$ and ${\bf n}_8$ and hence $\mathbf C_\mu \in \tilde{H}=U(1)\times U(1)$ has $2D$ components.
In contrast, $\mathbf X_\mu$ has no components in the direction of $\tilde{H}$, i.e., $\mathbf X_\mu\in G/\tilde{H}$  has  $(8-2)D=6D$ components. 
Therefore, there are 6 extra   degrees of freedom after the transformation $\mathbf A_\mu \rightarrow ({\bf n},\mathbf C_\mu,\mathbf X_\mu)$,  although the number of degrees of freedom agrees between 
$({\bf n},\mathbf A_\mu)$  and  $({\bf n},\mathbf C_\mu,\mathbf X_\mu)$.

By introducing a single color fields $\bm n$, the original Yang-Mills theory with the gauge symmetry $SU(3)$ is extended into a gauge theory with the enlarged gauge group $SU(3)\times[SU(3)/(U(1)\times U(1))]$, which we call the master Yang-Mills theory.  To eliminate the 6 extra degrees of freedom,  we must impose 6 independent conditions,  which we call the maximal version of the reduction condition.
Consequently, we can reduce the enlarged gauge symmetry to the original gauge symmetry, i.e., the equipollent gauge symmetry.
Thus, we can regard 
\begin{equation}
 \mathbf A_\mu \rightarrow ({\bf n},\mathbf C_\mu,\mathbf X_\mu)
\end{equation}
as the change of variables and obtain a new reformulation of $SU(3)$ Yang-Mills theory written in terms of the new variables $({\bf n},\mathbf C_\mu,\mathbf X_\mu)$.

\subsubsection{Reduction to $SU(3)$: maximal case}

To find the reduction condition for $SU(3)$, we calculate the $SU(3)$ version of the square of $\mathbf X^\mu$,  as suggested from the $SU(2)$ case. 
By introducing $\bm n_s= \{ \bm n_3, \bm n_8 \}$, the square of  $\mathbf X^\mu$ is expressed as
\begin{align}
g^2 \mathbf X^\mu\cdot\mathbf X_\mu
 &=\{{\bf n}_3\times(D_\mu[\mathbf A]{\bf n}_3)
    +{\bf n}_8\times(D_\mu[\mathbf A]{\bf n}_8)\}^2
   \nonumber\\
 &=\{{\bf n}_3\times(D_\mu[\mathbf A]{\bf n}_3)\}^2
   +2\{{\bf n}_3\times(D_\mu[\mathbf A]{\bf n}_3)\}\cdot
     \{{\bf n}_8\times(D_\mu[\mathbf A]{\bf n}_8)\}
   \nonumber\\
 &\quad
   +\{{\bf n}_8 \times(D_\mu[\mathbf A]{\bf n}_8)\}^2
   \nonumber\\
 &=\{{\bf n}_3\times(D_\mu[\mathbf A]{\bf n}_3)\}^2
   +\{{\bf n}_8\times(D_\mu[\mathbf A]{\bf n}_3)\}^2
   \nonumber\\
 &\quad
   +\{{\bf n}_3\times(D_\mu[\mathbf A]{\bf n}_8)\}^2
   +\{{\bf n}_8\times(D_\mu[\mathbf A]{\bf n}_8)\}^2
   \nonumber\\
 &=(D_\mu[\mathbf A]{\bf n}_3)^2
   +(D_\mu[\mathbf A]{\bf n}_8)^2
=: (D_\mu[\mathbf A]{\bf n}_s)^2
 ,
\end{align}
where in the third equality we have used the fact that 
\begin{equation}
\{{\bf n}_3\times(D_\mu[\mathbf A]{\bf n}_3)\}\cdot
\{{\bf n}_8\times(D_\mu[\mathbf A]{\bf n}_8)\}
 =\{{\bf n}_8\times(D_\mu[\mathbf A]{\bf n}_3)\}^2
 =\{{\bf n}_3\times(D_\mu[\mathbf A]{\bf n}_8)\}^2
 .
\end{equation}
From the transformation property of 
$
(D_\mu[\mathbf A]\bm n_s)^2
$
under the enlarged gauge transformation: 
\begin{align}
& \delta_{\omega,\theta} \left\{ \frac12(D_\mu[\mathbf A]{\bf n}_s)^2 \right\} 
 =g({\bf n}_s\times D_\mu[\mathbf A]{\bf n}_s)\cdot
   D_\mu[\mathbf A](\bm\omega_\perp-\bm\theta_\perp) 
 ,
\end{align}
it turns out that $(D_\mu[\mathbf A]\bm n_s)^2$ is invariant under the equipollent gauge transformation  (\ref{gt2}), i.e., the enlarged gauge transformation (\ref{enlarge-gt2}) when $\bm\omega_\perp=\bm\theta_\perp$.
Therefore, we impose the minimizing condition of the functional $R$ under the enlarged gauge transformation:
\begin{equation}
  R = \int d^Dx \frac12 (D_\mu[\mathbf A]{\bf n}_s(x))^2
   .
\end{equation}
From the observation that  
\begin{align}
 0=  \delta_{\omega,\theta} R =& \int d^Dx \frac12 \delta_{\omega,\theta} (D_\mu[\mathbf A]{\bf n}_s(x))^2
\nonumber\\
  =& - \int d^Dx g(\bm\omega_\perp-\bm\theta_\perp) \cdot ({\bf n}_s \times D_\mu[\mathbf A] D_\mu[\mathbf A]{\bf n}_s)
 ,
\end{align}
the minimizing condition reduces the enlarged gauge symmetry as follows:
\begin{align}
& \tilde G=SU(3)_\omega \times [SU(3)/(U(1)\times U(1))]_\theta 
\nonumber\\&
\rightarrow G'=SU(3)^\prime_\alpha , \quad 
\bm\alpha=(\bm\alpha_\parallel,\bm\alpha_\perp)
=(\bm\omega_\parallel,\bm\omega_\perp=\bm\theta_\perp) 
 .
\end{align}
The equipollent gauge transformation is
\begin{equation}
\delta_\alpha {\mathbf A}_\mu=D_\mu[\mathbf A]\bm\alpha,
\quad
\delta_\alpha {\bf n}=g{\bf n}\times\bm\alpha_\perp 
\label{gtII}
 ,
\end{equation}
and
\begin{align}
 \delta_{\alpha} \mathbf{V}_\mu(x) = D_\mu[\mathbf V] \bm\alpha(x)  , \quad
 \delta_{\alpha} \mathbf{X}_\mu(x) =  g\mathbf{X}_\mu(x) \times \bm\alpha(x)
  .
\end{align}
This is simply the $SU(3)$ version of the equipollent gauge symmetry, as discussed in the previous paper for $SU(2)$ \cite{KMS06}.
Note that the following quantities are also invariant under the equipollent gauge transformation  (\ref{gtII}): 
\begin{equation}
(D_\mu[\mathbf A_\mu]{\bf n}_8)^2,
\quad
(D_\mu[\mathbf A_\mu]{\bf n}_3)\cdot
(D_\mu[\mathbf A_\mu]{\bf n}_8)
 .
\end{equation}
This is also the case for their linear combinations. 

Thus,  minimizing the functional 
$\int d^4x(\mathbf X^\mu\cdot\mathbf X_\mu)$ 
yields the differential form of the $SU(3)$ version of the reduction condition: 
\begin{align}
 \bm{\chi}_{\rm rc}[\mathbf A,\bm n_s]
 := {\bf n}_s \times D_\mu[\mathbf A] D_\mu[\mathbf A]{\bf n}_s 
 ,
\end{align}
which is rewritten in terms of new variables as 
\begin{equation}
 \bm{\chi}_{\rm rc}[\bm n,\mathbf C,\mathbf X]
 :=D^\mu[\mathbf V]\mathbf X_\mu\equiv0
  .
  \label{rc2}
\end{equation}
This is called the differential reduction condition. 
 
Moreover, we can adopt the other functional written in terms of a single color field $\bm n$  to write  the reduction condition as
\begin{equation}
  F[\mathbf A,{\bf n}_3] = \int d^Dx \frac12 (D_\mu[\mathbf A]{\bf n}_3(x))^2
   .
\end{equation}
This  is a new option overlooked so far.  
We can repeat the same calculations as those given above.  
Then we obtain the differential form of the reduction condition: 
\begin{align}
 \bm{\chi}_{\rm rc}[\mathbf A,{\bf n}_3]
 = {\bf n}_3 \times D_\mu[\mathbf A] D_\mu[\mathbf A]{\bf n}_3 
 .
\end{align}
However, this cannot be rewritten into such a simple form as (\ref{rc2}). 

\subsection{Minimal case for $SU(3)$}

\subsubsection{Decomposition for $SU(3)$: minimal case}

According to the general discussion of $SU(N)$, the decomposition in the minimal case of $SU(3)$ is given by
 \begin{align}
\mathbf A_\mu
 &=\mathbf C_\mu
  +\mathbf B_\mu
  +\mathbf X_\mu,
  \\
& \mathbf C_\mu
 :=\mathbf A_\mu
    -\frac43{\bf n}_8\times(\mathbf A_\mu\times{\bf n}_8),
\quad
\\
& \mathbf B_\mu
 :=\frac4{3}g^{-1} \partial_\mu{\bf n}_8\times{\bf n}_8,
\\
& \mathbf X_\mu
 :=\frac4{3} g^{-1} {\bf n}_8\times D_\mu[\mathbf A_\mu]\bm n_8 
 ,
\end{align}
where ${\bf n}_8$ is the single color field ${\bf n}={\bf n}_8$. 
In a similar way to the $SU(2)$ case discussed in the previous paper \cite{KMS06}, the transformation properties of the decomposed fields $\mathbf B_\mu, \mathbf C_\mu, \mathbf X_\mu$ are uniquely determined once we specify those for $\mathbf A_\mu$ and ${\bf n}_8$.  
The infinitesimal enlarged gauge transformation is constructed from  $\delta_\omega{\mathbf A}_\mu$ and $\delta_\theta{\bf n}_8$ as
\begin{align}
\delta_\omega {\mathbf A}_\mu(x)=& D_\mu[\mathbf A]\bm\omega(x) ,
\quad
\delta_\theta {\bf n}_8(x) =g{\bf n}_8(x) \times\bm\theta_\perp(x)
\nonumber\\
& (\bm\omega \in SU(3),\bm\theta_\perp \in SU(3)/U(2))
\label{egt}
 ,
\end{align}
where  $\bm\theta_\perp \in \mathscr{L}(G/H)$. 
The new variables specified above are transformed according to
\begin{align}
  \delta_{\omega,\theta} \mathbf{V}_\mu(x) 
  =&  D_\mu[\mathbf{V}] \bm{\omega}_\parallel(x)  + D_\mu[\mathbf{V}] \bm{\theta}_\perp(x)  + g \mathbf{X}_\mu(x) \times (\bm{\omega}_\perp(x)-\bm{\theta}_\perp(x))
 ,
\nonumber\\
  \delta_{\omega,\theta} \mathbf{X}_\mu(x) 
  =& g \mathbf{X}_\mu(x) \times  (\bm{\omega}_\parallel(x)+\bm{\theta}_\perp(x)) + D_\mu[\mathbf{V}](\bm{\omega}_\perp(x)-\bm{\theta}_\perp(x))
  .
  \label{gt3}
\end{align}
In particular, for $\bm{\omega}_\perp(x)=\bm{\theta}_\perp(x)$, $ \mathbf{V}_\mu$ and $ \mathbf{X}_\mu$ are transformed as
\begin{align}
 \delta_{\omega} \mathbf{V}_\mu(x) = D_\mu[\mathbf V] \bm \omega(x)  , \quad
 \delta_{\omega} \mathbf{X}_\mu(x) =  g\mathbf{X}_\mu(x) \times \bm \omega(x)
.
\end{align}
This is the equipollent gauge transformation, which is to be obtained by imposing the reduction condition,   
 as in the maximal case.

The original gauge field $\mathbf A_\mu$ for $G=SU(3)$ has $8D$ components.  The color field ${\bf n}_8$ belongs to  $G/\tilde{H}=SU(3)/U(2)$ and hence has $8-4=4$ degrees of freedom.  The variable $\mathbf C_\mu$ commutes with ${\bf n}_8$, and hence $\mathbf C_\mu \in \tilde{H}=U(2)$ with $4D$ components. 
Therefore, $\mathbf C_\mu$ is a non-Abelian gauge field, in sharp contrast with the maximal case, where it is the maximal Abelian gauge field. 
Moreover, $\mathbf X_\mu$ has no components in the direction of $\tilde{H}$ (no $\tilde{H}$-commutative part), i.e., $\mathbf X_\mu\in G/\tilde{H}$  has  $(8-4)D=4D$ components. Note that $\mathbf B_\mu$ is written in terms  of ${\bf n}_8$. Therefore, there are 4 extra degrees of freedom associated with the transformation $\mathbf A_\mu \rightarrow ({\bf n}_8,\mathbf C_\mu,\mathbf X_\mu)$,  although  the number of degrees of freedom agrees between 
$({\bf n}_8,\mathbf A_\mu)$  and  $({\bf n}_8,\mathbf C_\mu,\mathbf X_\mu)$. 
\footnote{
Defining 
$
 {\bm n}_{(1)}
 :=U^\dagger T_1U,
$
$
 {\bm n}_{(2)}
 :=U^\dagger T_2U,
$
$
 {\bm n}_{(3)}
 :=U^\dagger T_3U 
$, 
 we find that
$
[ {\bm n}_{(a)} , {\bm n}_8]=0 
\quad(a=1,2,3,8) 
$,
since
$
[T_a,T_8]=0 \quad (a=1,2,3,8) 
$.
For an arbitrary element  $V\in{\cal H}=U(2)$ generated by $\{ T_1,T_2,T_3,T_8 \}$, we define  
$
W:=U^\dagger VU 
$.
Then we find that ${\bm n}_{(a)}$ is transformed as 
$$
 {\bm n}_{(a)}^\prime
 :=W^\dagger {\bm n}_{(a)} W
  =(U^\dagger V^\dagger U)(U^\dagger T_a U)(U^\dagger V U)
  =U^\dagger (V^\dagger T_a V)U
 .
$$
In particular,  $\bm n_8$ is invariant, i.e., $\bm n_8^\prime=\bm n_8$, since $V T_8V^\dagger=T_8$. Therefore,  $\tilde{H}=U(2)$ in $G=SU(3)$ leaves ${\bm n}_8$ invariant.  In other words,  ${\bm n}_8\in G/\tilde{H}$.  Using the degrees of freedom of $\tilde{H}$, we can change ${\bm n}_{(1)}$, ${\bm n}_{(2)}$ and ${\bm n}_{(3)}$ without changing ${\bm n}_8$.
}

\subsubsection{Reduction to $SU(3)$: minimal case}

By introducing a single color  field $\bm n_8$, the original Yang-Mills theory with gauge symmetry $SU(3)$ is extended  to the master Yang-Mills theory with the enlarged gauge group,  $SU(3)\times[SU(3)/U(2)]$.  Therefore, we impose a constraint which we call the minimal version of the reduction condition,  to eliminate the 4 extra   degrees of freedom.  By imposing the minimal version of the reduction condition with the same number of  independent conditions as the extra degrees of freedom, we reduce the enlarged gauge symmetry to the original gauge symmetry $SU(3)$, which is called the equipollent gauge symmetry. 
Thus, we can regard 
\begin{equation}
 \mathbf A_\mu \rightarrow ({\bf n}_8,\mathbf C_\mu,\mathbf X_\mu)
\end{equation}
as the change of variables and obtain a new reformulation of $SU(3)$ Yang-Mills theory written in terms of the new variables $({\bf n}_8,\mathbf C_\mu,\mathbf X_\mu)$.

It turns out that $ \mathbf X^\mu(x) \cdot \mathbf X_\mu(x)=\frac43 (D_\mu[\mathbf A]{\bf n}_8(x))^2$ is invariant under the enlarged gauge transformation (\ref{egt}) when $\bm\omega_\perp=\bm\theta_\perp$.
Then the minimizing functional for $SU(3)$ in the minimal case is given by
\begin{align}
  F 
= \int d^Dx \frac38 g^2\mathbf X^\mu(x) \cdot \mathbf X_\mu(x)
= \int d^Dx \frac12 (D_\mu[\mathbf A]{\bf n}_8(x))^2
 .
\end{align}
We impose the reduction condition as the minimizing condition of the functional:
\begin{align}
 0=  \delta_{\omega,\theta} F =& \int d^Dx \frac12 \delta_{\omega,\theta} \{(D_\mu[\mathbf A]{\bf n}_8(x))\}^2
\nonumber\\
  =& - \int d^Dx g(\bm\omega_\perp-\bm\theta_\perp) \cdot D_\mu[\mathbf A] ({\bf n}_8\times D_\mu[\mathbf A]{\bf n}_8)
\nonumber\\
  =& - \int d^Dx g(\bm\omega_\perp-\bm\theta_\perp) \cdot ({\bf n}_8 \times D_\mu[\mathbf A] D_\mu[\mathbf A]{\bf n}_8)
 .
\end{align}
By imposing the minimizing condition of the functional $F$ under the enlarged gauge transformation, the enlarged gauge symmetry $\tilde G$ is reduced to $G'$:
\begin{align}
& \tilde G=SU(3)_\omega \times [SU(3)/U(2)]_\theta
\nonumber\\&
\rightarrow G'=SU(3)^\prime_\alpha , \quad 
\bm\alpha=(\bm\alpha_\parallel,\bm\alpha_\perp)
=(\bm\omega_\parallel,\bm\omega_\perp=\bm\theta_\perp) 
 .
\end{align}
The equipollent gauge transformation is
\begin{equation}
\delta_{\alpha}{\mathbf A}_\mu(x)=D_\mu[\mathbf A] \bm\alpha(x),
\quad
\delta_{\alpha}{\bf n}_8(x)=g{\bf n}_8(x) \times\bm\alpha_\perp(x) 
 ,
\end{equation}
and
\begin{align}
 \delta_{\alpha} \mathbf{V}_\mu(x) = D_\mu[\mathbf V] \bm\alpha(x)  , \quad
 \delta_{\alpha} \mathbf{X}_\mu(x) =  g\mathbf{X}_\mu(x) \times \bm\alpha(x)
 .
\end{align}

Thus,  minimizing  the functional 
$F$ yields the differential form of the $SU(3)$ version of the reduction condition for the infinitesimal transformation: 
\begin{align}
  \bm{\chi}_{\rm rc}[\mathbf A,\bm n_8] 
 = {\bf n}_8 \times D_\mu[\mathbf A] D_\mu[\mathbf A]{\bf n}_8 
 .
\end{align}
This is rewritten in terms of the new variables as 
\begin{equation}
 \bm{\chi}_{\rm rc}[{\bf n}_8,\mathbf C,\mathbf X]
 :=D^\mu[\mathbf V]\mathbf X_\mu\equiv0
 .
\end{equation}

\subsection{Interpolation between maximal and minimal cases for $SU(3)$}

The maximal and minimal cases are interpolated by introducing a {\it global} parameter $\vartheta$. This is achieved by minimizing the following  functional with respect to the enlarged gauge transformation: 
\begin{align}
 F_\vartheta 
 :=& {\rm const.} \int d^Dx (D_\mu[\mathbf A]{\bf n}(x)) \cdot  (D_\mu[\mathbf A]{\bf n}(x)) 
,
\end{align}
where ${\bf n}(x)$ is introduced in (\ref{su3-n}):
\begin{align}
  {\bf n}(x) 
  =& (\cos \vartheta(x)){\bf n}_3(x)  + (\sin \vartheta(x)) {\bf n}_8(x) 
 .
\end{align}
The minimal case is realized for $\vartheta=\frac{1}{6}\pi,\frac{1}{2}\pi,\frac{5}{6}\pi,\frac{7}{6}\pi,\frac{3}{2}\pi,\frac{11}{6}\pi$ where $\vartheta=\pi/2$ is the simplest choice. 
The maximal case is realized for  other values of $\vartheta$, where  $\vartheta=0$ is the simplest choice. 
This  will be discussed in more detail elsewhere.

\section{Quantization of Yang-Mills theory}\label{appendix:quantization}

To construct a quantum version of  Yang-Mills theory, we adopt the functional integral quantization. 
We now reformulate a quantum Yang-Mills theory in terms of new variables. For this purpose, we must specify the action and the integration measure. 
We can immediately rewrite the Yang-Mills action $S_{\rm YM}[\mathscr{A}]$ in terms of the new variables  by substituting the transformation law (\ref{NLCV-maximal}) or (\ref{NLCV-minimal}) into the original gauge field $\mathscr{A}$ in the Yang-Mills Lagrangian $\mathscr{L}_{\rm YM}[\mathscr{A}]$. 
To write the integration measure explicitly in terms of the new variables, we need to know the Jacobian associated with the change of variables from the original Yang-Mills gauge field to the new variables in the reformulated Yang-Mills theory. 
 
\subsection{Jacobian for the nonlinear change of variables}\label{appendix:jacobian}

We consider the Jacobian associated with the change of variables for 
the decomposition
\begin{equation}
 \mathscr{A}_\mu(x) 
= \mathscr{V}_\mu(x) + \mathscr{X}_\mu(x) 
= \mathscr{C}_\mu(x) + \mathscr{B}_\mu(x) + \mathscr{X}_\mu(x)
  .
  \label{fdec0}
\end{equation}
In the following, we show that the Jacobian can be simplified by choosing appropriate bases in the group space without changing the physical content. 

For $G=SU(2)$,  we have already introduced \cite{Kondo06} the orthonormal basis 
$ 
 ({\bf e}_1(x), {\bf e}_2(x), {\bf n}(x))
$  
satisfying the following relations in  vector form using   ${\bf e}_a(x)=(e_a^A(x))$ , ${\bf n}(x) =(n^A(x))$:
\begin{align}
   {\bf e}_a(x) \cdot {\bf e}_b(x) =& \delta_{ab} , \quad 
    {\bf e}_a(x) \cdot {\bf n}(x)  = 0, \quad
   {\bf n}(x) \cdot {\bf n}(x)  = 1
    ,
 \nonumber\\
  {\bf e}_a(x) \times {\bf e}_b(x) =& \epsilon_{ab} {\bf n}(x) , \quad
 {\bf e}_a(x) \times  {\bf n}(x) = -\epsilon_{ab}  {\bf e}_b(x)   \ \quad (a,b=1,2) 
  ,
\label{basis}
\end{align} 
which are written in the Lie algebra form using $\bm{e}_a(x)= e_a^A(x) T^A$ , $\bm{n}(x) = n^A(x) T^A$ as
\begin{align}
   {\rm tr}[\bm{e}_a(x) \bm{e}_b(x)] =& \delta_{ab}/2 , \quad 
   {\rm tr}[\bm{e}_a(x) \bm{n}(x)]     = 0, \quad
   {\rm tr}[\bm{n}(x) \bm{n}(x)]  = 1/2
    ,
 \nonumber\\
  [ \bm{e}_a(x) , \bm{e}_b(x)] =& i\epsilon_{ab} \bm{n}(x) , \quad
 [ \bm{e}_a(x) , \bm{n}(x)   ] = -i\epsilon_{ab}  \bm{e}_b(x)   \ \quad (a,b=1,2).
\label{basis2}
\end{align} 
By definition, $\mathscr{C}_\mu(x)$ is parallel (or commutes with) $\bm{n}(x)$, i.e.,
\begin{equation}
  [ \mathscr{C}_\mu(x), \bm{n}(x)] = 0 
  .
\end{equation}
Hence, for SU(2), $\mathscr{C}_\mu(x)$ can be written  as
\begin{equation}
    \mathscr{C}_\mu(x) = C_\mu(x) \bm{n}(x) 
 ,
\end{equation}
which has the components 
\begin{equation}
    \mathscr{C}_\mu^A(x) = C_\mu(x) n^A(x)
 .
\end{equation}
On the other hand, $\mathscr{B}_\mu(x)$ and $\mathscr{X}_\mu(x)$ are orthogonal to (or noncommutative with) ${\bf n}(x)$,
\begin{equation}
    \mathscr{B}_\mu(x) \cdot {\bf n}(x)  = 0, \quad
    \mathscr{X}_\mu(x) \cdot {\bf n}(x)  = 0
, 
\end{equation}
and hence can be written using orthonormal bases $\{ \bm{e}_a(x) \}$ as
\begin{subequations}
\begin{equation}
    \mathscr{B}_\mu(x) = B_\mu^a(x) \bm{e}_a(x) ,
    \quad
    \mathscr{X}_\mu(x) = X_\mu^a(x) \bm{e}_a(x)  
  ,
\end{equation}
which yield the components 
\begin{equation}
    \mathscr{B}_\mu^A(x) = B_\mu^a(x) e_a^A(x) ,
    \quad
    \mathscr{X}_\mu^A(x) = X_\mu^a(x) e_a^A(x) 
  .
\end{equation}
\end{subequations}

For $G=SU(N)$ ($N \ge 3$), we introduce  complete orthonormal bases $\{  \bm{e}_a , \bm{u}_j  \}$ for $su(N)$ where 
\begin{align}
  \bm{e}_a \in \mathscr{L}(G/\tilde{H}), \quad
\bm{u}_j \in \mathscr{L}(\tilde{H})   ,
\
( a=1, \cdots, {\rm dim}G/\tilde{H};   
  j=1, \cdots, {\rm dim}\tilde{H} )
\end{align}
which satisfy  the relations in  vector form using   $\bm{e}_a(x)=(e_a^A(x))$ , $\bm{u}_j(x) =(\bm{u}_j^A(x))$:
\begin{align}
  \bm{e}_a(x) \cdot \bm{e}_b(x) = \delta_{ab}  ,   
    \quad 
   \bm{e}_a(x) \cdot \bm{u}_j(x)  = 0 
    \quad
  \bm{u}_j(x) \cdot \bm{u}_k(x) = \delta_{jk}  ,
\nonumber\\ 
( a,b=1, \cdots, {\rm dim}G/\tilde{H}; j,k=1, \dots, {\rm dim}\tilde{H})
 ,
\end{align}
which are written in the Lie algebra form using $\bm{e}_a(x)= e_a^A(x) T^A$ , $\bm{u}_j(x) = \bm{u}_j^A(x) T^A$:
\begin{align}
   {\rm tr}[\bm{e}_a(x) \bm{e}_b(x)] =& \delta_{ab}/2 , \quad 
   {\rm tr}[\bm{e}_a(x) \bm{u}_j(x)]     = 0, \quad
   {\rm tr}[\bm{u}_j(x) \bm{u}_k(x)]  = 1/2 \delta_{jk} 
    .
\end{align} 
It is instructive to note that $\bm{u}_j$ reduces to a single color field $\bm{n}$ in SU(2).

The fields $\mathscr{B}_\mu(x)$ and $\mathscr{X}_\mu(x)$ are orthogonal to (or noncommutative with) all $\bm{u}_j(x)$:  
\begin{equation}
    \mathscr{B}_\mu(x) \cdot \bm{u}_j(x)  = 0, \quad
    \mathscr{X}_\mu(x) \cdot \bm{u}_j(x)  = 0
    \ (j=1, \cdots, {\rm dim}\tilde{H})
 .
\end{equation}
Hence they can be written using the bases $\{ \bm{e}_a(x) \}$: 
\begin{align}
    \mathscr{B}_\mu(x) =& B_\mu^a(x) \bm{e}_a(x) \in \mathscr{L}(G/\tilde{H}) ,
 \nonumber\\
    \mathscr{X}_\mu(x) =& X_\mu^a(x) \bm{e}_a(x) \in \mathscr{L}(G/\tilde{H})  
    \ (a=1, \cdots, {\rm dim}G/\tilde{H})
  .
  \label{fdec2}
\end{align}
By definition, 
$\mathscr{C}_\mu(x)$ can be written as
\begin{equation}
    \mathscr{C}_\mu(x) = C_\mu^j(x) \bm{u}_j(x) \in \mathscr{L}(\tilde{H}) 
    \  (j=1, \cdots, {\rm dim}\tilde{H})
  .
    \label{fdec1}
\end{equation}
Thus, in the minimal case, $\tilde{H} =U(N-1)$, and the decomposition is 
\begin{align}
   \mathscr{A}_\mu^A(x) =   C_\mu^j(x) \bm{u}_j^A(x)   
+ B_\mu^a(x) \bm{e}_a^A(x)    
+ X_\mu^a(x) \bm{e}_a^A(x)  
\nonumber\\
    \ (j=1, \cdots, {\rm dim}\tilde{H}=(N-1)^2; 
\ a=1, \cdots, {\rm dim}G/\tilde{H}=2N-2)
  .
  \label{fdec3}
\end{align}
In the maximal case,  $\tilde{H}=H=U(1)^{N-1}$ and  
$\bm{u}_j(x)$ reduce to $\bm{n}_j(x)$ ($j=1, \cdots, {\rm dim}\tilde{H}$), which mutually commute, i.e., 
$
  [ \bm{n}_j(x), \bm{n}_k(x)] = 0
$ (\ref{com}).
Therefore, we obtain
\begin{align}
   \mathscr{A}_\mu^A(x) =   C_\mu^j(x) \bm{n}_j^A(x)   
+ B_\mu^a(x) \bm{e}_a^A(x)    
+ X_\mu^a(x) \bm{e}_a^A(x)  
\nonumber\\
    \ (j=1, \cdots, {\rm dim}\tilde{H}=N-1; 
\ a=1, \cdots, {\rm dim}G/\tilde{H}=N^2-N)
   .
  \label{fdec4}
\end{align}

We now consider the change of variables from 
$\{ n^\alpha, \mathscr{A}_\mu^A \}$ to the new variables 
$\{ n^\beta, C_\nu^k, X_\nu^b \}$ defined in the $su(N)$ bases $\{  \bm{e}_b , \bm{u}_k  \}$. 
Here $n^\alpha$ and $n^\beta$ denote  independent degrees of freedom after solving the constraint $n^A n^A=1$. Then the integration measure is transformed as
\begin{equation}
 \mathcal{D}n^\alpha \mathcal{D}\mathscr{A}_\mu^A  
 =   \mathcal{D}n^\beta \mathcal{D}C_\nu^k \mathcal{D}X_\nu^b J  
 .
\end{equation}
The Jacobian  $J$ is given by
\begin{align}
 J 
:=& \begin{vmatrix}
 \frac{\partial n^\alpha}{\partial n^\beta} & \frac{\partial n^\alpha}{\partial C_\nu^k} & \frac{\partial n^\alpha}{\partial X_\nu^b} \cr
 \frac{\partial \mathscr{A}_\mu^A}{\partial n^\beta} & \frac{\partial \mathscr{A}_\mu^A}{\partial C_\nu^k} & \frac{\partial \mathscr{A}_\mu^A}{\partial X_\nu^b} \cr
  \end{vmatrix}
=  \begin{vmatrix}
 \delta^\alpha_\beta & 0 & 0 \cr
 \frac{\partial \mathscr{A}_\mu^A}{\partial n^\beta} & \delta_{\mu\nu} u_k^A & \delta_{\mu\nu} e_b^A \cr
  \end{vmatrix}
\nonumber\\
=& \begin{vmatrix}
   \delta_{\mu\nu} u_k^A & \delta_{\mu\nu} e_b^A \cr
  \end{vmatrix} 
=  \begin{vmatrix}
    u_k^A &  e_b^A \cr
  \end{vmatrix}^D
= 1
 ,
\end{align}
where we have used the fact that $\{ n^\beta, C_\nu^k, X_\nu^b \}$ are independent variables in the second equality,  that $\mu$ and $\nu$ run from $1$ to $D$ in the fourth equality, and that $\{ u_k^A ,  e_b^A\}$ constitute  the orthonormal base in the last equality. 
Thus, we obtain
\begin{equation}
 \mathcal{D}n^\alpha \mathcal{D}\mathscr{A}_\mu^A  
 =   \mathcal{D}n^\beta \mathcal{D}C_\nu^k \mathcal{D}X_\nu^b J, \quad J = 1
 .
\end{equation}
For the Jacobian $J$ to be simplified to $J=1$, the fact that $C_\nu^k, X_\nu^b$ are independent of $n^\beta$ is important.

\subsection{Functional integration}\label{appendix:integration}

We now reformulate the quantum Yang-Mills theory by the functional integral method. 
The following steps can be clearly understood by comparing them with the corresponding ones in Fig.~\ref{fig:enlarged-YM}.

First, the original Yang-Mills (YM) theory in the Euclidean space is defined by the partition function (see the middle left  part in Fig.~\ref{fig:enlarged-YM}):
\begin{align}
 Z_{{\rm YM}} = \int \mathcal{D}\mathscr{A}_\mu^A \exp (-S_{{\rm YM}}[\mathscr{A}]) .
\end{align}
Of course,  to obtain a well-defined Yang-Mills theory by completely fixing the gauge, we must impose a gauge fixing condition for the $SU(N)$ gauge symmetry, for example, the Landau gauge $\partial^\mu \mathscr{A}_\mu^A(x)=0$, which is called the overall gauge fixing condition  (see the down arrows in Fig.~\ref{fig:enlarged-YM}). 
For a while, we disregard this procedure, as it can be carried out following the standard method.

Second, the color field  $\bm{n}(x)$ is introduced to extend the original YM theory (see the up arrow in Fig.~\ref{fig:enlarged-YM})   to the enlarged Yang-Mills theory called the master Yang-Mills  (M-YM) theory, which is defined by a partition function written in terms of both $n^\alpha(x)$ and $\mathscr{A}_\mu^A(x)$ (see the top part in Fig.~\ref{fig:enlarged-YM}),
\begin{align}
 \tilde{Z}_{{\rm YM}}  = \int \mathcal{D}n^\alpha
\int \mathcal{D}\mathscr{A}_\mu^A \exp (-S_{{\rm YM}}[\mathscr{A}]) 
 . 
\label{Z}
\end{align}

Third, we regard (\ref{Z}) as 
\begin{align}
 \tilde{Z}_{{\rm YM}}  
= \int \mathcal{D}n^\beta
 \int \mathcal{D}C_\nu^k \int 
\mathcal{D}X_\nu^b 
J 
 \exp (-\tilde S_{\rm YM}[\bm n, \mathscr{C},\mathscr{X}]) 
 , 
 \label{Z2}
\end{align}
where $J$ is the Jacobian associated with the change of variables,  
and the action $\tilde S_{\rm YM}[\bm n, \mathscr{C},\mathscr{X}]$ is obtained by substituting the  decomposition (\ref{fdec3}) of $\mathscr A_\mu^A$
 into $S_{{\rm YM}}[\mathscr{A}]$: 
\begin{align}
\tilde S_{\rm YM}[\bm n, \mathscr{C},\mathscr{X}]
 =S_{\rm YM}[\mathscr A] .
\end{align}
Starting with this form, we wish to obtain a new Yang-Mills (YM') theory written in terms of new variables 
$
(n^\beta, C_\nu^k, X_\nu^b)
$
(see the middle right part in Fig.~\ref{fig:enlarged-YM}), which is equipollent to the original Yang-Mills theory.
Here, as already discussed in the previous subsection, 
\begin{align}
 (n^\alpha(x), \mathscr{A}_\mu^A(x)) \rightarrow (n^\beta(x), C_\nu^k(x), X_\nu^b(x))
\end{align}
should  be considered as a change of variables in the sense that $n^\beta(x), C_\nu^k(x), X_\nu^b(x)$ become {\it independent} field variables in the new YM' theory.

However, the M-YM has enlarged gauge symmetry, which is larger than the original gauge symmetry $G=SU(N)$. Therefore,  to obtain a YM' theory that is equipollent to the original YM theory, the extra gauge degrees of freedom must be eliminated. 
To fix the enlarged gauge symmetry in the M-YM theory and retain only the same gauge symmetry as that in the original YM, therefore, 
we impose the constraint $\bm{\chi}[ \mathscr{A},\bm{n}]=0$ (the reduction condition), 
which is also written in terms of the new variables as 
$
\tilde{\bm\chi} 
 :=\tilde{\bm\chi} [\bm n, \mathscr{C},\mathscr{X}]=0
$.
(see the diagonal down arrow in Fig.~\ref{fig:enlarged-YM}).

The constraint $\bm{\chi}[ \mathscr{A},\bm{n}]=0$ is introduced in the functional integral as follows.
  We write  unity in the form 
\begin{align}
  1 = \int \mathcal{D} \bm{\chi}^\theta \delta(\bm{\chi}^\theta)
=   \int\! \mathcal{D}\bm\theta\delta(\bm\chi^\theta)
   \det\left(\frac{\delta\bm\chi^\theta}{\delta{\bm\theta}}\right) ,
\end{align}
where $\bm{\chi}^\theta$ is the constraint written in terms of the gauge-transformed variables, i.e., 
$\bm{\chi}^\theta:=\bm{\chi}[ \mathscr{A},\bm{n}^\theta]$.
As an intermediate step for moving from M-YM to YM', we insert  this into the functional integral   (\ref{Z2}), 
and cast the partition function of the M-YM theory into an intermediate form 
(see the middle right part in Fig.~\ref{fig:enlarged-YM})
\begin{align}
 \tilde{Z}_{{\rm YM}}  
=& \int \mathcal{D}n^\beta
 \int \mathcal{D}C_\nu^k \int 
\mathcal{D}X_\nu^b 
J 
\nonumber\\
 & \times \int\! \mathcal{D}\bm\theta\delta(\bm\chi^\theta)
   \det\left(\frac{\delta\bm\chi^\theta}{\delta{\bm\theta}}\right)
 \exp (-\tilde S_{\rm YM}[\bm n, \mathscr{C},\mathscr{X}])
 . 
\end{align}
We next perform the change of variables $\bm{n} \rightarrow \bm{n}^{\theta}$ obtained through a local rotation by   angle $\theta$ and the corresponding gauge   transformations for the other new variables $\mathscr{C}_\mu$ and $\mathscr X_\mu$: 
$\mathscr{C}_\mu, \mathscr{X}_\mu \rightarrow \mathscr{C}_\mu^{\theta}, \mathscr{X}_\mu^{\theta}$. 
From the gauge invariance of the action $\tilde S_{\rm YM}[\bm n, \mathscr{C},\mathscr{X}]$ and the measure 
$
 \mathcal{D}n^\beta  \mathcal{D}C_\nu^k \mathcal{D}X_\nu^b 
$, 
we can rename the dummy integration variables $\bm{n}^{\theta}, \mathscr{C}_\mu^{\theta}, \mathscr{X}_\mu^{\theta}$  as $\bm{n}, \mathscr{C}_\mu, \mathscr{X}_\mu$, respectively.
Thus, the integrand does not depend on $\theta$, and the gauge volume $\int\! \mathcal{D}\bm\theta$ can be removed: 
\begin{align}
 \tilde{Z}_{{\rm YM}}  
=& \int\! \mathcal{D}\bm\theta 
\int \mathcal{D}n^\beta
 \int \mathcal{D}C_\nu^k \int 
\mathcal{D}X_\nu^b 
J  
\nonumber\\
 & \times \delta(\bm\chi)
   \det\left(\frac{\delta\bm\chi}{\delta{\bm\theta}}\right)
 \exp (-\tilde S_{\rm YM}[\bm n, \mathscr{C},\mathscr{X}]) .
\end{align}
Thus, we have arrived at the reformulated  YM' theory with the partition function:
\begin{align}
 Z_{{\rm YM}}^\prime  
=  \int \mathcal{D}n^\beta
 \int \mathcal{D}C_\nu^k \int 
\mathcal{D}X_\nu^b 
J 
\delta(\tilde{\bm\chi})  
   \Delta_{FP}^{rc}
 \exp (-\tilde S_{\rm YM}[\bm n, \mathscr{C},\mathscr{X}]) , 
\end{align}
where the reduction condition is written in terms of the new  variables:
\begin{equation}
\tilde{\bm\chi} 
 :=\tilde{\bm\chi} [\bm n, \mathscr{C},\mathscr{X}]
 :=D^\mu[\mathscr{V}]\mathscr{X}_\mu 
 , 
\end{equation}
and 
$
 \Delta_{FP}^{rc}
:= \det\left(\frac{\delta\bm\chi}{\delta{\bm\theta}}\right)_{\bm{\chi}=0}
=   \det\left(\frac{\delta\bm\chi}{\delta\bm n^\theta}\right)_{\bm{\chi}=0} 
$
is  the Faddeev-Popov determinant associated with the reduction condition.
It is important to note that the independent variables are regarded as $n^\beta(x)$, $C_\nu^k(x)$ and $X_\nu^b(x)$ in the reformulated Yang-Mills theory to simplify the Jacobian.

As already mentioned above, in order to obtain a well-defined Yang-Mills theory by completely fixing the gauge, we must impose the  overall gauge fixing condition  (see the down arrows in Fig.~\ref{fig:enlarged-YM}) in addition to the reduction condition. 
However, we have omitted to write this procedure while simplifying the presentation, since the overall gauge fixing can be performed according to standard procedures. 
Moreover, a systematic treatment of  gauge fixing and the associated Faddeev-Popov determinant or the introduction of ghost fields can be carried out using the BRST symmetry of the new theory. 
Although the BRST treatment can also be performed for the $SU(N)$ case following the method already performed for $SU(2)$ \cite{KMS06}, 
such a treatment is rather involved.  Therefore, we have given a heuristic explanation based on the Faddeev-Popov trick for $SU(N)$ in the above, even though it is possible to develop the BRST approach.
Some applications of this reformulation will be given in a subsequent paper.

\section{Conclusion and discussion}\label{appendix:conclusion}

In this paper we have proposed a new version of the $SU(N)$ Yang-Mills theory  reformulated in terms of new field variables, which are obtained by a nonlinear change of variables from the original Yang-Mills gauge field. 
The basic idea of our reformulation originates from the Cho-Faddeev-Niemi decomposition of the gauge potential.
However, our reformulation differs from other proposals \cite{Cho80c,FN99a} for the gauge group $SU(N), N \ge 3$, including the $SU(3)$ case, which is most interesting  from the physical  viewpoint, although it agrees with the conventional approach for the $SU(2)$ gauge  group.
In particular, we have mainly studied the maximal and minimal cases.

This reformulation of  Yang-Mills theory has already been applied to a few physical problems. 
It  enables us to study the low-energy dynamics of   Yang-Mills theory by explicitly extracting the topological degrees of freedom such as magnetic monopoles and vortices \cite{Kondo08b}.
Such topological configurations are believed to provide the dominant degrees of freedom responsible for quark confinement from the viewpoint of  dual superconductivity.
Using the reformulation,  the dual superconductivity in Yang-Mills theory can indeed be understood in a gauge-invariant way, as demonstrated recently by  a non-Abelian Stokes theorem for the Wilson loop operator \cite{Kondo08}.  
The non-Abelian Stokes theorem has played a crucial role in understanding the gauge-invariant meaning of the Abelian dominance in the Wilson loop operator \cite{KS08}, which was confirmed using the Abelian projection approach. 
From this viewpoint, the reformulation given in this paper will give a  useful framework for other applications, which will be discussed in subsequent papers. 
 

\appendix
\section{$SU(N)$ group}\label{appendix:SU(N)-group}

\par
 The group $G=SU(N)$ has rank $N-1$, and the Cartan subalgebra is constructed from $N-1$ diagonal generators $H_k$.  Hence, there must be $N(N-1)$ off-diagonal shift operators $E_\alpha$, since ${\rm dim} SU(N):=N^2-1=(N-1)+N(N-1)$. Therefore, the total number of roots is $N(N-1)$, of which there are $N-1$ simple roots,  and the other roots are constructed as linear combinations of the simple roots.  

An element of $SU(N)$ in the fundamental representation is expressed by the $N\times N$ unitary matrices with determinant $1$ that are generated by traceless Hermitian matrices, i.e., $N^2-1$ linearly independent generators $T^A(A=1,\cdots,N^2-1)$.  The generators are normalized as 
$
 {\rm tr}(T_A T_B) = {1 \over 2} \delta_{AB} .
$
The diagonal generator $H_m$ is defined by
\begin{eqnarray}
 (H_m)_{ab} &=& {1 \over \sqrt{2m(m+1)}}
 (\sum_{k=1}^{m} \delta_{ak}\delta_{bk} - m \delta_{a,m+1}\delta_{b,m+1})
 \\
 &=& {1 \over \sqrt{2m(m+1)}}{\rm diag}(1,\cdots,1,-m,0,\cdots,0)
  ,
\label{H}
\end{eqnarray}
where for $m=1$ to $N-1$, the first $m$ diagonal elements (beginning from the upper left-hand corner) of $H_m$ are $1$, the next element is $-m$, and the rest of the diagonal elements are $0$.  Thus, $H_m$ is traceless.
Each off-diagonal generator $E_\alpha$ has a single nonzero element $1/\sqrt{2}$. 

The weight vector $\nu_j$ is  defined as the eigenvector of  $H_j$, i.e., 
\begin{equation}
H_j |\nu \rangle = \nu^j |\nu \rangle
 .
\end{equation} 
We have $N$ weight vectors in the fundamental representation ${\bf N}$ ($N$-dimensional irreducible representation of $SU(N)$) given  by 
\begin{eqnarray}
  \nu^1 &=& \left({1 \over 2}, {1 \over 2\sqrt{3}}, \cdots,
  {1 \over \sqrt{2m(m+1)}}, \cdots, {1 \over \sqrt{2(N-1)N}}\right) ,
  \nonumber\\
  \nu^2 &=& \left(-{1 \over 2}, {1 \over 2\sqrt{3}}, \cdots,
  {1 \over \sqrt{2m(m+1)}}, \cdots, {1 \over \sqrt{2(N-1)N}}\right) ,
  \nonumber\\
  \nu^3 &=& \left(0, -{1 \over \sqrt{3}}, {1
\over 2\sqrt{6}},\cdots,  {1 \over \sqrt{2(N-1)N}}\right) ,
  \nonumber\\
\vdots
  \nonumber\\
  \nu^{m+1} &=& \left(0,0, \cdots,0,
  -{m \over \sqrt{2m(m+1)}}, \cdots, {1 \over \sqrt{2(N-1)N}}
\right) ,
  \nonumber\\
\vdots
  \nonumber\\
  \nu^N &=& \left(0, 0, \cdots, 0, {-N+1 \over \sqrt{2(N-1)N}}
\right) .
\end{eqnarray}
All the weight vectors have the same length, and the angles between different weights are the same:
\begin{equation}
 \nu^i \cdot \nu^i = {N-1 \over 2N} .
 \quad \nu^i \cdot \nu^j = -{1 \over 2N} . \quad ({\rm for}\  i \not= j) 
 \label{wr}
\end{equation}
The weights constitute a polygon in the $(N-1)$-dimensional space.  This implies that any weight can be used as the highest weight. A weight is called {\bf positive} if its {\it last} nonzero component is positive. With this definition, the weights satisfy 
\begin{eqnarray}
 \nu^1 > \nu^2 > \cdots > \nu^N .
\label{order}
\end{eqnarray}
The simple roots are given by  
\begin{equation}
 \alpha^i = \nu^i - \nu^{i+1}   ,
\quad (i=1, \cdots, N-1) 
\end{equation}
and are written explicitly as
\begin{eqnarray}
  \alpha^1 &=& \left(1, 0, \cdots, 0 \right),
  \nonumber\\
  \alpha^2 &=& \left(-{1 \over 2}, {\sqrt{3} \over 2}, 0, \cdots,0 \right),
  \nonumber\\
  \alpha^3 &=& \left(0, -{1 \over \sqrt{3}}, \sqrt{{2 \over 3}},0, \cdots, 0 \right),
  \nonumber\\
\vdots
  \nonumber\\
  \alpha^{m} &=& \left(0,0, \cdots, 
  -\sqrt{{m-1 \over 2m}}, \sqrt{{m+1 \over 2m}}, 0, \cdots, 0  \right),
  \nonumber\\
\vdots
  \nonumber\\
  \alpha^{N-1} &=& \left(0, 0, \cdots,  -\sqrt{{N-2 \over 2(N-1)}},
\sqrt{{N \over 2(N-1)}} \right) .
\label{simple root}
\end{eqnarray}
As can be shown from (\ref{wr}),
all these roots have length 1, the angles between successive roots are the same, $2\pi/3$, and other pairs of roots are orthogonal:
\begin{eqnarray}
   \alpha^j \cdot \alpha^j = 1,
  \quad \alpha^i \cdot \alpha^j = - {1 \over 2}, \quad (j=i\pm1) 
  \nonumber\\
  \alpha^i \cdot \alpha^j = 0 . \quad (j \not= i, i\pm 1)
  \label{rr}
\end{eqnarray}


If we choose $\nu^1$ as the highest-weight $\vec \Lambda$ of the fundamental representation ${\bf N}$, some of the roots are orthogonal to $\nu^1$.    From the above construction, it is easy to see that only one simple root
$\alpha_1$ is non-orthogonal to $\nu^1$, and that all the other simple roots are orthogonal:
\begin{equation}
 \nu^1 \cdot \alpha^1 \not= 0,
 \quad \nu^1 \cdot \alpha^2 = \nu^1 \cdot \alpha^3
 = \cdots = \nu^1 \cdot \alpha^{N-1} = 0 .
\end{equation}
Therefore, all the linear combinations constructed from  $\alpha^2, \cdots, \alpha^{N-1}$ are also orthogonal to $\nu^1$. Nonorthogonal roots are only  obtained when $\alpha^1$ is included in the linear combinations.  It is not difficult to show that the total number of non-orthogonal roots is $2(N-1)$, and hence there are $N(N-1)-2(N-1)=(N-2)(N-1)$ orthogonal roots.  The $(N-2)(N-1)$
shift operators $E_\alpha$ corresponding to these orthogonal roots together with the $N-1$ diagonal generators of the Cartan subalgebra $H_k$ constitute  the maximal stability subgroup
$\tilde H=U(N-1)$, since $(N-2)(N-1)+(N-1)=(N-1)^2={\rm dim} U(N-1)$.  Thus, for the fundamental representation, the stability subgroup $\tilde H$ of $SU(N)$ is given  by 
\begin{equation}
  \tilde H=U(N-1) 
 . 
\end{equation}
To describe the coset space $G/\tilde H$, we need only $N-1$ complex numbers, since
\begin{equation}
 G/\tilde H = SU(N)/U(N-1) =CP^{N-1}
\end{equation}
has ${\rm dim} G/\tilde H= 2(N-1)$,  
where $CP^{N-1}=P^{N-1}(\mathbb{C})$ is the $(N-1)$-dimensional complex projective space, a submanifold of the flag manifold $F_{N-1}$.

\section{Derivation of identities}\label{appendix:idv}

\subsection{Maximal case}\label{section:maximal-id}

The identity (\ref{idv}) is proved as follows.
We consider any $su(N)$ Lie algebra valued function $\mathscr{V}$.
Using the adjoint rotation, $\mathscr{V}'=U\mathscr{V}U^\dagger$, we have only to prove
\begin{align}
  \mathscr{V}' = \sum_{j=1}^{N-1} H_j (H_j, \mathscr{V}') + \sum_{j=1}^{N-1} [H_j, [H_j, \mathscr{V}']] .
\label{idd}
\end{align}
The Cartan decomposition for $\mathscr{V}'$ is
\begin{align}
  \mathscr{V}' 
=  \sum_{k=1}^{N-1} V^k H_k + \sum_{\alpha=1}^{(N^2-N)/2} (W^*{}^{\alpha} \tilde{E}_{\alpha} + W^{\alpha} \tilde{E}_{-\alpha})  ,
\end{align}
where the Cartan basis is given by
\begin{align}
 \vec{H} =& (H_1, H_2, H_3, \cdots, H_{N-1}) = (T^3, T^8, T^{15}, \cdots, T^{N^2-1}) ,
\nonumber\\
  \tilde{E}_{\pm 1} =& {1 \over \sqrt{2}}(T^1 \pm  i T^2) , \quad
  \tilde{E}_{\pm 2} = {1 \over \sqrt{2}}(T^4 \pm  i T^5) ,  
  \nonumber\\ &
  \cdots,  \quad
  \tilde{E}_{\pm (N^2-N)/2} =  {1 \over \sqrt{2}}(T^{N^2-3} \pm i T^{N^2-2}) ,
\end{align}
and the complex field is defined by
\begin{align}
  W^1 =& {1 \over \sqrt{2}}(V^1+ i V^2) , \quad
  W^2 = {1 \over \sqrt{2}}(V^4+ i V^5) , \quad 
  \nonumber\\ &
  \cdots,  \quad
  W^{(N^2-N)/2} = {1 \over \sqrt{2}}(V^{N^2-3}+ i V^{N^2-2})  .
\end{align}

We now calculate the double commutator as
\begin{align}
&  [H_j, [H_j, \mathscr{V}']]
\nonumber\\
 =& \sum_{j=1}^{N-1} V_k [H_j, [H_j, H_k]] + \sum_{\alpha=1}^{(N^2-N)/2} (W^*{}^{\alpha} [H_j, [H_j,  \tilde{E}_{\alpha}]] + W^{\alpha} [H_j, [H_j,  \tilde{E}_{-\alpha}]]) 
 \nonumber\\
 =& 
  \sum_{\alpha=1}^{(N^2-N)/2} (W^*{}^{\alpha} [H_j, \alpha_j \tilde{E}_{\alpha}] + W^{\alpha} [H_j, -\alpha_j   \tilde{E}_{-\alpha}]) 
 \nonumber\\
 =& 
   \alpha_j  \alpha_j \sum_{\alpha=1}^{(N^2-N)/2} (W^*{}^{\alpha}  \tilde{E}_{\alpha}  + W^{\alpha}  \tilde{E}_{-\alpha} )
 .
\end{align}
On the other hand, we have 
\begin{align}
& (H_j, \mathscr{V}') 
 \nonumber\\
 =&  \sum_{k=1}^{N-1} V_k (H_j, H_k) + \sum_{\alpha=1}^{(N^2-N)/2} (W^*{}^{\alpha} (H_j, \tilde{E}_{\alpha}) + W^{\alpha} (H_j, \tilde{E}_{-\alpha}))  
 \nonumber\\
 =&   V^j  ,
\end{align}
since 
$(H_j, \tilde{E}_{\alpha})={\rm tr}(H_j \tilde{E}_{\alpha})=0$. 
Thus the RHS of (\ref{idd}) reduces to 
\begin{align}
 \sum_{j=1}^{N-1} V_j H_j  + \sum_{j=1}^{N-1} \alpha_j  \alpha_j \sum_{\alpha=1}^{(N^2-N)/2} (W^*{}^{\alpha}  \tilde{E}_{\alpha}  + W^{\alpha}  \tilde{E}_{-\alpha} ) .
\end{align}
This is equal to the Cartan decomposition of $\mathscr{V}$ itself, since 
\begin{equation}
\sum_{j=1}^{N-1} \alpha_j  \alpha_j=1
 . 
\end{equation}

\subsection{Minimal case}\label{section:minimal-id}

Any $su(N)$ Lie algebra valued function $\mathscr{F}(x)$ is decomposed into the $\tilde{H}$-commutative part $\mathscr{F}_{\tilde{H}}$ and the remaining part $\mathscr{F}_{G/\tilde{H}}$ as 
\begin{align}
  \mathscr{F} 
=  \mathscr{F}_{\tilde{H}} + \mathscr{F}_{G/\tilde{H}} 
, 
\quad 
   \mathscr{F}_{\tilde{H}}  =   \tilde{\mathscr{F}} + \bm{h} (\bm{h},\mathscr{F}) 
, \quad
\mathscr{F}_{G/\tilde{H}} =  2\frac{N-1}{N}  [\bm{h} , [\bm{h} , \mathscr{F}]]
 ,
\label{idv2}
\end{align}
where $\bm{h} (\bm{h},\mathscr{F})=2{\rm tr}(\mathscr{F} \bm{h})\bm{h} $
and we have defined the matrix $\tilde{\mathscr{F}}$  in which all the elements in both the last column and the last row are zero.
It is important to remark that
\begin{equation}
 [\tilde{\mathscr{F}},\bm{h}]=0 .
\end{equation}
This identity shows that the $\tilde{H}$-commutative part is not necessarily written in a form that is proportional to $\bm{h}$ and there are additional contributions of $\tilde{\mathscr{F}}$ to the $\tilde{H}$-commutative part. 
This forces us to use (\ref{defXL2}) for the $\tilde{H}$-commutative part.

For any $su(N)$ Lie algebra valued function $\mathscr{M}(x)$, the following identity holds:
\begin{align}
  \mathscr{V} 
= \sum_{A=1}^{N^2-1} \mathscr{V}^A T_A
=  \tilde{\mathscr{V}} + \bm{h} (\bm{h},\mathscr{V}) 
+  2\frac{N-1}{N}  [\bm{h} , [\bm{h} , \mathscr{V}]]
 ,
\label{id2}
\end{align}
where $(\bm{h},\mathscr{V}) :=2{\rm tr}(\mathscr{V} \bm{h})$
and we have defined the matrix $\tilde{\mathscr{V}}$  in which all the elements in the last column and the last row are zero:
\begin{equation}
 \tilde{\mathscr{V}} = \sum_{A=1}^{(N-1)^2-1} \mathscr{V}^A T_A
= \sum_{A=1}^{(N-1)^2-1} (\mathscr{V},T_A) T_A 
= \sum_{A=1}^{(N-1)^2-1} 2{\rm tr}(\mathscr{V}T_A)T_A
 . 
\end{equation}
Note that $\tilde{\mathscr{V}} + \bm{h} (\bm{h},\mathscr{V})$ commutes with $\bm n$, since $[\tilde{\mathscr{V}},\bm{h}]=0$.

The Cartan decomposition for $\mathscr{V}_N \in su(N)$ is
\begin{align}
  \mathscr{V}_N 
=&   \sum_{k=1}^{N-1} V_k H_k + \sum_{\alpha=1}^{(N^2-N)/2} (W^*{}^{\alpha} \tilde{E}_{\alpha} + W^{\alpha} \tilde{E}_{-\alpha})  
\nonumber\\
=& \tilde{\mathscr{V}}_{N} + M_{N-1} H_{N-1} 
+ \sum_{\alpha=[(N-1)^2-(N-1)]/2+1}^{(N^2-N)/2} (W^*{}^{\alpha} \tilde{E}_{\alpha} + W^{\alpha} \tilde{E}_{-\alpha})  
 .
\end{align}

We now calculate the double commutator ($r=N-1$) as
\begin{align}
&  [H_r, [H_r, \mathscr{V}_N]]
\nonumber\\
 =& \sum_{j=1}^{N-1} V_k [H_r, [H_r, H_k]] + \sum_{\alpha=1}^{(N^2-N)/2} (W^*{}^{\alpha} [H_r, [H_r,  \tilde{E}_{\alpha}]] + W^{\alpha} [H_r, [H_r,  \tilde{E}_{-\alpha}]]) 
 \nonumber\\
 =& 
  \sum_{\alpha=[(N-1)^2-(N-1)]/2+1}^{(N^2-N)/2} (W^*{}^{\alpha} [H_r, \alpha_r \tilde{E}_{\alpha}] + W^{\alpha} [H_r, -\alpha_r   \tilde{E}_{-\alpha}]) 
 \nonumber\\
 =& 
   \alpha_r  \alpha_r \sum_{\alpha=[(N-1)^2-(N-1)]/2+1}^{(N^2-N)/2} (W^*{}^{\alpha}  \tilde{E}_{\alpha}  + W^{\alpha}  \tilde{E}_{-\alpha} )
 .
\end{align}
On the other hand, we have 
\begin{align}
& (H_r, \mathscr{V}_N) 
 \nonumber\\
 =&  \sum_{k=1}^{N-1} V_k (H_r, H_k) + \sum_{\alpha=1}^{(N^2-N)/2} (W^*{}^{\alpha} (H_r, \tilde{E}_{\alpha}) + W^{\alpha} (H_r, \tilde{E}_{-\alpha}))  
 \nonumber\\
 =&  V_r  ,
\end{align}
since 
$(H_j, \tilde{E}_{\alpha})={\rm tr}(H_j \tilde{E}_{\alpha})=0$. 

For any Lie algebra valued function $\mathscr{V}_N(x)$, we obtain the identity 
\begin{align}
  \mathscr{V}_N 
=& \sum_{A=1}^{N^2-1} \mathscr{V}_A T_A
\nonumber\\
=&  \tilde{\mathscr{V}_N} + (\mathscr{V}_N,H_r) H_r 
+  \frac{1}{\alpha_r^2}  [H_r , [H_r , \mathscr{V}_N]]
\nonumber\\
=&  \tilde{\mathscr{V}_N} + (\mathscr{V}_N,H_r) H_r 
+  \frac{2(N-1)}{N}  [H_r , [H_r , \mathscr{V}_N]]
 .
\label{id4}
\end{align}

\section{More $SU(3)$ cases}\label{appendix:moreSU3}

Recall that ${\bf n}_3$ and ${\bf n}_8$ are representatives of the maximal case and minimal case, respectively, which are distinguished by the  value  of $\vartheta$.  To take into account all the degrees of freedom of ${\bf n}$, it is necessary to introduce another commutative and orthogonal unit vector ${\bf n}'$ in addition to ${\bf n}$, i.e., ${\bf n}\cdot{\bf n}^\prime  =0$, since $\{ {\bf n}, {\bf n}^\prime\}$ constitutes the complete set for   bases of the maximal Abelian subgroup $U(1)\times U(1)$.
\footnote{
This does not necessarily mean that the analogous relationship for  $\{{\bf n}_3,{\bf n}_8\}$ also  holds for $\{ {\bf n}, {\bf n}^\prime\}$.
}
Such an example of ${\bf n}^\prime$ for ${\bf n}$ is given   by
\begin{align}
 {\bf n}
 &= \cos\vartheta {\bf n}_3
   +\sin\vartheta {\bf n}_8 ,
   \\
 {\bf n}^\prime
 &=\cos(\vartheta+\pi/2)  {\bf n}_3
   +\sin(\vartheta+\pi/2)  {\bf n}_8
   \nonumber\\
 &=-\sin\vartheta  {\bf n}_3
   +\cos\vartheta  {\bf n}_8
 ,
\end{align}
which has the matrix representation 
\begin{equation}
\left(
\begin{array}{c}
  {\bf n} \cr
  {\bf n}^\prime \cr
\end{array}
\right)
=
\left(
\begin{array}{cc}
\cos\vartheta & \sin\vartheta \cr
-\sin\vartheta & \cos\vartheta \cr
\end{array}
\right)
\left(
\begin{array}{c}
 {\bf n}_3 \cr
 {\bf n}_8 \cr
\end{array}
\right).
\end{equation}

In fact, the set $\{ {\bf n},{\bf n}^\prime\}$ satisfies the key relation 
\begin{equation}
\mathbf{V}
  ={\bf n} (\mathbf{V}\cdot{\bf n})
    + {\bf n}^\prime (\mathbf{V}\cdot{\bf n}^\prime) 
+ {\bf n}\times(\mathbf{V}\times{\bf n})
 + {\bf n}^\prime\times(\mathbf{V}\times{\bf n}^\prime)
 \label{cid}
 ,
\end{equation}
which is derived by combining 
\begin{align}
{\bf n}\times(\mathbf{V}\times{\bf n})
+{\bf n}^\prime\times\mathbf{V} \times {\bf n}^\prime)
 ={\bf n}_3\times(\mathbf{V}\times {\bf n}_3)
   +{\bf n}_8\times(\mathbf{V}\times {\bf n}_8)
  ,
\end{align}
and 
\begin{align}
(\mathbf{V}\cdot{\bf n}){\bf n}
+(\mathbf{V}\cdot{\bf n}^\prime){\bf n}^\prime
 =(\mathbf{V}\cdot{\bf n}_3){\bf n}_3
  +(\mathbf{V}\cdot{\bf n}_8){\bf n}_8
 .
\end{align}
Thus, we obtain the decomposition in the vector form as
\begin{align}
\mathbf A_\mu
 &=\mathbf V_\mu
  +\mathbf X_\mu 
  =\mathbf C_\mu
  +\mathbf B_\mu
  +\mathbf X_\mu
 ,
\nonumber\\
\mathbf C_\mu
 &=({\bf n}\cdot\mathbf A_\mu){\bf n}
    +({\bf n}^\prime\cdot\mathbf A_\mu){\bf n}^\prime,
    \nonumber\\
\mathbf B_\mu
 &=g^{-1} \partial_\mu{\bf n}\times{\bf n}
    +g^{-1} \partial_\mu{\bf n}^\prime\times{\bf n}^\prime,
    \nonumber\\
\mathbf X_\mu
 &=g^{-1} {\bf n}\times D_\mu[\mathbf A]{\bf n}
    + g^{-1} {\bf n}^\prime\times D_\mu[\mathbf A]{\bf n}^\prime
  .
\end{align}
This decomposition is obtained by solving the following defining equations: (a) the covariant constantness of ${\bf n}$ and ${\bf n}^\prime$ in the background $\mathbf V_\mu$:
\begin{align}
  0 = D_\mu[\mathbf{V}] \bm{n}_s(x) 
:=\partial_\mu {\bf n}_s(x) + g  \mathbf{V}_\mu(x) \times {\bf n}_s(x) 
\quad ({\bf n}_s={\bf n},{\bf n}^\prime) 
 ,
\label{defVL3}
\end{align}
and (b) the orthogonality of $\bm n$ and $\bm n^\prime$ to $\mathbf X_\mu$:
\begin{align}
   0 = {\bf n}_s(x) \cdot \mathbf{X}_\mu(x)   := {\bf n}_s^A(x)  X_\mu^A(x)  
\quad ({\bf n}_s={\bf n}, {\bf n}^\prime) 
\label{defXL33}
 . 
\end{align}
These defining equations can be solved using the completeness identity (\ref{cid}). 
It is easy to see that we can obtain the same reduction condition as that given in the text. 

We consider the determinant of the color field $\bm n$ as a function of $\vartheta$, which is invariant under the similarity transformation (adjoint rotation) $\bm n \rightarrow U^\dagger \bm n U$.  Therefore, it can be a character for classification. 
The minimal case, i.e., $\vartheta=\frac{1}{6}\pi,\frac{1}{2}\pi,\frac{5}{6}\pi,\frac{7}{6}\pi,\frac{3}{2}\pi,\frac{11}{6}\pi$ is realized if and only if $|\det({\bm n})|$ takes the maximum value, i.e., $\det({\bm n})=\pm 1$: 
\begin{align}
\det({\bm n})
  =-\frac1{3\sqrt3}\sin\vartheta
    \left(\sin\vartheta-\frac{\sqrt3}2\right)
    \left(\sin\vartheta+\frac{\sqrt3}2\right)
 =\frac1{12\sqrt3}\sin3\vartheta
 , 
\end{align}
or equivalently, if and only if
$\det( {\bm n}^\prime)=0$, since 
\begin{align}
\det( {\bm n}^\prime)
=-\frac1{3\sqrt3}\cos\vartheta
    \left(\cos\vartheta-\frac{\sqrt3}2\right)
    \left(\cos\vartheta+\frac{\sqrt3}2\right)
=\frac1{12\sqrt3}\cos3\vartheta 
 .
\end{align}

On the other hand, $\det({\bm n})=0$ is obtained for  
$\vartheta=0,\pi/3, 2\pi/3, \pi, 4\pi/3, 5\pi/3$.
It turns out that these cases represent the conventional approach. 
In fact,  we find that $\det({\bm n}_3)=0$. 

\section*{Acknowledgments}
This work is financially supported by a Grant-in-Aid for Scientific Research (C) No. 18540251  from the Japan Society for the Promotion of Science (JSPS).


\end{document}